\documentclass[twocolumn,astrosymb,trackchanges,twocolappendix]{aastex631}

\begin{document}

\title{COBIPLANE: A Systematic Search for Compact Binary Millisecond Pulsars at Low Galactic Latitudes}

\author[0000-0003-0438-4956]{Marco Turchetta}
\affiliation{Department of Physics, Norwegian University of Science and Technology, NO-7491 Trondheim, Norway}
\author[0000-0002-0237-1636]{Manuel Linares}
\affiliation{Department of Physics, Norwegian University of Science and Technology, NO-7491 Trondheim, Norway}
\affiliation{Departament de F{\'i}sica, EEBE, Universitat Polit{\`e}cnica de Catalunya, Av. Eduard Maristany 16, E-08019 Barcelona, Spain}
\author{Karri Koljonen}
\affiliation{Department of Physics, Norwegian University of Science and Technology, NO-7491 Trondheim, Norway}
\author[0000-0003-2446-8882]{Paulo A. Miles-P\'aez}
\affiliation{Centro de Astrobiolog\'ia, CSIC-INTA, Camino Bajo del Castillo s/n, 28692 Villanueva de la Ca\~nada, Madrid, Spain}
\author[0000-0001-6841-0725]{Jordan A. Simpson}
\affiliation{Department of Physics, Norwegian University of Science and Technology, NO-7491 Trondheim, Norway}





\begin{abstract}
We present the main results obtained from the COmpact BInary Pulsar search in the low-LAtitude NEighborhood (COBIPLANE), an optical photometric survey designed to find new `spider' binary millisecond pulsars. We conducted observations targeting $30$ unidentified sources from the 4FGL-DR3 \textit{Fermi Large Area Telescope} (\textit{Fermi}-LAT) catalog, selected for their pulsar-like $\gamma$-ray properties. Extending to Galactic latitudes as low as $\pm3^{\circ}$, this survey reaches closer to the Galactic plane than its predecessor survey, the COmpact BInary PULsar SEarch (COBIPULSE). We report the discovery of five optical variables coincident with the localizations of 4FGL~J0821.5$-$1436, 4FGL~J1517.9$-$5233, 4FGL~J1639.3$-$5146, 4FGL~J1748.8$-$3915, and 4FGL~J2056.4$+$3142. These systems show optical flux modulation at the presumed orbital periods of $0.41576(6) \ \mathrm{d}$, $0.305(2) \ \mathrm{d}$, $0.204(7) \ \mathrm{d}$, $0.3(2) \ \mathrm{d}$, and $0.4395(1) \ \mathrm{d}$, respectively, and photometric temperatures of $4000$–$6000 \ \mathrm{K}$, consistent with the companion stars of `redback' subtype of spider pulsar binaries. Based on their optical light curve shapes and X-ray properties characteristic for spider systems---namely, a luminosity of $1.5 \times 10^{32} \ (D / 3.9 \ \mathrm{kpc})^2 \ \mathrm{erg} \ \mathrm{s}^{-1}$ ($0.3$–$10 \ \mathrm{keV}$) for 4FGL~J1748.8$-$3915, and upper limits of $\sim10^{31}$--$10^{33} \ \mathrm{erg} \ \mathrm{s}^{-1}$ ($0.2$–$12 \ \mathrm{keV}$) for the others---we classify these sources as new spider candidate systems.

\end{abstract}



\section{Introduction} \label{sec:intro}
Millisecond pulsars (MSPs) are old neutron stars that have been spun-up to ms periods through the transfer of mass and angular momentum from a low-mass companion star \citep{1991PhR...203....1B}. When the mass transfer rate declines, the pulsar's magnetic pressure sweeps away the accretion disk, ending the low-mass X-ray binary (LMXB) phase and activating the system as a rotation-powered, or `radio', MSP \citep{2013A&A...558A..39T}.

Over $20\%$ of known radio MSPs are hosted in compact binaries with orbital periods shorter than one day \citep[according to the ATNF Pulsar Catalog\footnote{\url{https://www.atnf.csiro.au/research/pulsar/psrcat/}},][]{2005AJ....129.1993M}. Their close orbital separations enable the pulsar’s high-energy particle wind to irradiate and gradually consume the companion star. Their destructive behavior following the accretion-driven `mating' phase has earned these systems cannibalistic spider nicknames: \textit{black widows} (BWs), which host extremely low-mass companions \citep[$\lesssim 0.1 \ \mathrm{M}_{\sun}$; e.g.,][]{1988Natur.333..237F,2019ApJ...883..108D,2023ApJ...942....6K}, and \textit{redbacks} (RBs), with companion masses in the range of $0.1$–$0.7 \ \mathrm{M}_{\sun}$ \citep[e.g.,][]{2009Sci...324.1411A,2016ApJ...823..105D,2025PASA...42..139P}.
Additionally, two spider systems with giant companions and orbital periods of $5$–$10 \ \mathrm{d}$, known as \textit{huntsmen},  have been discovered \citep{2015ApJ...804L..12S,2016ApJ...820....6C,2025ApJ...980..124S}.

Among spider systems, three rare RBs known as transitional MSPs have been observed switching between the disk/LMXB and radio pulsar states on timescales of weeks to months \citep{2009Sci...324.1411A,2013Natur.501..517P,2014MNRAS.441.1825B}. These rapid transitions provide direct observational evidence for the `recycling' scenario, confirming that compact binary MSPs are formed via sustained accretion over Gyr timescales \citep{1982CSci...51.1096R}. Thus, spiders constitute a promising environment for hosting the most massive neutron stars, with the heaviest reaching $\sim2.3 \ \mathrm{M}_{\sun}$ in these systems\footnote{We note that the neutron star mass estimate can be affected by systematics from the offset between the companion's center of light and center of mass in irradiated systems, poorly constrained temperatures near inferior conjunction in faint systems, and uncertainties in orbital inclination.} \citep{2018ApJ...859...54L,2022ApJ...934L..17R}.

Spider MSPs frequently show eclipses of their radio pulsations over a wide range of orbital phases, caused by absorption in the outflowing material from the companion star \citep{2001ApJ...561L..89D,2013IAUS..291..127R}. This makes them particularly challenging to detect in blind radio surveys without prior knowledge of their sky locations and orbital parameters. Since its launch, the \textit{Fermi}-LAT has played a central role in the discovery of 62 confirmed spiders (see \citealt{2023ApJ...958..191S} and references therein) out of the 84 currently known in our Galaxy (see SpiderCat version 1.9.1\footnote{\url{https://astro.phys.ntnu.no/SpiderCAT}} for an updated compilation of the spider population; \citealt{2025ApJ...994....8K}). These discoveries were enabled not only by the bright $\gamma$-ray emission of spiders, but also by targeted radio searches of previously unassociated \textit{Fermi} sources \citep{2023ApJ...958..191S}. An additional 31 systems have been identified as spider candidates based on their multi-wavelength properties (e.g., \citealt{2015ApJ...803L..27B}; \citealt{2021ApJ...911...92L}; \citealt{2023MNRAS.524.3020K}; \citealt{2024ApJ...977...65T}; \citealt{2025ApJ...978..106L}; see also \citealt{2025ApJ...994....8K} for a complete compilation of candidates).

The optical emission from spider systems is dominated by the companion star's flux and exhibits clear orbital modulation \citep[e.g.,][]{2014ApJ...795..115L,2015ApJ...812L..24R}. When the companion is strongly irradiated by the pulsar wind, its optical light curve shows a single flux maximum per orbit, with peak-to-peak amplitudes $\gtrsim1 \ \mathrm{mag}$ \citep[e.g.,][]{2011ApJ...743L..26R,2013ApJ...769..108B,2023MNRAS.520.2217M}. In contrast, systems with weak or no irradiation are dominated by ellipsoidal modulation, with two peaks per orbit and smaller amplitudes ($\simeq0.3 \ \mathrm{mag}$), primarily due to tidal distortion of the companion \citep[e.g.,][]{2016ApJ...816...74B,2024ApJ...973..121S}. The degree of irradiation depends largely on the companion's intrinsic or `base' temperature, $T_\mathrm{base}$. All known BWs---typically with $T_\mathrm{base}\simeq1000$--$3000 \ \mathrm{K}$---show strongly irradiated light curves, featuring a single bright peak and sharp minima per orbit. In comparison, roughly half of the RB population, which host hotter companions ($T_\mathrm{base}\simeq4000$--$6000 \ \mathrm{K}$), show sinusoidal, double-peaked light curves with little or no evidence of irradiation \citep{2023MNRAS.525.2565T}.

Despite recent efforts, which have established pulsars as the most common class of Galactic $\gamma$-ray emitters \citep{2023ApJ...958..191S}, over $2100$ sources in the latest 4FGL \textit{Fermi}-LAT catalog remain unidentified \citep{2020ApJS..247...33A,2023arXiv230712546B}.
Optical observations followed by radio pulsation searches of these unidentified objects with pulsar-like $\gamma$-ray characteristics provide a promising avenue for discovering new spider systems. Indeed, many radio-obscured spiders have only been detected as MSPs after the identification of their variable optical counterparts, pointing deep radio follow-ups at their precise sky positions (see, e.g., \citealt{2017MNRAS.465.4602L} and \citealt{2023ApJ...952..150P} for PSR~J0212$+$5321, \citealt{2017MNRAS.471.2902R} and \citealt{2024MNRAS.530.4676T} for PSR~J0838$-$2827, \citealt{2023ApJ...943..103A} and \citealt{2024MNRAS.528.4337D} for PSR~J1910$-$5320).

In this paper we present the COmpact BInary Pulsar search in the low-LAtitude NEighborhood (COBIPLANE), a robotic optical photometric survey targeting $30$ unidentified $\gamma$-ray sources, selected as promising pulsar candidates from the 4FGL-DR3 \textit{Fermi}-LAT catalog \citep{2022ApJS..260...53A}. The COBIPLANE analysis framework builds on the methodology of the COBIPULSE survey \citep{2024ApJ...977...65T}---which discovered four RB MSP candidates---while extending coverage to lower Galactic latitudes and improving data sampling through consecutive nights of observations.

\section{Observations and data analysis} \label{sec:data}
\begin{figure}[t!]
\centering
\includegraphics[width=\columnwidth]{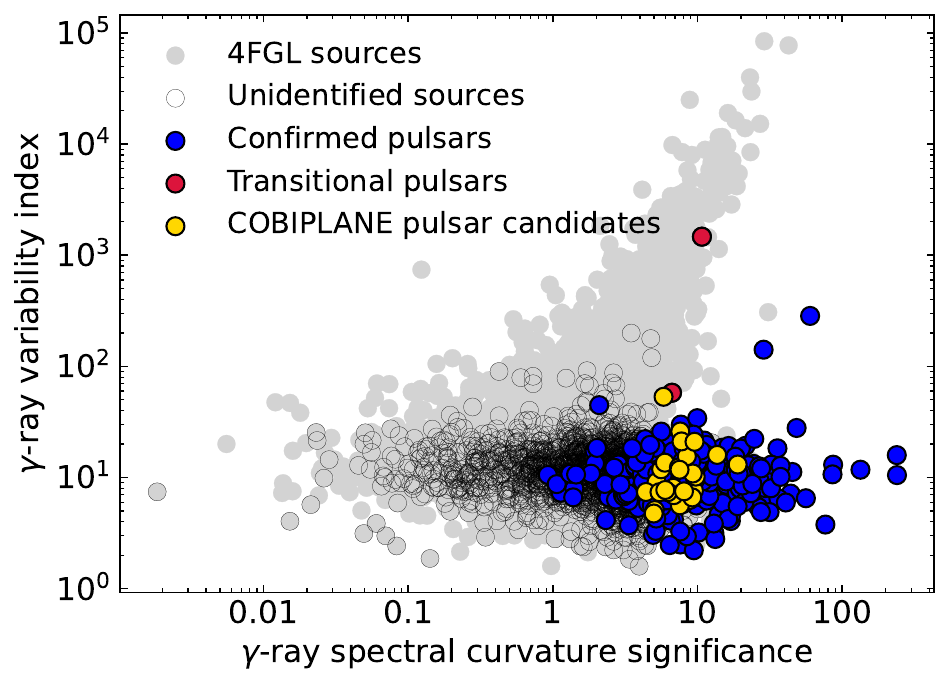}
\caption{$\gamma$-ray spectral curvature significance versus $\gamma$-ray variability index significance plot. All \textit{Fermi}-LAT 4FGL sources are shown as grey filled circles, unidentified sources as black open circles, confirmed pulsars as blue circles, the transitional MSPs PSR~J1023$+$0038 and PSR~J1227$-$4853 as red circles and pulsar candidates selected in this work as yellow circles.}
\label{fig:varvscurv}
\end{figure}
\begin{figure*}[t!]
\centering
\includegraphics[width=0.97\textwidth]{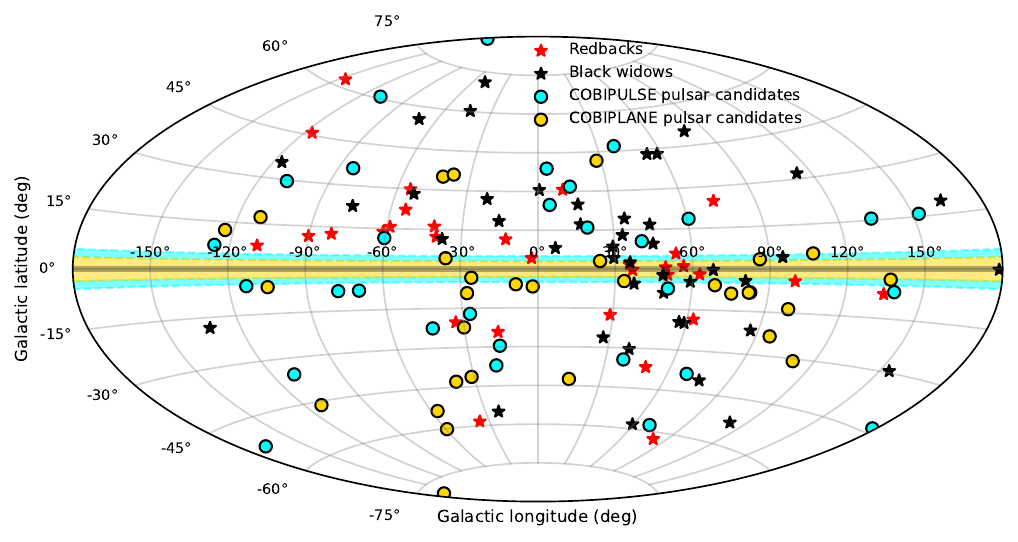}
\caption{Full-sky Aitoff projection in Galactic coordinates showing confirmed spider systems (RBs as red stars, BWs as black stars) and pulsar candidates from the COBIPULSE (cyan circles, $|b| > 5^{\circ}$) and COBIPLANE (yellow circles, $|b| > 3^{\circ}$) surveys. Color-matched bands highlight each survey's Galactic latitude cutoffs.}
\label{fig:Galmap}
\end{figure*}
\begin{table*}
\raggedright
\small
    \caption{Log of COBIPLANE observations.}
    \setlength{\tabcolsep}{4.0pt}
    \begin{tabular}{lccccc}
    \hline\hline
        Field & Telescope & Dates$^a$ & Images in \textit{g'} & Images in \textit{r'} & Images in \textit{i'}\\
        4FGL source & (diameter/instrument) & (yr: mm/dd) & ($\mathrm{nr}\times\mathrm{exp.}\ \mathrm{time}$) & ($\mathrm{nr}\times\mathrm{exp.}\ \mathrm{time}$) & ($\mathrm{nr}\times\mathrm{exp.}\ \mathrm{time}$)\\ 
        \hline
        {J0003.6$+$3059}& STELLA-1.2m/WiFSIP & 2022: 09/22, 10/18--19 & $41\times120 \ \mathrm{s}$ & $81\times120 \ \mathrm{s}$ & $41\times120 \ \mathrm{s}$\\
        {J0048.6$-$6347}& LCO-1m/Sinistro & 2022: 09/21--22 & $74\times120 \ \mathrm{s}$ & $141\times120 \ \mathrm{s}$ & $73\times120 \ \mathrm{s}$\\
        {J0059.4$-$5654}& LCO-1m/Sinistro & 2022: 09/24--27 & $77\times120 \ \mathrm{s}$ & $154\times120 \ \mathrm{s}$ & $76\times120 \ \mathrm{s}$\\
        {J0139.5$-$2228}& LCO-1m/Sinistro & 2022: 10/21--22 & $40\times120 \ \mathrm{s}$ & $80\times120 \ \mathrm{s}$ & $40\times120 \ \mathrm{s}$\\
        {J0235.3$+$5650}& STELLA-1.2m/WiFSIP & 2022: 11/16 & $20\times120 \ \mathrm{s}$ & $40\times120 \ \mathrm{s}$ & $20\times120 \ \mathrm{s}$\\
        {J0414.7$-$4300}& LCO-1m/Sinistro & 2022: 11/20, 11/22, 11/24, 12/01 & $84\times120 \ \mathrm{s}$ & $160\times120 \ \mathrm{s}$ & $80\times120 \ \mathrm{s}$\\
        {J0754.9$-$3953}& LCO-1m/Sinistro & 2023: 01/18--21 & $80\times120 \ \mathrm{s}$ & $160\times120 \ \mathrm{s}$ & $80\times120 \ \mathrm{s}$\\
        {J0821.5$-$1436}& P48-1.2m/ZTF$^b$ & 2018: 03/27--2024: 10/23 & $329\times30 \ \mathrm{s}$ & $466\times30 \ \mathrm{s}$ & --\\
        {J0906.8$-$2122}& LCO-1m/Sinistro & 2022: 02/24--27 & $80\times120 \ \mathrm{s}$ & $160\times120 \ \mathrm{s}$ & $80\times120 \ \mathrm{s}$\\
        {J1345.9$-$2612}& LCO-1m/Sinistro & 2022: 04/21--24 & $91\times120 \ \mathrm{s}$ & $181\times120 \ \mathrm{s}$ & $91\times120 \ \mathrm{s}$\\
        {J1400.0$-$2415}& LCO-1m/Sinistro & 2022: 05/03--05, 05/08 & $75\times120 \ \mathrm{s}$ & $148\times120 \ \mathrm{s}$ & $75\times120 \ \mathrm{s}$\\
        {J1517.9$-$5233}& LCO-1m/Sinistro & 2022: 05/24--27 & $80\times120 \ \mathrm{s}$ & $160\times120 \ \mathrm{s}$ & $80\times120 \ \mathrm{s}$\\
        {J1612.1$+$1407}& LCO-1m/Sinistro & 2023: 05/16--18, 05/21--22 & $62\times120 \ \mathrm{s}$ & $120\times120 \ \mathrm{s}$ & $60\times120 \ \mathrm{s}$\\
        {J1639.3$-$5146}& LCO-1m/Sinistro & 2022: 06/02--06 & $82\times120 \ \mathrm{s}$ & $161\times120 \ \mathrm{s}$ & $81\times120 \ \mathrm{s}$\\
        {J1702.7$-$5655}& LCO-1m/Sinistro & 2022: 06/07--09, 06/12 & $61\times120 \ \mathrm{s}$ & $121\times120 \ \mathrm{s}$ & $60\times120 \ \mathrm{s}$\\
        {J1748.8$-$3915}& LCO-1m/Sinistro & 2022: 06/24--27 & $62\times120 \ \mathrm{s}$ & $122\times120 \ \mathrm{s}$ & $61\times120 \ \mathrm{s}$\\
        {J1808.4$-$3358}& LCO-1m/Sinistro & 2022: 06/29, 07/01 & $40\times120 \ \mathrm{s}$ & $78\times120 \ \mathrm{s}$ & $40\times120 \ \mathrm{s}$\\
        {J1824.2$-$0621}& LCO-1m/Sinistro & 2022: 07/04, 07/06--07 & $61\times120 \ \mathrm{s}$ & $120\times120 \ \mathrm{s}$ & $61\times120 \ \mathrm{s}$\\
        {J1831.1$-$6503}& LCO-1m/Sinistro & 2023: 06/18, 06/22, 06/23, 06/25 & $61\times120 \ \mathrm{s}$ & $120\times120 \ \mathrm{s}$ & $61\times120 \ \mathrm{s}$\\
        {J1908.8$-$0131}& LCO-1m/Sinistro & 2023: 07/11--13 & $60\times120 \ \mathrm{s}$ & $120\times120 \ \mathrm{s}$ & $60\times120 \ \mathrm{s}$\\
        {J2027.0$+$2811}& LCO-1m/Sinistro & 2022: 07/27--28, 2022: 08/08--09 & $69\times120 \ \mathrm{s}$ & $136\times120 \ \mathrm{s}$ & $68\times120 \ \mathrm{s}$\\
        {J2041.1$+$4736}& LCO-1m/Sinistro & 2022: 08/10, 08/12 & $40\times120 \ \mathrm{s}$ & $80\times120 \ \mathrm{s}$ & $40\times120 \ \mathrm{s}$\\
        {J2056.4$+$3142}& LCO-1m/Sinistro & 2022: 07/29, 08/14, 08/16, 08/26 & $52\times120 \ \mathrm{s}$ & $102\times120 \ \mathrm{s}$ & $51\times120 \ \mathrm{s}$\\
        {J2112.5$-$3043}& LCO-1m/Sinistro & 2022: 08/17--18 & $40\times120 \ \mathrm{s}$ & $80\times120 \ \mathrm{s}$ & $40\times120 \ \mathrm{s}$\\
        {J2116.2$+$3701}& LCO-1m/Sinistro & 2022: 08/17--19 & $59\times120 \ \mathrm{s}$ & $120\times120 \ \mathrm{s}$ & $59\times120 \ \mathrm{s}$\\
        {J2133.1$-$6432}& LCO-1m/Sinistro & 2022: 08/23--24 & $40\times120 \ \mathrm{s}$ & $80\times120 \ \mathrm{s}$ & $40\times120 \ \mathrm{s}$\\
        {J2219.7$-$6837}& LCO-1m/Sinistro & 2022: 08/27--30 & $82\times120 \ \mathrm{s}$ & $163\times120 \ \mathrm{s}$ & $81\times120 \ \mathrm{s}$\\
        {J2220.8$+$6319}& STELLA-1.2m/WiFSIP & 2022: 09/07--08, 09/10 & $60\times120 \ \mathrm{s}$ & $120\times120 \ \mathrm{s}$ & $60\times120 \ \mathrm{s}$\\
        {J2241.2$+$4303}& STELLA-1.2m/WiFSIP & 2022: 09/14 & $10\times120 \ \mathrm{s}$ & $21\times120 \ \mathrm{s}$ & $10\times120 \ \mathrm{s}$\\
        {J2250.5$+$3305}& LCO-1m/Sinistro & 2022: 08/27--29 & $47\times120 \ \mathrm{s}$ & $94\times120 \ \mathrm{s}$ & $47\times120 \ \mathrm{s}$\\
        \hline
    \end{tabular}
    \tablenotetext{a}{Consecutive nights are listed as ranges (e.g., 09/24--27 to indicate 09/24, 09/25, 09/26, 09/27).}
    \tablenotetext{b}{Our analysis was restricted to ZTF public data for this field (see Section~\ref{subsec:ZTFdata}), as the planned STELLA/WiFSIP observations could not be completed due to maintenance on the camera's critical components.}
    \label{tab:observationlog}
\end{table*}
\subsection{Pulsar candidates from the \textit{Fermi}-4FGL catalog} \label{subsec:gammaselection}
We selected the COBIPLANE candidates by leveraging the two key properties commonly exhibited by pulsars at \textit{Fermi}-LAT $\gamma$-ray energies: a steady, non-variable emission and significant spectral curvature across the 0.1--300 GeV range \citep{Ackermann_2012}. These characteristics are quantified by the variability index and curvature significance parameters, respectively (see Eqs. (3) and (4) in \citealt{2012ApJS..199...31N}).

Accordingly, we searched the 4FGL-DR3 catalog for unidentified sources that met all of the following requirements\footnote{Our target selection and optical campaigns were conducted in 2022–2023, based on the 4FGL-DR3 catalog \citep{2022ApJS..260...53A} available in 2022, which preceded the release of 4FGL-DR4 \citep{2023arXiv230712546B}.}:
\begin{enumerate}
\item Spectral curvature significance exceeding $4\sigma$.
\item Variability index below $60$.
\item Galactic latitude $|b|>3^{\circ}$, to mitigate contamination from diffuse $\gamma$-ray emission near the Galactic plane.
\item A 95\% confidence error ellipse with semi-major axis smaller than $13'$ (matching the field of view of STELLA and LCO; see Section~\ref{subsec:optphotometry}).
\end{enumerate}
This selection yielded 30 COBIPLANE targets (listed in Table~\ref{tab:observationlog}), which occupy the lower-right region of the $\gamma$-ray curvature-variability diagram (yellow circles), alongside the majority of known pulsars (blue circles), as shown in Figure~\ref{fig:varvscurv}. The two transitional MSPs with a \textit{Fermi}-LAT association, PSR~J1023$+$0038 \citep{2009Sci...324.1411A} and PSR~J1227$-$4853 \citep{2014MNRAS.441.1825B} (red circles), exhibit variability indices of $\mathbf{\simeq 1470}$ and $\mathbf{\simeq 58}$, respectively, both higher than the main cluster of pulsars. This reflects their observed transitions between disk and radio pulsar states, during which the $\gamma$-ray emission level changes \citep{2014ApJ...790...39S,2015ApJ...806...91J,2017ApJ...836...68T}. While PSR~J1023$+$0038 stands out clearly from the pulsar cluster, the effect for PSR~J1227$-$4853 is much less pronounced, as the emission changes by factors of $\simeq10$ and $\simeq3$, respectively. Therefore, the variability threshold of 60 applied in our selection does not necessarily exclude finding new transitional MSPs in our search.

While following the same selection criteria adopted in our previous COBIPULSE survey \citep[][Section 2.1]{2024ApJ...977...65T}, we relaxed the Galactic latitude cutoff from $|b| > 5^{\circ}$ to $|b| > 3^{\circ}$ (see Figure~\ref{fig:Galmap}). This is justified by improved diffuse emission modeling and reduced systematic uncertainties in the 4FGL catalog \citep{2020ApJS..247...33A}, allowing us to probe regions closer to the Galactic plane---where the spider population is known to be concentrated \citep{2025ApJ...994....8K}.
%
%

\subsection{Observations and optical photometry} \label{subsec:optphotometry}
Our optical campaign targeted the selected \textit{Fermi}-LAT localization areas using two main facilities: the $1.2$-$\mathrm{m}$ STELLA telescope equipped with the WiFSIP camera, and the Las Cumbres Observatory (LCO) global network of $1$-$\mathrm{m}$ telescopes with Sinistro instruments. We conducted observations between 2022 February 24 and 2023 July 13, cycling through the SDSS \textit{g'}, \textit{r'}, and \textit{i'} filters each night. To optimize sampling in the \textit{r'}-band while maintaining color information, we 
cycled through a sequence of \textit{g'}-\textit{r'}-\textit{i'}-\textit{r'}, taking 2-min exposures in each filter. 
The instrumental configurations are summarized in Table~\ref{tab:observationlog}.

We carried out data reduction using the dedicated pipelines for STELLA/WiFSIP and LCO/Sinistro (see \citealt{2016SPIE.9910E..0NW,2018SPIE10707E..0KM}), which perform standard preprocessing steps including bad-pixel masking, bias subtraction, and flat-field correction. For each field, we produced deep median-combined images in all three bands (\textit{g'}, \textit{r'}, and \textit{i'}) to maximize source detection sensitivity. Representative \textit{r'}-band fields are shown in Appendix \ref{sec:appA} (Figures~\ref{fig:FoVs}--\ref{fig:FoVs_cont3}).

We performed source extraction using the \textsc{SEP}\footnote{\url{https://github.com/kbarbary/sep}} package \citep{2016zndo....159035B}, a Python implementation of \textsc{SExtractor} \citep{1996A&AS..117..393B}, with a signal-to-noise ratio threshold of $\geq2$ for detection. This yielded a survey depth of $\textit{r'}\simeq21 \ \mathrm{mag}$.
We note that some of the faintest sources detected in the median-combined images may remain undetected in individual exposures, limiting the quality of their light curves. The number of detected sources varied from $\sim500$ to $80000$ per field, depending on Galactic location and filter (Table~\ref{tab:results}).

We then carried out circular aperture photometry on all identified sources across the field applying the \textsc{SEP} routines. After testing multiple aperture radii, we determined that setting the radius to $1.2\times$ the average full-width at half maximum (FWHM) effectively minimizes sky background noise and reduces contamination from nearby stars in crowded regions. For differential photometry, we selected three independent sets of $7$--$10$ comparison stars, one for each filter (\textit{g'}, \textit{r'}, and \textit{i'}), using the \textsc{astrosource}\footnote{\url{https://github.com/zemogle/astrosource}} package \citep{2021JOSS....6.2641F}, which identifies the most stable stars in a given field. These reference stars show low variability, with rms amplitudes typically between $0.005$ and $0.02$ mag\footnote{Combining the fluxes of $N$ comparison stars improves the signal-to-noise ratio of the target light curves by approximately a factor of $\sqrt{N}$ \citep{1992PASP..104..435H}.}. To ensure consistency and avoid saturation, we restricted the reference star selection to sources with magnitudes in the range $\sim15$–$17$ mag, comparable to or brighter than our targets of interest (for reference, the brightest known spider companion, PSR~J0212$+$5321, has $\textit{r'}\simeq14.3$; \citealt{2017MNRAS.465.4602L}). 

\subsection{Variable selection and periodicity search} \label{subsec:variablesandperiods}
\begin{table*}[t!]
\raggedright
    \caption{Photometry and periodicity search results. The $5^{\mathrm{th}}$ through $8^{\mathrm{th}}$ columns list the number of \textit{Fermi}-field sources detected in the \textit{r'} band, along with those identified as photometric variables, periodic variables, and spider candidates, respectively, based on the selection criteria described in Section~\ref{subsec:variablesandperiods}.}
    \setlength{\tabcolsep}{3.0pt}
    \begin{tabular}{lccccccc}
    \hline\hline
        Field & Srcs. in \textit{g'} & Srcs. in \textit{r'} & Srcs. in \textit{i'} & \textit{Fermi}-field srcs. in \textit{r'} & Phot. variables & Per. variables & Spider candidates\\
        4FGL & (nr) & (nr) & (nr) & (nr) & (nr) & (nr) & (nr)\\ 
        \hline
        {J0003.6$+$3059}& 941 & 1136 & 766 & 200 & 95 & 1 &--\\
        {J0048.6$-$6347}& 2819 & 3487 & 3412 & 106 & 55 & -- &--\\
        {J0059.4$-$5654}& 3233 & 4401 & 3340 & 199 & 118 & -- &--\\
        {J0139.5$-$2228}& 3716 & 4426 & 4974 & 163 & 92 & -- &--\\
        {J0235.3$+$5650}& 6826 & 7839 & 2581 & 1015 & 312 & 1 &--\\
        {J0414.7$-$4300}& 4263 & 4418 & 3901 & 196 & 108 & -- &--\\
        {J0754.9$-$3953}& 20237 & 22023 & 23790 & 1259 & 636 & 2 &--\\
        {J0821.5$-$1436}\tablenotemark{a}& -- & -- & -- & 546 & 427 & 1 & 1\\
        {J0906.8$-$2122}& 4813 & 4457 & 7511 & 870 & 449 & 1 &--\\
        {J1345.9$-$2612}& 3075 & 4445 & 4791 & 244 & 118 & -- &--\\
        {J1400.0$-$2415}& 4473 & 6992 & 6512 & 181 & 95 & -- &--\\
        {J1517.9$-$5233}& 38718 & 55358 & 51825 & 1530 & 798 & 2 & 1\\
        {J1612.1$+$1407}& 3731 & 5639 & 4915 & 210 & 116 & -- &--\\
        {J1639.3$-$5146}& 34600 & 43784 & 40695 & 474 & 239 & 1 & 1\\
        {J1702.7$-$5655}\tablenotemark{b}& 16820 & 21159 & 22687 & 254 & 156 & -- &--\\
        {J1748.8$-$3915}& 38662 & 38967 & 51477 & 2381 & 1060 & 5 & 1\\
        {J1808.4$-$3358}& 67296 & 76859 & 80634 & 3924 & 1964 & 3 &--\\
        {J1824.2$-$0621}\tablenotemark{c} & 33018 & 55303 & 58019 & 903 & 380 & 1 &--\\
        {J1831.1$-$6503}& 8063 & 9571 & 8062 & 190 & 102 & -- &--\\
        {J1908.8$-$0131}& 48261 & 60833 & 60702 & 1655 & 818 & -- &--\\
        {J2027.0$+$2811}& 21912 & 25881 & 26359 & 1178 & 625 & 1 & --\\
        {J2041.1$+$4736}& 9062 & 22407 & 29663 & 251 & 127 & 1 &--\\
        {J2056.4$+$3142}& 15961 & 20504 & 21655 & 713 & 498 & 1 & 1\\
        {J2112.5$-$3043}& 1986 & 3973 & 4325 & 18 & 8 & -- &--\\
        {J2116.2$+$3701}\tablenotemark{d} & 11423 & 15037 & 18835 & 1754 & 802 & 3 &--\\
        {J2133.1$-$6432}& 2065 & 3699 & 3776 & 68 & 39 & -- & --\\
        {J2219.7$-$6837}& 5119 & 5079 & 4940 & 81 & 46 & -- &--\\
        {J2220.8$+$6319}& 532 & 924 & 634 & 53 & 34 & 1 &--\\
        {J2241.2$+$4303}& 1839 & 2702 & 1973 & 852 & 411 & 1 &--\\
        {J2250.5$+$3305}& 4121 & 5766 & 6002 & 292 & 129 & -- &--\\
        \hline
    \end{tabular}
    \tablenotetext{a}{We obtained ZTF data for the \textit{Fermi}-field sources in \textit{r'} via the IRSA light curve service in this case, relying on preprocessed images and source detections provided by the survey.}
    \tablenotetext{b}{Previously proposed as a RB candidate by \citet{2022ApJ...935....2C}, based on the detection of $\gamma$-ray modulation at a period of $\sim5.85 \ \mathrm{hr}$. In Section~\ref{subsec:J1824J2116J1702disc} we explain the absence of a variable optical counterpart in our survey.}
    \tablenotetext{c}{Already associated with PSR~J1824$-$0621, a MSP–He white dwarf binary with an orbital period of $100.9 \ \mathrm{d}$ discovered in the CRAFTS survey \citep{2018IMMag..19..112L}. The absence of an optical counterpart in our data is discussed in Section~\ref{subsec:J1824J2116J1702disc}.}
    \tablenotetext{d}{Already identified as PSR~J2116$+$3701, a young isolated radio pulsar detected by \citet{2023MNRAS.524.5132D}. We did not detect an optical counterpart in our survey (details in Section~\ref{subsec:J1824J2116J1702disc}).}
    \label{tab:results}
\end{table*}

We computed the mean differential magnitude ($\Delta m$)
and its standard deviation ($\sigma$) for each source detected in the \textit{r'}-band, which provides the densest data sampling
(see Section~\ref{subsec:optphotometry}). The resulting $\sigma$-$\Delta m$ diagrams are shown in Appendix \ref{sec:appB} (Figures~\ref{fig:sigmavsdm}--\ref{fig:sigmavsdm_cont4}). For each COBIPLANE field, we selected sources for further analysis based on the following criteria:
\begin{enumerate}
    \item The source lies within a square region centered on the \textit{Fermi}-4FGL position, with side length equal to twice the semi-major axis of the 95\% confidence ellipse. The number of such sources---reported in the $5^\mathrm{th}$ column of Table~\ref{tab:results} as ``\textit{Fermi}-field sources in \textit{r'}"---ranges from roughly $20$ to $4000$ per field.
    \item The source's magnitude standard deviation $\sigma$ exceeds the median $\sigma$ in its corresponding $\Delta m$ bin (bin width = $0.1 \ \mathrm{mag}$).
    When fewer than $5$ sources are present in a given bin, we skip this variability-based filtering due to insufficient statistics and apply only criterion (1).
\end{enumerate}
Depending on the field, between $\sim10$ and $2000$ sources passed both criteria. These are listed in the $6^\mathrm{th}$ column of Table~\ref{tab:results}, labeled as ``photometric variables".

We then ran periodicity searches on the \textit{r'}-band light curves of all sources identified in the previous step, using both the Lomb-Scargle (LS; \citealt{1976Ap&SS..39..447L}; \citealt{1982ApJ...263..835S}) and phase-dispersion minimization (PDM; \citealt{1978ApJ...224..953S}) algorithms. The period search covered the $0.02$--$2.5$ day range, which is typical for spider binary orbits \citep[see Table 2 in][]{2025ApJ...994....8K}, with a resolution of $\simeq2 \ \mathrm{min}$. Each light curve was folded at the periods corresponding to the strongest LS peak and the deepest PDM minimum (marked with solid red lines in Figure~\ref{fig:periodograms}).

To assess the significance of detected periods, we applied a Fisher randomization test \citep{1985AJ.....90.2317L}. This involves shuffling the magnitudes $m_{i}$ across their associated time stamps $t_{i}$ to break any inherent temporal structure. For each source, we generated $\sim10000$ randomized light curves and computed periodograms for each. We then determined empirical LS and PDM thresholds corresponding to a 0.1\% false-alarm probability (FAP) (shown as dashed orange lines in Figure~\ref{fig:periodograms}), equivalent to 99.9\% confidence level or $>3\sigma$ significance.

We visually inspected all periodograms and folded light curves to rule out spurious detections caused by source blending or other artifacts. Only sources with significant periodicities (FAP$<$0.1\%) and clear, repeating \textit{r'}-band features---such as sinusoidal modulations, eclipses, or pulsations---were retained. We also folded their \textit{g'} and \textit{i'}-band light curves using the best period $P_{\mathrm{best}}$ in the \textit{r'}-band. We classified a source as a ``periodic variable” if its phase-folded light curves showed consistent variability across all three filters. The number of such sources per field ranges from $0$ to $5$, as listed in the $7^\mathrm{th}$ column of Table~\ref{tab:results}. This final filtering step helped remove flaring stars, planetary transits, and other various contaminants.

To estimate uncertainties on the derived periods, we generated $\sim1000$ perturbed light curves for each case by adding Gaussian noise to the original data. For each point in the light curve, we sampled a new magnitude $m'_{i}$ from a normal distribution centered at $m_{i}$ with standard deviation $\sigma_{m_{i}}$. We repeated the period search on each of these mock light curves, recording the period with the highest power from the LS or PDM method. The standard deviation of this distribution, $\sigma_{P_{\mathrm{best}}}$, was adopted as the uncertainty on the original period.

\subsection{Supplementary data from the ZTF optical survey} \label{subsec:ZTFdata}
To extend the light curves of northern sky sources previously identified as “periodic variables,” we incorporated data from the latest public release of the Zwicky Transient Facility (ZTF) survey \citep[March 2018–October 2024;][]{2019PASP..131a8002B}, obtained with the 48-inch Schmidt telescope at Palomar Observatory.

For 16 of our fields located within the ZTF sky coverage (declination $\delta > -31^{\circ}$), we retrieved \textit{g'} and \textit{r'}-band photometry from the IRSA light curve service\footnote{\url{https://irsa.ipac.caltech.edu/docs/program_interface/ztf_lightcurve_api.html}}. The ZTF photometry is based on circular apertures with radii equal to $0.5\times$ the PSF FWHM \citep{2019PASP..131a8002B}. We analyzed these light curves using the same periodicity search techniques described in Section~\ref{subsec:variablesandperiods}.

The extended temporal coverage provided by ZTF, which far exceeds that of COBIPLANE, enabled us to refine the photometric periods, particularly for systems with incomplete phase coverage in our observations (see, e.g., Section~\ref{subsec:J2056res}). In the case of 4FGL~J0821.5$-$1436, our analysis was solely based on ZTF data (details in Section~\ref{subsec:J0821res}), as the corresponding STELLA/WiFSIP observations were canceled due to servicing of critical camera components.

\subsection{Classification of spider candidates} \label{subsec:spiderclass}
We adopt the following multi-wavelength approach to classify a periodic variable (see Section~\ref{subsec:variablesandperiods}) as a spider candidate:
\begin{enumerate}
    \item \textit{Unclassified}: The source has no prior firm identification and no published optical light curve available in the literature.
    \item \textit{$\gamma$-ray association}: The source lies within a square centered on the corresponding \textit{Fermi}-4FGL position, with side lengths equal to twice the semi-major axis of the 95\% confidence ellipse. This conservative criterion accounts for potential positional uncertainties arising from diffuse background modeling in crowded sky regions near the Galactic plane \citep{2025A&A...695A.187M}.
    \item \textit{Optical variability}: The light curves show either smooth, sinusoidal variations with peak-to-peak amplitudes of $\simeq0.3 \ \mathrm{mag}$, broad minima and no significant color change, as observed from spiders in the ellipsoidal modulation regime; larger modulations of $\gtrsim1 \ \mathrm{mag}$ with sharp flux minima and color variations peaking near the light curve maximum, as expected from irradiated companions (Section~\ref{sec:intro}); or moderate amplitudes of $\simeq0.5$--$0.6 \ \mathrm{mag}$ with a broad flat maximum, typical of spiders in the intermediate irradiation regime \citep{2025MNRAS.538..380T}. We consider only sources with peak-to-peak amplitudes $\geq0.1 \ \mathrm{mag}$---a conservative threshold near the lowest values observed in spiders---to accommodate low-inclination systems \citep[e.g.,][]{beronya23,2025MNRAS.536.2169S}.
\end{enumerate}
Spider candidate classifications are reported in the $8^{\mathrm{th}}$ column of Table~\ref{tab:results}.

In addition, we searched for known X-ray sources within a $8''$ radius of each optical source using the latest catalogs from \textit{Chandra}, \textit{eROSITA}, \textit{Swift}, and \textit{XMM-Newton}, \citep[][respectively]{2024ApJS..274...22E,2024A&A...682A..34M,2020ApJS..247...54E,2020A&A...641A.136W}. The detection of an X-ray counterpart strengthens the spider classification, as many of them (over $50\%$) are relatively bright X-ray emitters \citep[$L_{\mathrm{X}}\simeq10^{30}$--$10^{33} \ \mathrm{erg} \ \mathrm{s}^{-1}$; see e.g.,][]{2011ApJ...742...97B,2014ApJ...783...69G,2018ApJ...861...89A,2018ApJ...866...83S}. This emission typically comes from the intrabinary shock region, where the pulsar’s relativistic wind collides with the companion’s outflow \citep[e.g.,][]{2018ApJ...869..120W} accelerating particles to relativistic speeds and producing high-energy synchrotron radiation \citep[e.g.,][]{2014ApJ...783...69G,2025MNRAS.tmp..266C}. Additional X-ray emission can also arise from thermal radiation at the polar caps, heated by particles returning from the outer magnetospheric gaps \citep{2001ApJ...556..987H}.
However, the non-uniform depth of X-ray coverage across different fields and distance uncertainties imply that the absence of an X-ray counterpart does not rule out a source as a spider candidate.

\section{Results} \label{sec:results}
\begin{table*}[t!]
\raggedright
    \caption{Spider candidate properties. The columns list \textit{Fermi} source name, optical coordinates, \textit{g'}, \textit{r'}, and \textit{i'}-band magnitudes ranges and best estimate of the photometric period $P_{\mathrm{best}}$. The period uncertainties, reported in brackets, are statistical and do not account for possible aliasing effects.}
    \setlength{\tabcolsep}{7.pt}
    \begin{tabular}{lccccccc}
    \hline\hline
        Name & R.A. (J2000)\tablenotemark{a} & Decl. (J2000)\tablenotemark{a} & Error radius\tablenotemark{b} & \textit{g'} band & \textit{r'} band & \textit{i'} band & $P_{\mathrm{best}}$\\
        4FGL & (h:m:s) & ($^{\circ}$: $'$: $''$) & ($''$) & (mag) & (mag) & (mag) & (d)\\ 
        \hline
    \hline
        J0821.5$-$1436 & 08:21:43.66 & $-$14:37:48.0 & 1.0 & [15.7-15.9] & [15.3-15.6] & [15.2-15.4] & 0.20788(3)\\
        J1517.9$-$5233 & 15:17:42.96 & $-$52:32:20.3 & 0.6 & [19.9-20.2] & [18.9-19.2] & [18.3-18.6] & 0.153(1)\\
        J1639.3$-$5146 & 16:39:20.74 & $-$51:45:03.8 & 0.8 & [17.9-18.5] & [17.2-17.6] & [16.6-17.1] & 0.204(7)\\
        J1748.8$-$3915 & 17:48:53.75 & $-$39:17:44.8 & 1.0 & [19.2-19.5] & [18.6-19.0] & [18.2-18.6] & 0.3(2)\\
        J2056.4$+$3142 & 20:56:34.61 & $+$31:42:37.0 & 0.9 & [17.9-18.1] & [17.3-17.5] & [17.0-17.2] & 0.21976(6)\\
        \hline
    \end{tabular}
    \tablenotetext{a}{The equatorial coordinates have been obtained using our astrometry-corrected combined image in the \textit{r'}-band of the corresponding field of view.}
    \tablenotetext{b}{The error radius on the optical location has been estimated as FWHM/2 of the corresponding source profile.}
    \label{tab:candresults}
\end{table*}
\begin{figure*}[t!]
\gridline{\fig{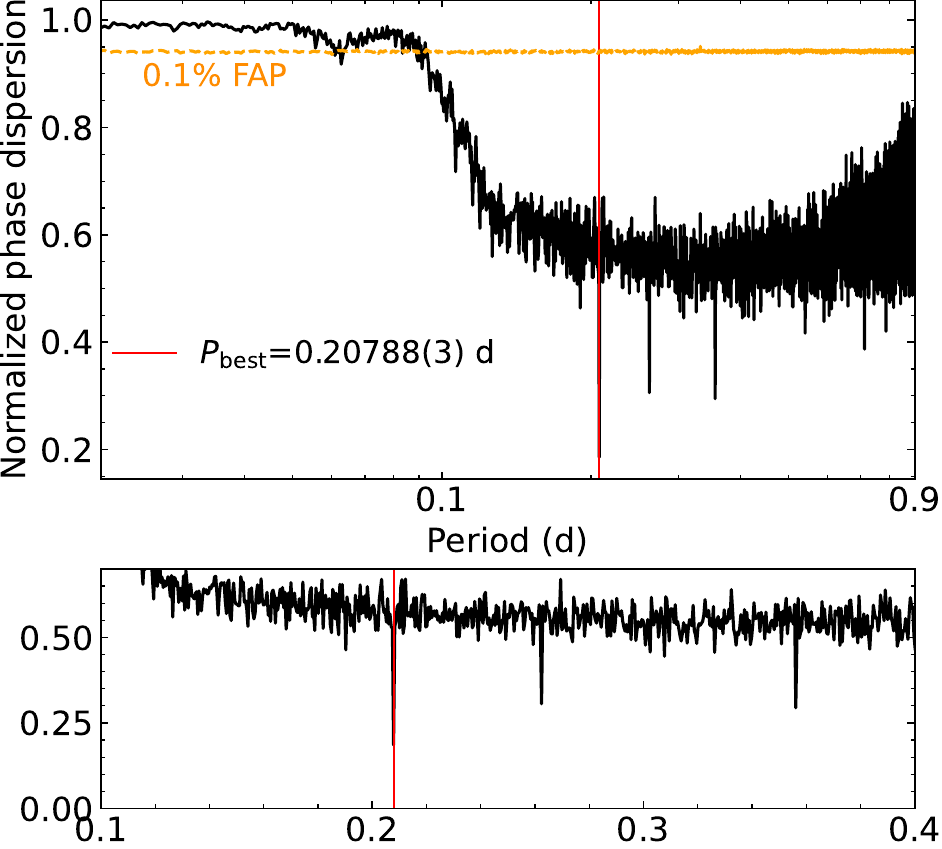}{0.33\textwidth}{(a) 4FGL~J0821.5$-$1436 -- P48/ZTF}
          \fig{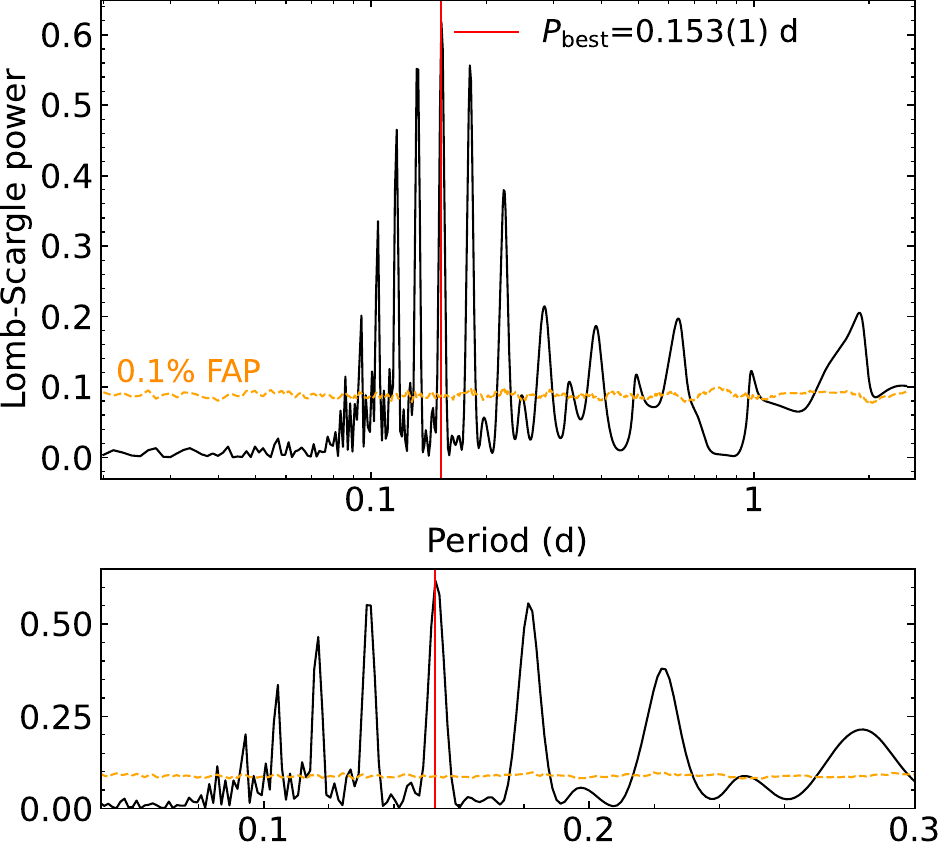}{0.33\textwidth}{(b) 4FGL~J1517.9$-$5233 -- LCO/Sinistro}
          \fig{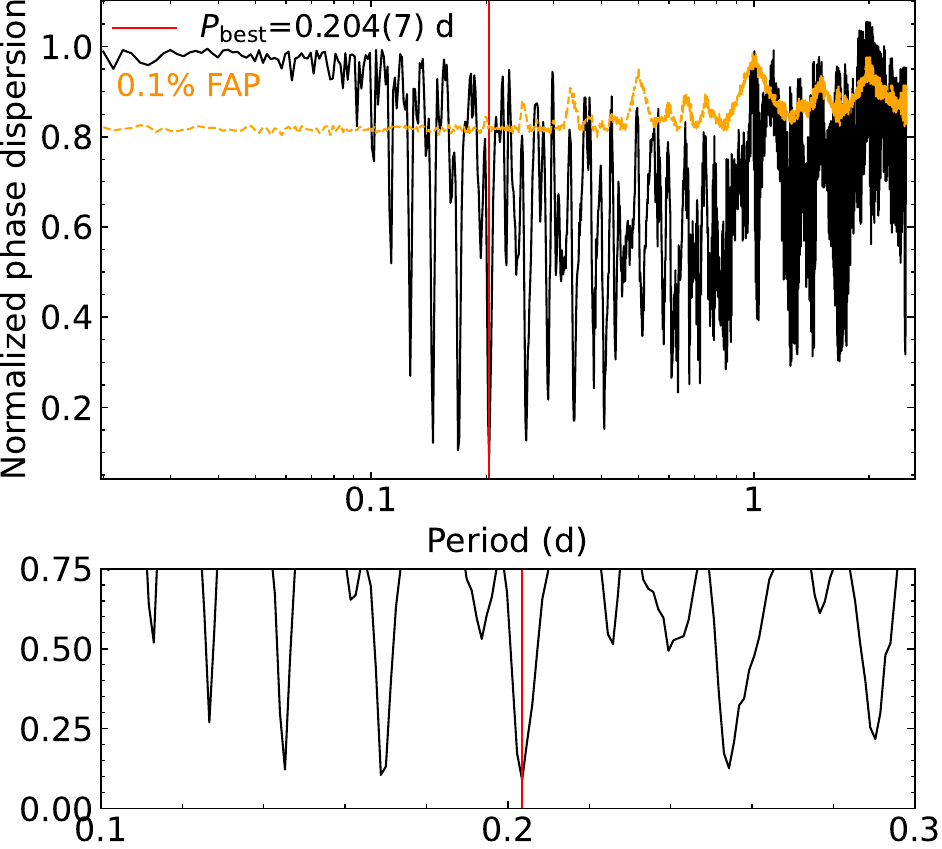}{0.33\textwidth}{(c) 4FGL~J1639.3$-$5146 -- LCO/Sinistro}}
\gridline{\fig{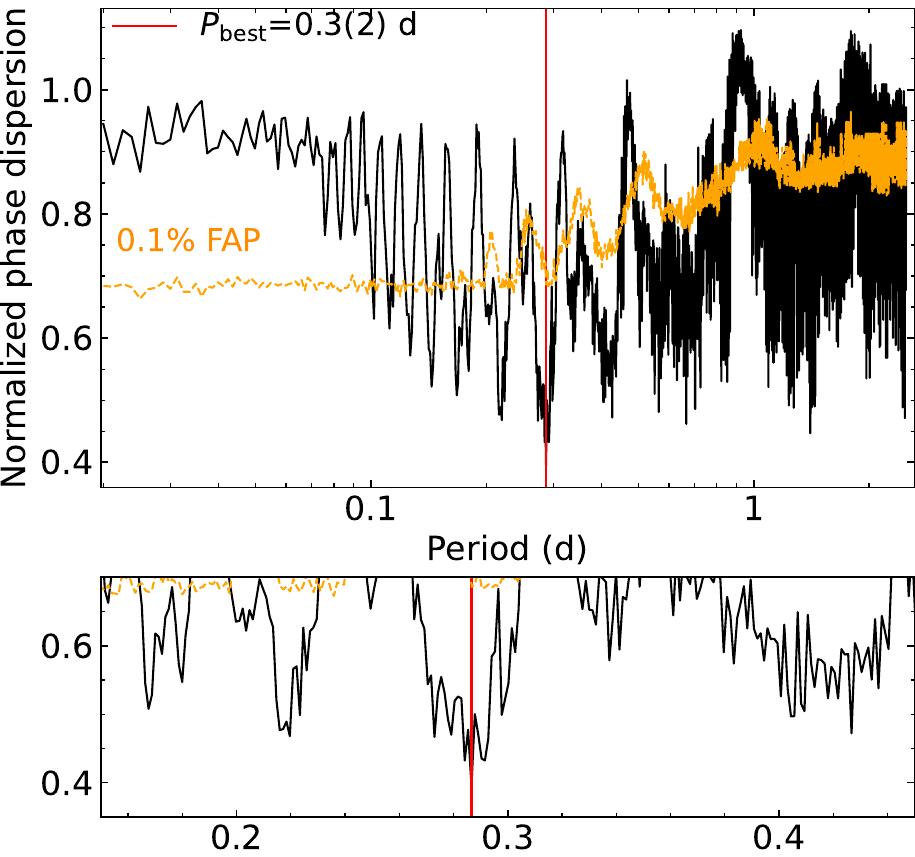}{0.33\textwidth}{(d) 4FGL~J1748.8$-$3915 -- LCO/Sinistro}
          \fig{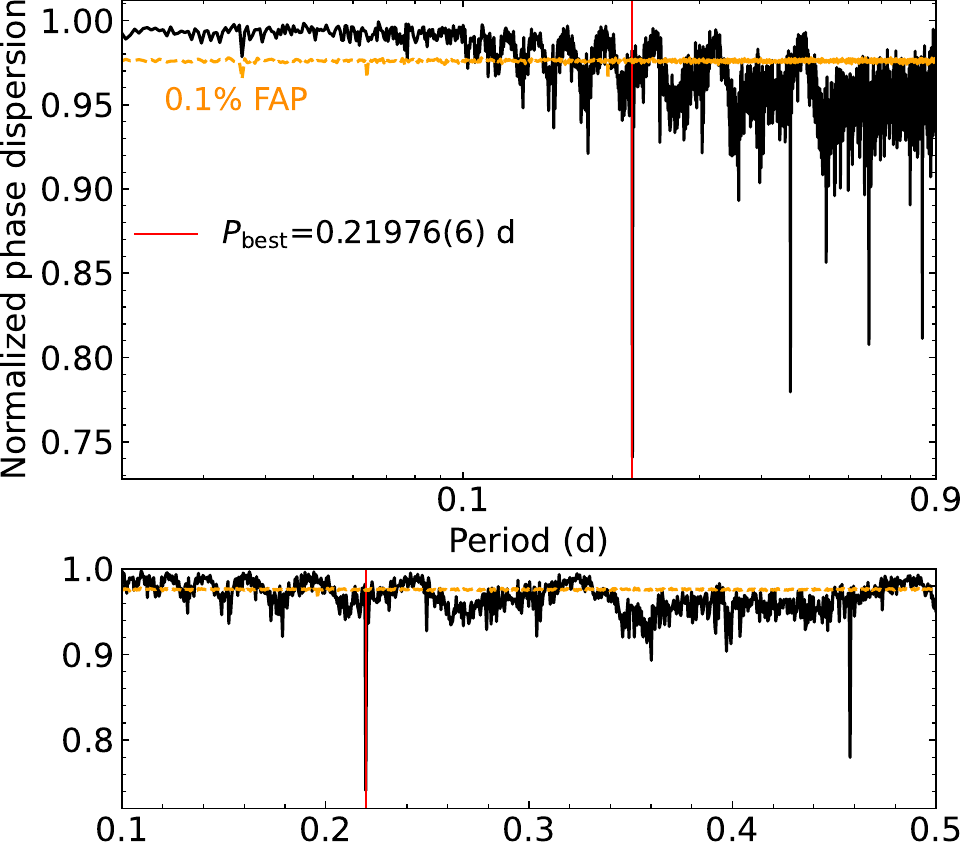}{0.35\textwidth}{(e) 4FGL~J2056.4$+$3142 -- P48/ZTF}}
\caption{LS and PDM periodograms of the \textit{r'}-band light curves for each spider candidate, based on observations from the respective telescopes. The identified photometric periods $P_{\mathrm{best}}$ are marked with solid red lines, while the 0.1\% FAP thresholds are shown as dashed orange lines. A zoom-in around the detected period is shown in the bottom panels for clarity.}
\label{fig:periodograms}
\end{figure*}

We summarize the main results of our analysis for each \textit{Fermi}-4FGL pulsar candidate in Table~\ref{tab:results}. This wide-field optical survey led to the identification of 26 periodic variables (as defined in Section~\ref{subsec:variablesandperiods}). Using the classification criteria described in Section~\ref{subsec:spiderclass}, we identified five spider candidates associated with 4FGL~J0821.5$-$1436, 4FGL~J1517.9$-$5233, 4FGL~J1639.3$-$5146, 4FGL~J1748.8$-$3915 and 4FGL~J2056.4$+$3142 (see Sections~\ref{subsec:J0821res}–-\ref{subsec:J2056res} for details). Their key properties are listed in Table~\ref{tab:candresults}, including the photometric periods, $P_{\mathrm{best}}$, as inferred from the corresponding periodograms shown in Figure~\ref{fig:periodograms}. As discussed in Section~\ref{sec:discussion}, the photometric period $P_{\mathrm{best}}$ corresponds to the presumed orbital period $P_{\mathrm{orb}}$ of the spider candidate only when the optical emission is dominated by irradiation modulation. If irradiation is weak or absent, ellipsoidal modulation prevails and the presumed orbit is then $P_{\mathrm{orb}}=2\times P_{\mathrm{best}}$. The zoomed-in fields of view and optical light curves folded on $P_{\mathrm{best}}$ of our spider candidates are presented in the left and right panels of Figures~\ref{fig:J0821_FoV&lc}--\ref{fig:J2056_FoV&lc}, respectively. The other 21 periodic variables identified in this survey---classified as eclipsing binaries, pulsating stars, or W UMa binaries---are listed in Appendix \ref{sec:appC} (Table~\ref{tab:periodicresults}), with classifications based on cross-matches with the
\textit{ATLAS} \citep{2018AJ....156..241H} and \textit{Gaia} Data Release 3 \citep[DR3;][]{refId0,2022gdr3.reptE..10R} catalogs, and their light curves are reported in (Figures~\ref{fig:pervarlightcurves}--\ref{fig:pervarlightcurves_cont3}). For a general overview of these variable types, see \citet{Chambliss_1992}. 

\subsection{4FGL~J0821.5--1436} \label{subsec:J0821res}
\begin{figure*}[t!]
\gridline{\fig{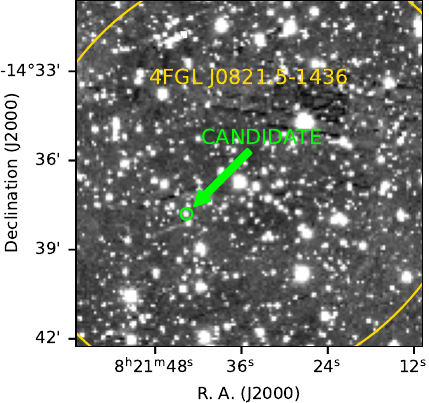}{0.47\textwidth}{(a)}
          \fig{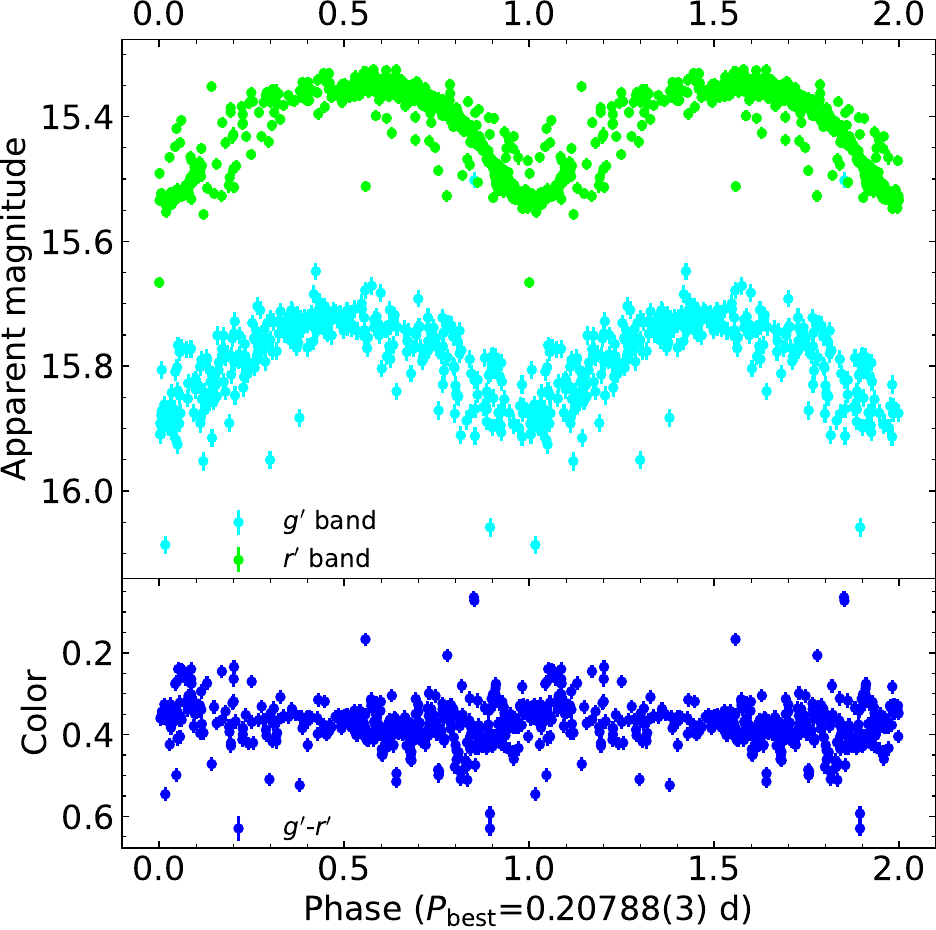}{0.47\textwidth}{(b)}}
\caption{(a) Zoomed-in field around 4FGL~J0821.5$-$1436 as observed by Pan-STARRS \citep{2016arXiv161205560C} in the \textit{r'}-band. The 4FGL 95\% error ellipse is shown in yellow, while the spider candidate J0821 is highlighted in green. (b) \textit{Top panel}: P48/ZTF light curves of J0821 in the \textit{g'} and \textit{r'} optical bands folded on the photometric period $P_{\mathrm{best}}=0.20788 \ \mathrm{d}$} and reference epoch $T_{0}=58539.3137 \ \mathrm{MJD}$. Two cycles are shown for clarity. \textit{Bottom panel}: Folded color curve of J0821 using the same period.
\label{fig:J0821_FoV&lc}
\end{figure*}

We analyzed public P48/ZTF \textit{g'}- and \textit{r'}-band data within a $15'\times15'$ square region centered on the position of 4FGL~J0821.5$-$1436, according to the criterion (1) described in Section~\ref{subsec:variablesandperiods}. This led to the identification of an optical variable and spider candidate (hereafter referred to as J0821), marked by a green circle in Figure~\ref{fig:J0821_FoV&lc}(a), located inside the 95\% error ellipse---highlighted in yellow.

We detect a photometric period of $P_{\mathrm{best}} = 0.20788 \pm 0.00003 \ \mathrm{d}$ from the deepest minimum in the PDM periodogram of J0821's \textit{r'}-band light curve, marked by a red solid line in Figure~\ref{fig:periodograms}(a). This detection exceeds the 0.1\% FAP threshold---shown as an orange dashed line---corresponding to a confidence level higher than $3\sigma$. We fold the \textit{g'} and \textit{r'} light curves, as well as the color curve, on this period using the time of lowest flux in the \textit{r'} band as the phase-zero epoch, $T_{0} = 58539.3137 \pm 0.0002 \ \mathrm{MJD}$\footnote{In this work, we use the same convention to define phase zero for all folded light curves. The error on $T_{0}$ is determined as half the exposure time corresponding to the lowest flux point in the \textit{r'}-band.
}.

As shown in Figure~\ref{fig:J0821_FoV&lc}(b), the light curves of J0821 show an evident periodic modulation with peak-to-peak amplitudes of $\simeq0.2 \ \mathrm{mag}$ in both bands, while the (\textit{g'}$-$\textit{r'}) color curve remains flat with no significant variation. The average color over a full cycle is $(\textit{g}'-\textit{r}') = 0.37 \pm 0.06 \ \mathrm{mag}$. After correcting for extinction using a color excess of $E(g-r) = 0.04 \pm 0.01$ \citep{2019ApJ...887...93G}, we derive an intrinsic color of $(\textit{g}'-\textit{r}') = 0.33 \pm 0.06 \ \mathrm{mag}$. Comparing this value to the low-mass spectral templates of \citet{allard11}, we estimate a mean effective temperature of $T_\mathrm{eff} = 6300 \pm 300 \ \mathrm{K}$ for the companion star. Hereinafter, we use this method to estimate the effective temperatures from the optical colors.

In Section~\ref{subsec:J0821disc}, we discuss J0821 as the potential optical counterpart to a spider MSP whose emission is dominated by ellipsoidal modulation, and infer its presumed orbital period as $P_{\mathrm{orb}}=2\times P_{\mathrm{best}}$.
\subsection{4FGL~J1517.9--5233} \label{subsec:J1517res}
\begin{figure*}[t!]
\gridline{\fig{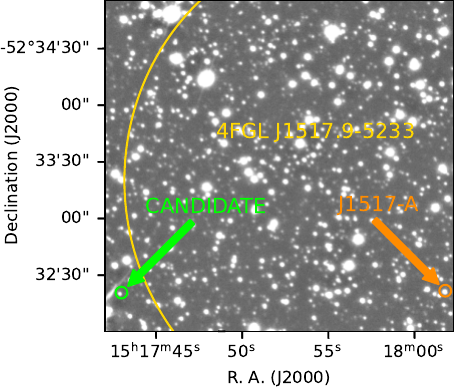}{0.515\textwidth}{(a)}
          \fig{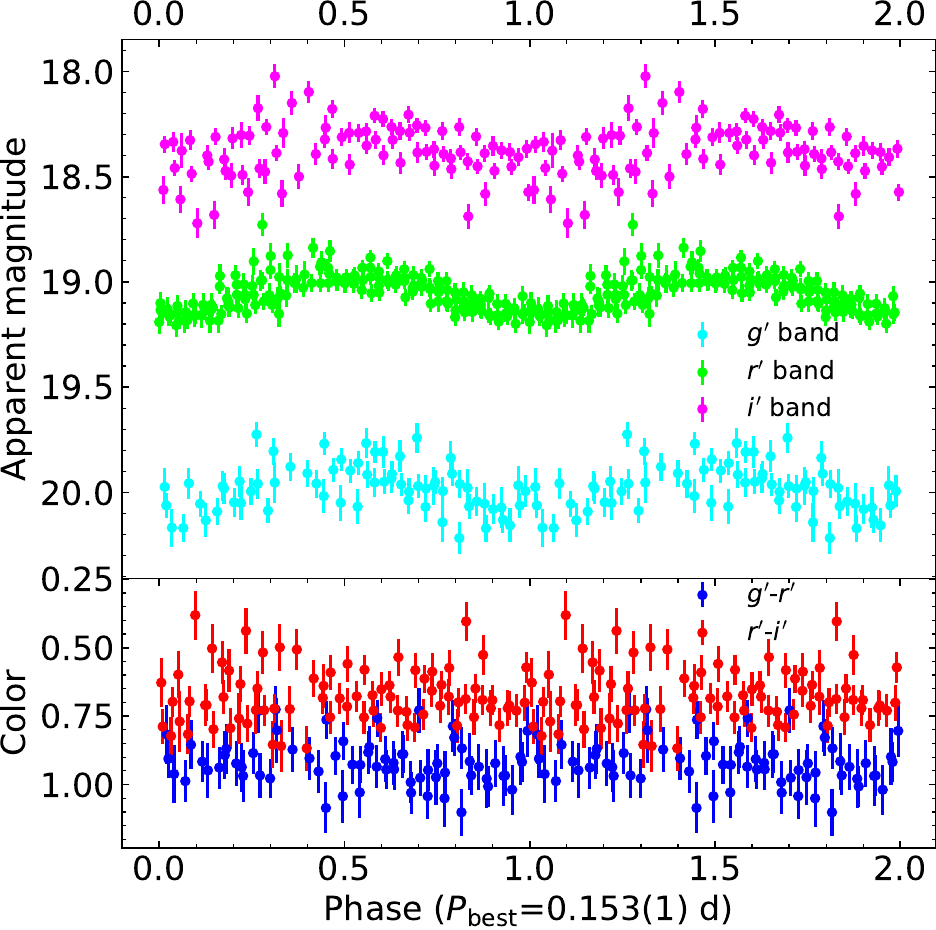}{0.47\textwidth}{(b)}}
\caption{(a) Zoomed-in field around 4FGL~J1517.9$-$5233 observed with LCO/Sinistro in the \textit{r'}-band. The 4FGL 95\% error ellipse is shown in yellow, while the spider candidate J1517 is highlighted in green and the other periodic variable is reported in orange. (b) \textit{Top panel}: LCO/Sinistro light curves of J1517 in the \textit{g'}, \textit{r'}, and \textit{i'} optical bands folded on the photometric period $P_{\mathrm{best}}=0.153 \ \mathrm{d}$} and reference epoch $T_{0}=59725.020 \ \mathrm{MJD}$, with two cycles shown for displaying purposes. \textit{Bottom panel}: Observed color curves of J1517 folded on the same period.
\label{fig:J1517_FoV&lc}
\end{figure*}
\begin{figure*}[t!]
\gridline{\fig{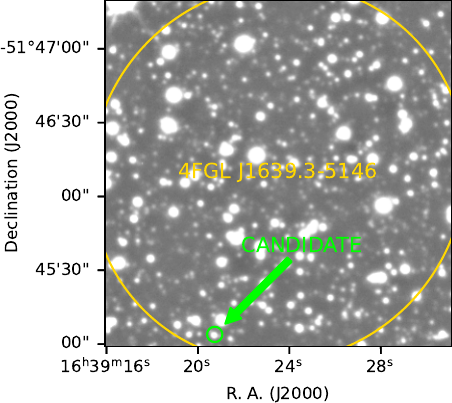}{0.5\textwidth}{(a)}
          \fig{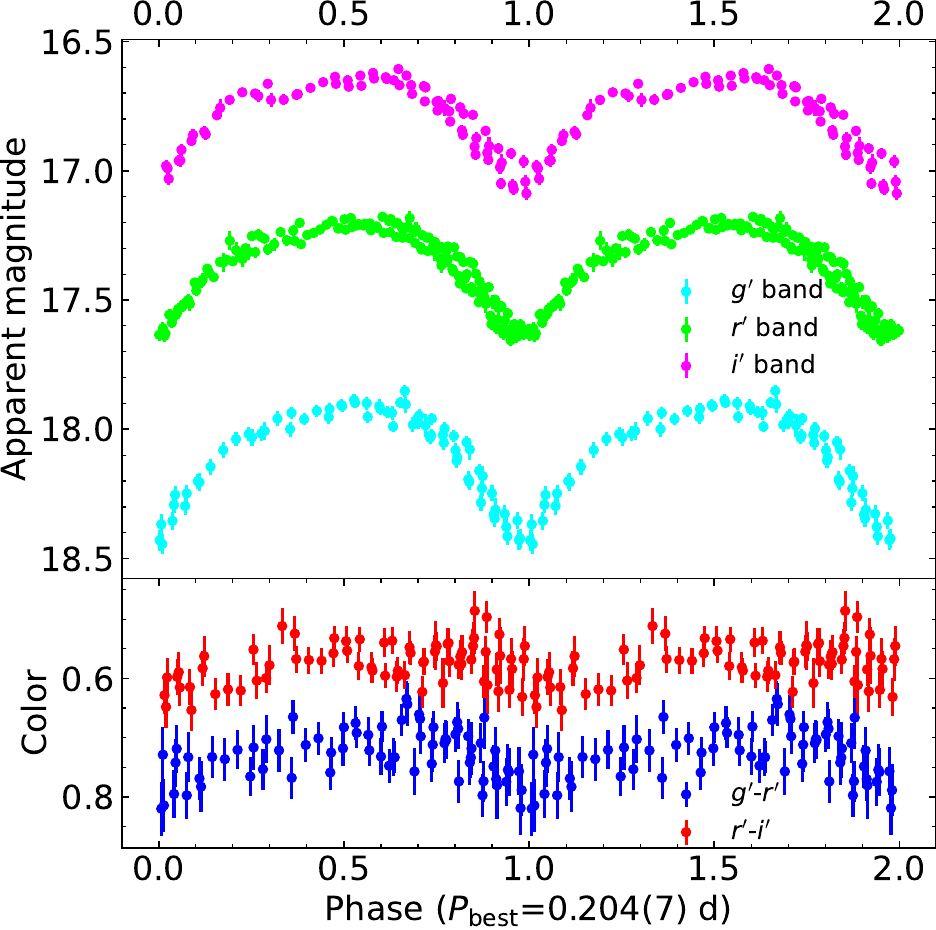}{0.47\textwidth}{(b)}}
\caption{(a) Zoomed-in field around 4FGL~J1639.3$-$5146 observed with LCO/Sinistro in the \textit{r'}-band. The 4FGL 95\% error ellipse is shown in yellow, while the spider candidate J1639 is highlighted in green. (b) \textit{Top panel}: LCO/Sinistro light curves of J1639 in the \textit{g'}, \textit{r'} and \textit{i'} optical bands folded on the photometric period $P_{\mathrm{best}}=0.204 \ \mathrm{d}$, corresponding to the presumed orbital period, and reference epoch $T_{0}=59737.087 \ \mathrm{MJD}$, with two cycles shown for clarity. \textit{Bottom panel}: Folded color curves of J1639 using the same period.}
\label{fig:J1639_FoV&lc}
\end{figure*}
\begin{figure*}[t!]
\gridline{\fig{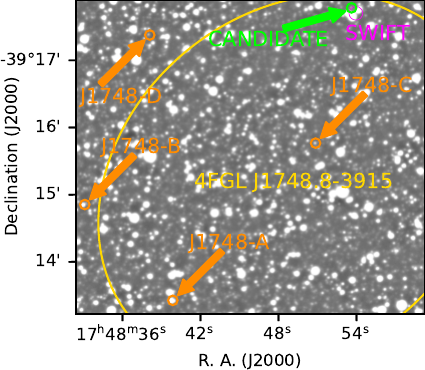}{0.51\textwidth}{(a)}
          \fig{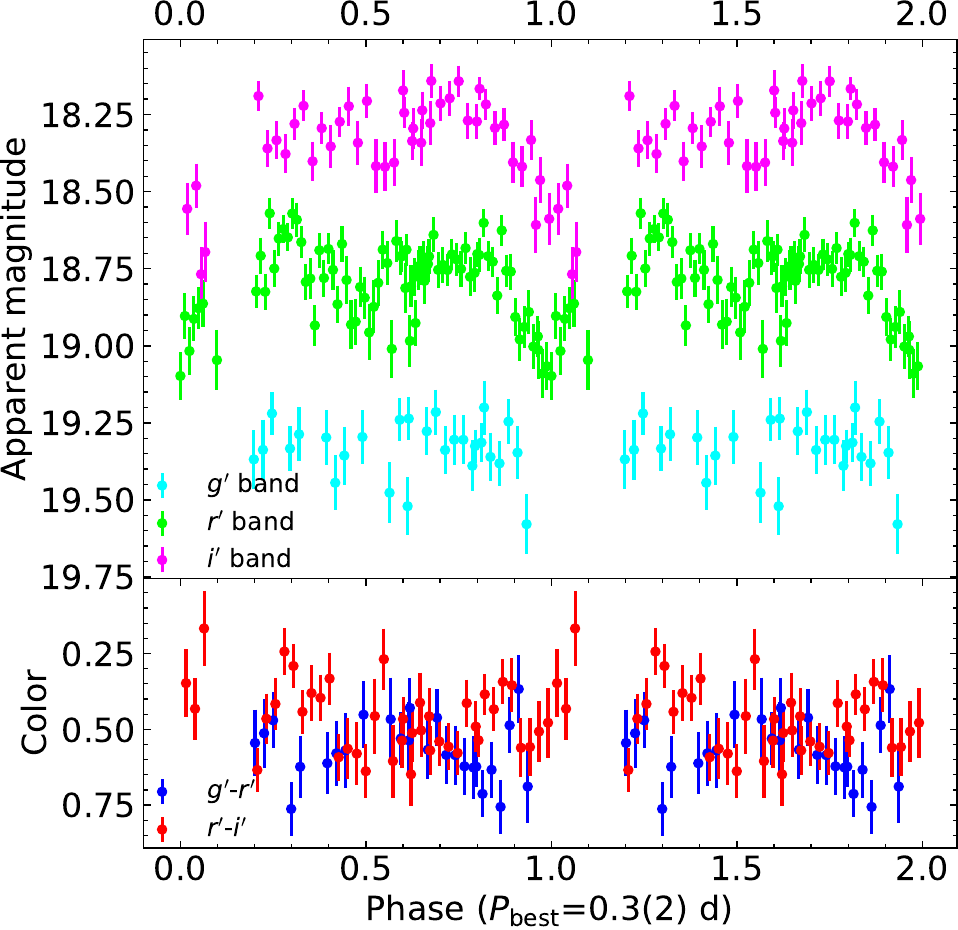}{0.48\textwidth}{(b)}}
\caption{(a) Zoomed-in field around 4FGL~J1748.8$-$3915 observed with LCO/Sinistro in the \textit{r'}-band. 4FGL 95\% error ellipse is shown in yellow, while the spider candidate J1517 is highlighted in green, its \textit{Swift} X-ray counterpart 2SXPS~J174854.0-391739 in magenta and the other periodic variables are reported in orange. (b) \textit{Top panel}: LCO/Sinistro light curves of J1748 in the \textit{g'}, \textit{r'}, and \textit{i'} optical bands folded on the photometric period $P_{\mathrm{best}}=0.3 \ \mathrm{d}$ and reference epoch $T_{0}=59757.006 \ \mathrm{MJD}$, with two cycles shown for displaying purposes. \textit{Bottom panel}: Observed color curves of J1748 folded on the same period.}
\label{fig:J1748_FoV&lc}
\end{figure*}
\begin{figure*}[t!]
\gridline{\fig{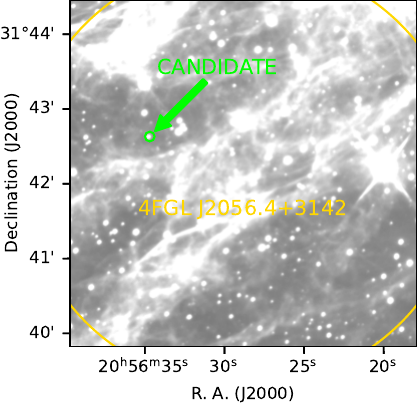}{0.465\textwidth}{(a)}
          \fig{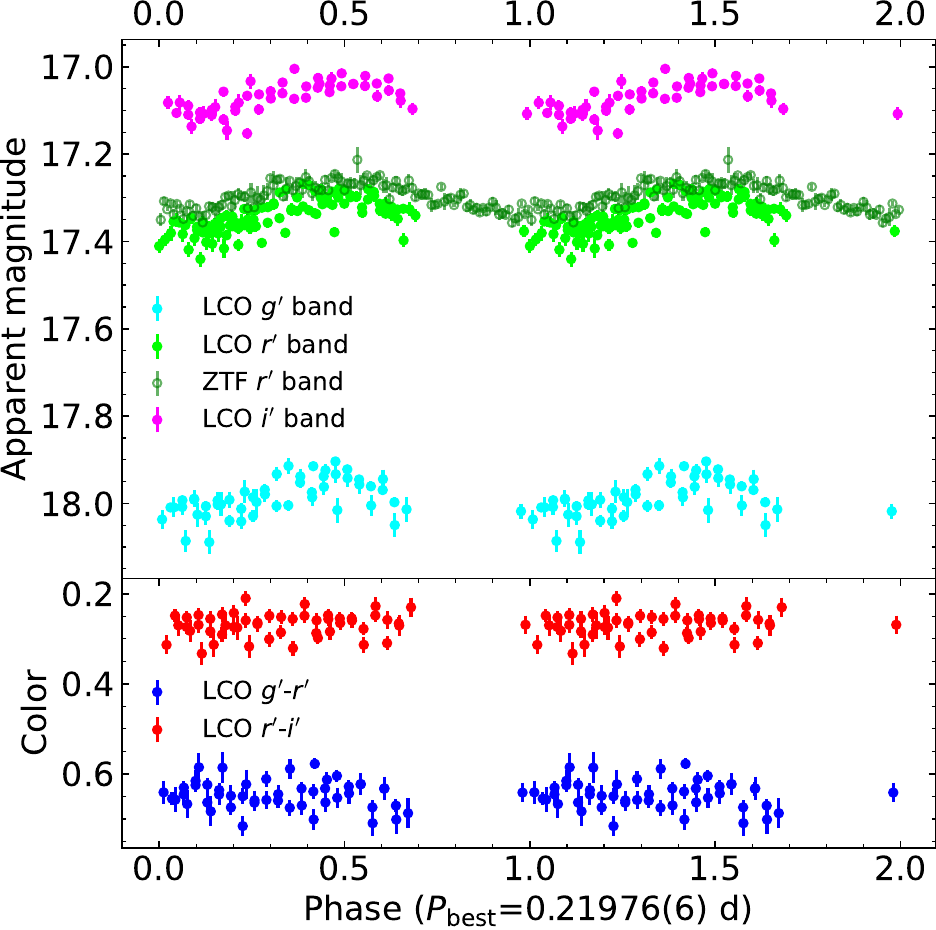}{0.48\textwidth}{(b)}}
\caption{(a) Zoomed-in field around 4FGL~J2056.4$+$3142 observed with LCO/Sinistro in the \textit{r'}-band. The 95\% confidence ellipse from the 4FGL catalog is shown in yellow, with the spider candidate J2056 highlighted in green. (b) \textit{Top panel}: LCO/Sinistro light curves in the \textit{g'}, \textit{r'}, and \textit{i'} bands folded at the photometric period $P_{\mathrm{best}} = 0.21976 \ \mathrm{d}$ and reference epoch $T_{0} = 60606.1330$ MJD, as derived from the ZTF \textit{r'}-band data overplotted on the LCO light curve. Two cycles are shown for clarity. \textit{Bottom panel}: Folded color curves of J2056 using the same period.}
\label{fig:J2056_FoV&lc}
\end{figure*}

We detect two optical variables within a $4'\times4'$ square box surrounding 4FGL~J1517.9$-$5233 from our analysis of LCO/Sinistro observations. One of them is identified as a W UMa binary (light curve reported in Figure~\ref{fig:pervarlightcurves}) and is marked in orange as J1517-A in Figure~\ref{fig:J1517_FoV&lc}(a). We classify the other source, highlighted with a green circle, as a spider candidate (hereafter J1517), despite lying just outside the 4FGL 95\% confidence ellipse (see criterion (2) in Section~\ref{subsec:spiderclass}).

We determine a photometric period of $P_{\mathrm{best}} = 0.153 \pm 0.001 \ \mathrm{d}$ for J1517, corresponding to the strongest peak in its LS periodogram with a detection significance exceeding the $3\sigma$ level, as shown in Figure~\ref{fig:periodograms}(b). The light curves have been folded on this period using $T_{0}=59725.020 \pm 0.001\ \mathrm{MJD}$ as the phase-zero epoch, adopting the same convention as in Section~\ref{subsec:J0821res}.

The optical light curves of J1517 show the same sinusoidal modulation across all bands, with smooth, broad flux maximum and minima and amplitudes of $0.4$, $0.3$, and $0.4 \ \mathrm{mag}$ in the \textit{g'}, \textit{r'}, and \textit{i'} bands, respectively (see Figure~\ref{fig:J1517_FoV&lc}(b)). 
The color curves show no variation over the phase cycle, with average values of $(\textit{g}'-\textit{r}') = 0.92 \pm 0.08 \ \mathrm{mag}$ and $(\textit{r}'-\textit{i}') = 0.68 \pm 0.10 \ \mathrm{mag}$. Adopting a color excess of $E(g-r) = 0.77 \pm 0.01$ \citep{2011ApJ...737..103S}, we obtain dereddened colors of $(\textit{g}'-\textit{r}') = 0.15 \pm 0.08 \ \mathrm{mag}$ and $(\textit{r}'-\textit{i}') = 0.23 \pm 0.10 \ \mathrm{mag}$, corresponding to a mean temperature of $T_{\mathrm{eff}} = 5100 \pm 500 \ \mathrm{K}$.

The light curve and color features of J1517 make it a likely spider system in the ellipsoidal regime, with $P_{\mathrm{orb}}=2\times P_{\mathrm{best}}$, as we discuss in Section~\ref{subsec:J1517disc}.

\subsection{4FGL~J1639.3--5146} \label{subsec:J1639res}

We find an optical variable within the 95\% error ellipse of 4FGL~J1639.3$-$5146, which we classify as a spider candidate (henceforth called J1639), as shown in Figure~\ref{fig:J1639_FoV&lc}(a).

The photometric period identified from the \textit{r'}-band PDM periodogram of J1639 is $P_{\mathrm{best}}=0.204 \pm 0.007 \ \mathrm{d}$, with a detection significance greater than $3\sigma$ (see Figure~\ref{fig:periodograms}(c)). The light and color curves have been phase-folded on this period using a reference epoch of $T_{0} = 59737.087 \pm 0.001 \ \mathrm{MJD}$.

We observe a clear periodic modulation in the optical emission of J1639, characterized by a broad maximum and sharp minima, with peak-to-peak amplitudes of $0.6$, $0.5$, and $0.5 \ \mathrm{mag}$ in \textit{g'}, \textit{r'}, and \textit{i'} bands, respectively (see Figure~\ref{fig:J1639_FoV&lc}(b)). Additionally, the colors vary in phase with the light curves, peaking near the flux maximum. Correcting for extinction using a color excess of $E(g-r) = 0.89 \pm 0.01$ \citep{2011ApJ...737..103S}, we derive average dereddened colors of $(\textit{g}'-\textit{r}') = -0.07 \pm 0.04 \ \mathrm{mag}$ and $(\textit{r}'-\textit{i}') = -0.002 \pm 0.03 \ \mathrm{mag}$, which correspond to an effective temperature of $T_{\mathrm{eff}} = 6700 \pm 300 \ \mathrm{K}$.

In Section~\ref{subsec:J1639disc}, we discuss J1639 as the likely optical counterpart to an irradiated spider system, with $P_{\mathrm{orb}}=P_{\mathrm{best}}$.

\subsection{4FGL~J1748.8--3915} \label{subsec:J1748res}

We detect five optical variables within a $3'\times3'$ square region centered on 4FGL~J1748.8$-$3915. Four of them---J1748-A, J1748-B, J1748-C, and J1748-D---are shown in orange in Figure~\ref{fig:J1748_FoV&lc}(a) and identified as an eclipsing binary, an RR Lyrae pulsator, a W UMa system and another eclipsing binary, respectively, based on the shapes of their light curves presented in Figure~\ref{fig:pervarlightcurves_cont1}. We classify the fifth optical variable---located within the 95\% confidence ellipse of 4FGL~J1748.8$-$3915---as a spider candidate (green circle in Figure~\ref{fig:J1748_FoV&lc}(a), hereafter J1748). Moreover, we find a likely X-ray counterpart in the \textit{Swift} 2SXPS catalog \citep{2020ApJS..247...54E}, located $6.2''$ from J1748 and listed with a location error radius $6.1''$, which further supports its spider classification (see final paragraph of Section~\ref{subsec:spiderclass}).

Although J1748 shows variability with amplitudes of approximately $0.3$, $0.4$, and $0.4 \ \mathrm{mag}$ in \textit{g'}, \textit{r'}, and \textit{i'} bands, respectively, the light curves folded using the PDM-estimated period of $P_{\mathrm{best}} = 0.3 \pm 0.2 \ \mathrm{d}$ (Figure~\ref{fig:periodograms}(d)) do not exhibit a clear sinusoidal shape. This is likely due to contamination from three blended bright stars of $15$--$16 \ \mathrm{mag}$, the closest located just $\simeq5''$ far from J1748, whose broad profiles (FWHM up to $\simeq3''$) graze the edges of the target aperture in most images. This effect, which could not be mitigated by using smaller aperture radii, affects the photometric precision of our target (see Figure~\ref{fig:J1748_FoV&lc}(b)). Using a color excess of $E(g-r) = 0.41 \pm 0.01$ \citep{2011ApJ...737..103S}, we obtain mean intrinsic colors of $(\textit{g}'-\textit{r}') = 0.16 \pm 0.10 \ \mathrm{mag}$ and $(\textit{r}'-\textit{i}') = 0.23 \pm 0.11 \ \mathrm{mag}$, consistent with a temperature of $T_{\mathrm{eff}} = 5100 \pm 600 \ \mathrm{K}$.

The optical variability of J1748, combined with the identification of an X-ray counterpart, suggest this source as a potential spider system with $P_{\mathrm{orb}}=P_{\mathrm{best}}$ (details in Section~\ref{subsec:J1748disc}).

\subsection{4FGL~J2056.4+3142} \label{subsec:J2056res}

We identify an optical variable and spider candidate (green circle in Figure~\ref{fig:J2056_FoV&lc}(a), hereinafter J2056) matching the 95\% confidence ellipse of 4FGL~J2056.4$+$3142.

We extend the incomplete phase coverage of J2056 in our LCO/Sinistro observations by retrieving ZTF \textit{r'}-band data of the same source, to estimate its photometric period as accurately as possible (see Section~\ref{subsec:ZTFdata}). We obtain $P_{\mathrm{best}}=0.21976 \pm 0.00006 \ \mathrm{d}$ from the deepest minimum in the PDM periodogram of the ZTF light curve, with a confidence level higher than $3\sigma$, as shown in Figure~\ref{fig:periodograms}(e). Both LCO and ZTF optical light curves of J2056 have been folded on this period, using as phase-zero epoch the time $T_{0} = 60606.1330 \pm 0.0002 \ \mathrm{MJD}$, corresponding to the flux minimum in ZTF data.

As shown in Figure~\ref{fig:J2056_FoV&lc}(b), the light curves of J2056 show the same sinusoidal modulation both in LCO and ZTF datasets, with broad flux maximum and minima and amplitudes of $\simeq0.2 \ \mathrm{mag}$ across all bands. The color curves remain flat with no significant changes over the phase cycle. Adopting a color excess of $E(g-r) = 0.18 \pm 0.02$ \citep{2019ApJ...887...93G}, we estimate average dereddened colors of $(\textit{g}'-\textit{r}') = 0.48 \pm 0.03 \ \mathrm{mag}$ and $(\textit{r}'-\textit{i}') = 0.15 \pm 0.03 \ \mathrm{mag}$, which correspond to an effective temperature of $T_{\mathrm{eff}} = 5400 \pm 200 \ \mathrm{K}$.

In Section~\ref{subsec:J2056disc}, we discuss J2056 as a potential spider system dominated by ellipsoidal modulation, with $P_{\mathrm{orb}}=2\times P_{\mathrm{best}}$.

\section{Discussion} \label{sec:discussion}
\begin{table*}[t!]
\raggedright
    \caption{Galactic longitude and latitude, \textit{Gaia} DR3 distance estimate, orbital period, \textit{Gaia} absolute magnitude $M_{\mathrm{G}}$, X-ray luminosity estimate or upper limit and $\gamma$-ray luminosity estimate for each spider candidate.}
    \setlength{\tabcolsep}{12.pt}
    \begin{tabular}{lccccccc}
    \hline\hline
        Name & $l$ & $b$ & $D$ & $P_{\mathrm{orb}}$ & $M_{\mathrm{G}}$ & $L_{\mathrm{X}}$ & $L_{\mathrm{\gamma}}$\\
        4FGL & (deg) & (deg) & (kpc) & (d) & (mag) & $10^{32} \ \mathrm{erg} \ \mathrm{s}^{-1}$ & $10^{33} \ \mathrm{erg} \ \mathrm{s}^{-1}$\\ 
        \hline
    \hline
        J0821.5$-$1436 & 236.736 & $+$12.438 & $3.1_{-0.3}^{+0.4}$ & 0.41576(6) & $2.9_{-0.2}^{+0.2}$ & $<42$ & 1.4\\
        J1517.9$-$5233 & 324.237 & $+$4.175 & $3.9_{-1.8}^{+3.8}$ & 0.305(2) & $4.4_{-1.5}^{+1.3}$ & $<2.7$ & 16\\
        J1639.3$-$5146 & 334.260 & $-$3.315 & $1.9_{-0.2}^{+0.6}$ & 0.204(7) & $3.9_{-0.6}^{+0.3}$ & $<0.5$ & 9.9\\
        J1748.8$-$3915 & 351.457 & $-$5.921 & $3.9_{-1.1}^{+2.2}$ & 0.3(2) & $4.6_{-1.0}^{+0.7}$ & 1.5 & 12\\
        J2056.4$+$3142 & 75.562 & $-$8.829 & $3.5_{-0.6}^{+0.9}$ & 0.4395(1) & $4.3_{-0.5}^{+0.4}$ & $<7.7$ & 1.6\\
        \hline
    \end{tabular}
    \label{tab:candparameters}
\end{table*}

We have discovered a spider candidate in 5 out of the 30 \textit{Fermi}-4FGL fields searched, which represents a 17\% success rate, nearly double the 9\% of the COBIPULSE survey \citep{2024ApJ...977...65T}. Their main properties are summarized in Table~\ref{tab:candparameters}, which also reports distances estimated from \textit{Gaia} DR3 parallax measurements (see Section~\ref{subsec:J0821disc}) and absolute \textit{G}-band magnitudes (see Section~\ref{subsec:GAIAcolmag}).
The newly identified candidates span Galactic latitudes from $-8.8^\circ$ to $12.4^\circ$, excluding the inner region between $-3^\circ$ to $+3^\circ$.
Our selection criteria (Section~\ref{subsec:gammaselection}) allowed us to find two candidates close to the Galactic plane: J1517 at $b=4.18^\circ$ and J1639 at $b=-3.32^\circ$.

Here we discuss the optical light curves folded on the derived orbital periods and multi-wavelength properties of each spider candidate, and place them on the \textit{Gaia} color-magnitude diagram to compare their locations with those of confirmed spiders.

\subsection{4FGL~J0821.5--1436} \label{subsec:J0821disc}
\begin{figure}[ht!]
\centering
\includegraphics[width=\columnwidth]{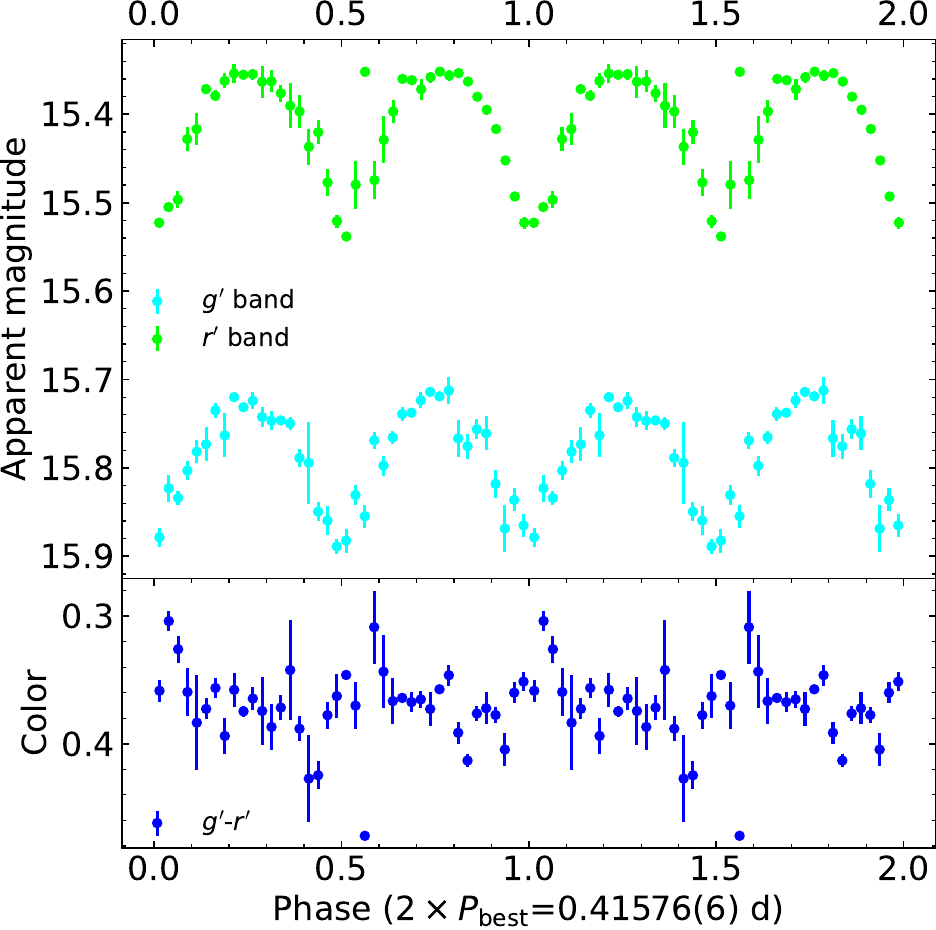}
\caption{\textit{Top panel}: P48/ZTF light curves of J0821 folded using the assumed orbital period $P_{\mathrm{orb}}=0.41576 \ \mathrm{d}$ and reference epoch $T_{0}=58539.314 \ \mathrm{MJD}$, where two cycles are shown for displaying purposes. \textit{Bottom panel}: Observed color curve of J0821 folded on $P_{\mathrm{orb}}$.}
\label{fig:J0821_lcdoubleP}
\end{figure}

The mean effective temperature $T_{\mathrm{eff}}=6300 \pm 300 \ \mathrm{K}$ of J0821 estimated from its colors is consistent with the hotter RB subtype of spider systems. The light curve shape, its $\simeq0.2 \ \mathrm{mag}$ amplitude and lack of color changes presented in Section~\ref{subsec:J0821res} indicate a non-irradiated companion exhibiting only ellipsoidal modulation. Therefore, we folded the ZTF light and color curves on the presumed orbital period $P_{\mathrm{orb}}=2\times P_{\mathrm{best}}=0.41576 \pm 0.00006 \ \mathrm{d}$, adopting twice the photometric period, and rebinned the data into 40 phase bins to better visualize the variability pattern\footnote{We computed the mean magnitude within each bin, with the standard deviation as the associated uncertainty.}. As we can see in Figure~\ref{fig:J0821_lcdoubleP}, the resulting light curves show two minima at $\phi=0$ and $\phi=0.5$, and two maxima at the ascending ($\phi=0.25$) and descending nodes ($\phi=0.75$), with colors remaining constant within uncertainties across the orbit. The same phenomenology is commonly observed in non- or mildly-irradiated RBs dominated by ellipsoidal modulation (e.g., PSR~J2129$-$0429, \citealt{2016ApJ...816...74B}; PSR~J1622$-$0315, \citealt{2024ApJ...973..121S}).

We also identified an optical source coincident with J0821 in the \textit{Gaia} DR3 catalog \citep{brown2016gaia,refId0}, flagged as a variable source with ID 5723165628410609664.
Using the \textit{Gaia} parallax measurement of $0.28\pm0.03 \ \mathrm{mas}$ and adopting the distance prior from \citet{2021AJ....161..147B}, based on stellar population models, we estimate a geometric distance of $D=3.1_{-0.3}^{+0.4} \ \mathrm{kpc}$ for J0821\footnote{Computed using the median and $1\sigma$ asymmetric confidence intervals from the \textit{Gaia} DR3 Lite Distances service available at \url{https://dc.g-vo.org/gedr3dist/q/cone/form}.}.

For J0821, no X-ray counterpart is reported in the most recent point-source catalogs from \textit{Chandra}, \textit{Swift}, \textit{XMM-Newton}, or \textit{eROSITA}. The most stringent X-ray upper limit is provided by a \textit{XMM-Newton} slew-mode observation (0.2--12 keV), which covered the \textit{Fermi} field containing J0821 for only $8 \ \mathrm{s}$. Given that spiders X-ray emission is typically characterized by a power-law photon index in the range $\Gamma\simeq1$--$1.5$ \citep{2014ApJ...795...72L}, we assumed $\Gamma = 1.2$. Using this value together with an interstellar hydrogen column density\footnote{Estimated using the H\,{\sc i} column density tool at \url{https://heasarc.gsfc.nasa.gov/cgi-bin/Tools/w3nh/w3nh.pl}.} of $N_{\mathrm{H}} = 5.3 \times 10^{20} \ \mathrm{cm}^{-2}$, we derive a $3\sigma$ upper limit on the unabsorbed X-ray energy flux\footnote{Computed using the upper limit service at \url{http://xmmuls.esac.esa.int/upperlimitserver/}.} of $3.7 \times 10^{-12} \ \mathrm{erg} \ \mathrm{cm}^{-2} \ \mathrm{s}^{-1}$. Adopting the previously inferred parallax distance, this corresponds to an upper limit on the X-ray luminosity of $4.2\times10^{33} \ (D/3.1 \ \mathrm{kpc})^2 \ \mathrm{erg} \ \mathrm{s}^{-1}$ in the 0.2--12 keV band. This value lies within the typical X-ray luminosity range of $10^{31}$--$10^{33} \ \mathrm{erg} \ \mathrm{s}^{-1}$ for known Galactic RBs (see \citealt{2025ApJ...994....8K} and references therein). Thus, our shallow upper limit is not sensitive enough to exclude the presence of X-ray emission from J0821.

The match between our optical variable and the pulsar-like \textit{Fermi} source 4FGL~J0821.5$-$1436 strengthens the identification of J0821 as a RB MSP candidate (for details, see Section \ref{subsec:gammaselection}). The integrated 0.1--100 GeV energy flux from 4FGL is $(1.3\pm0.3)\times10^{-12}\ \mathrm{erg} \ \mathrm{cm}^{-2} \ \mathrm{s}^{-1}$, which, using the parallax distance, corresponds to a $\gamma$-ray luminosity of $L_{\gamma}=1.4\times10^{33} \ (D/3.1 \ \mathrm{kpc})^2 \ \mathrm{erg} \ \mathrm{s}^{-1}$. This value lies within the range of $\gamma$-ray luminosities typically observed from MSPs \citep[$\sim10^{32}$--$10^{34} \ \mathrm{erg} \ \mathrm{s}^{-1}$;][]{2023ApJ...958..191S}.

To extend the multi-wavelength investigation, we searched for radio counterparts to J0821 but found no matches in the 1.4 GHz \textit{NRAO VLA Sky Survey} \citep[NVSS;][]{1998AJ....115.1693C}, the 1.4 GHz \textit{Faint Images of the Radio Sky at Twenty Centimeters} survey \citep[FIRST;][]{2015ApJ...801...26H}, the 3 GHz \textit{Very Large Array Sky Survey} \citep[VLASS;][]{2021ApJS..255...30G}, or the 0.1--20 GHz Combined Radio Multi-Survey Catalog of Fermi Unassociated Sources compiled by \citet{2023ApJ...943...51B}. Additionally, the other candidates discussed in Sections~\ref{subsec:J1517disc}, \ref{subsec:J1639disc}, \ref{subsec:J1748disc}, and \ref{subsec:J2056disc} also lack radio counterparts, likely due to eclipses caused by intrabinary material and/or dispersion in the interstellar medium.

\subsection{4FGL~J1517.9--5233} \label{subsec:J1517disc}
\begin{figure}[ht!]
\centering
\includegraphics[width=\columnwidth]{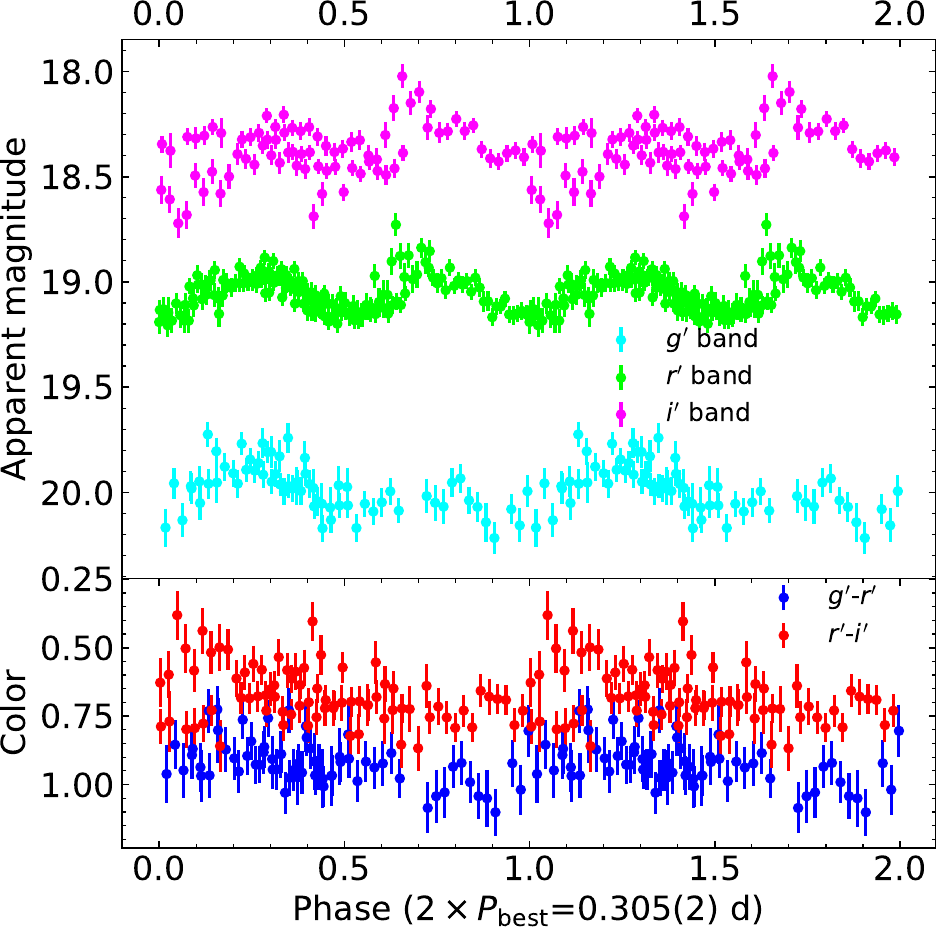}
\caption{\textit{Top panel}: LCO/Sinistro light curves of J1517 folded on the assumed orbital period $P_{\mathrm{orb}}=0.305 \ \mathrm{d}$ and reference epoch $T_{0}=59725.020 \ \mathrm{MJD}$, where two cycles are shown for displaying purposes. \textit{Bottom panel}: Observed color curves of J1517 folded using the same period.}
\label{fig:J1517_lcdoubleP}
\end{figure}

The shape of J1517's optical light curves and its average temperature $T_{\mathrm{eff}} = 5100 \pm 500 \ \mathrm{K}$ (see Section~\ref{subsec:J1517res}) suggest a RB system dominated by ellipsoidal modulation, with little to no irradiation. Indeed, when the light curves are folded on the presumed orbital period $P_{\mathrm{orb}}=2\times P_{\mathrm{best}}=0.305 \pm 0.002 \ \mathrm{d}$, they show two minima around $\phi=0$ and $\phi=0.5$, and two maxima at quadratures, with flat colors throughout the orbit (Figure~\ref{fig:J1517_lcdoubleP}).

The data scattering shown by the multi-band light curves might indicate flaring activity, previously observed in some spider systems (e.g., the BW PSR~J1311$-$3430, \citealt{2012ApJ...754L..25R}; the RBs PSR~J1048$+$2339 and PSR~J0838$-$2827, \citealt{2018ApJ...866...71C}; and the RB candidate 1FGL~J0523.5$-$2529, \citealt{2022ApJ...935..151H}). Such flares presumably arise from magnetic events on the companion star or from the intrabinary shock between the pulsar and companion's wind \citep{2018ApJ...866...71C}. An additional contribution may come from a bright star, located $\sim10''$ from J1517, that fully saturates the CCD and could contaminate our photometry. Its diffraction spikes nearly reach our target, as we can see in Figure~\ref{fig:J1517_FoV&lc}(a). Although the \textit{g'} and \textit{i'} light curves are more scattered than that in the \textit{r'} band (see Section~\ref{subsec:J1517res}), the \textit{g'} data show a potential asymmetry, with the peak at $\phi = 0.25$ appearing brighter than the one at $\phi = 0.75$. This may be caused by the intrabinary shock, 
which wraps around the pulsar and heats the trailing side of the companion star, leading to enhanced irradiation that is more prominent at bluer wavelengths. The same feature has been observed in other RBs, such as PSR~J1628$-$3205 \citep{2014ApJ...795..115L} and PSR~J2039$-$5617 \citep{2015ApJ...814...88S}.

Our candidate J1517 matches the variable optical source reported in \textit{Gaia} DR3 with ID 5888147588733054208. Using the distance prior distribution from \cite{2021AJ....161..147B}, we convert the parallax measurement $0.54\pm0.36 \ \mathrm{mas}$ of J1517 to a geometric distance of $D=3.9_{-1.8}^{+3.8} \ \mathrm{kpc}$.

We found no X-ray counterpart to J1517 in the most recent X-ray point-source catalogs. The deepest upper limit on its X-ray luminosity comes from a \textit{Swift}/XRT observation (0.3--10 keV) performed on 2010 September 29, which covered the target for $3.7 \ \mathrm{ks}$. Using the \textit{Swift}-SXPS upper limit server\footnote{\url{https://www.swift.ac.uk/LSXPS/ulserv.php}}, we derived a $3\sigma$ count rate upper limit of $1.9\times10^{-3} \ \mathrm{ct} \ \mathrm{s}^{-1}$ at the position of J1517. Assuming a photon index of $\Gamma=1.2$ and an interstellar hydrogen column density of $N_{\mathrm{H}}=3.6\times10^{21} \ \mathrm{cm}^{-2}$, this corresponds to an unabsorbed X-ray flux upper limit\footnote{Converted using the \textsc{WebPIMMS} tool at \url{https://heasarc.gsfc.nasa.gov/cgi-bin/Tools/w3pimms/w3pimms.pl}.} of $1.5 \times 10^{-13} \ \mathrm{erg} \ \mathrm{cm}^{-2} \ \mathrm{s}^{-1}$. Adopting the geometric distance estimate above, we infer an upper limit on the X-ray luminosity of $2.7 \times 10^{32} \ (D/3.9 \ \mathrm{kpc})^2 \ \mathrm{erg} \ \mathrm{s}^{-1}$, consistent with undetected high-energy emission from the intrabinary shock.

We deem J1517 as a possible optical counterpart to the pulsar-like unidentified source 4FGL~J1517.9$-$5233, even though it lies slightly outside the 95\% confidence region of the \textit{Fermi} localization (see Figure~\ref{fig:J1517_FoV&lc}(a) and Section~\ref{subsec:spiderclass}). The $\gamma$-ray energy flux of this source is $(8.5 \pm 0.7) \times 10^{-12} \ \mathrm{erg} \ \mathrm{cm}^{-2} \ \mathrm{s}^{-1}$ (0.1--100 GeV), which translates into a luminosity of $L_\gamma = 1.6 \times 10^{34} \ (D/3.9 \ \mathrm{kpc})^2 \ \mathrm{erg} \ \mathrm{s}^{-1}$, consistent with the typical range observed in MSPs.

\subsection{4FGL~J1639.3--5146} \label{subsec:J1639disc}
The temperature $T_{\mathrm{eff}}=6700 \pm 300 \ \mathrm{K}$ of J1639 is consistent with typical RB values. Its optical modulation and phase-dependent color variability point to a companion star that is likely irradiated by the pulsar wind. The light curves folded on the presumed orbital period $P_{\mathrm{orb}}=P_{\mathrm{best}}=0.204 \pm 0.007 \ \mathrm{d}$ show a single flux maximum per orbit at the companion's superior conjunction ($\phi=0.5$), as seen in Figure~\ref{fig:J1639_FoV&lc}(b).

However, the light curves exhibit peak-to-peak amplitudes of $0.5$--$0.6\ \mathrm{mag}$, lower than the $\gtrsim1$-mag modulations typically observed in irradiated RB systems (e.g., PSR~J2215$+$5135, \citealt{2018ApJ...859...54L}; PSR~J2339$-$0533, \citealt{2011ApJ...743L..26R}). We attribute this moderate amplitude to an intermediate-to-low orbital inclination for J1639, such that significant fractions of both the irradiated `day-side' and cooler `night-side' of the companion remain visible over the orbit.

We infer a temperature variation of approximately $1000 \ \mathrm{K}$, from $T_{\mathrm{inf}}=6000 \pm 400 \ \mathrm{K}$ at inferior conjunction to $T_{\mathrm{sup}}=7100 \pm 300 \ \mathrm{K}$ at superior conjunction. The light curves also show a slightly asymmetric peak, consistent to what is observed in some spiders and likely arising from the complex geometry of the intrabinary shock (e.g., PSR~J1311$-$3430, \citealt{2012ApJ...760L..36R}; PSR~J2055$+$1545, \citealt{2025MNRAS.538..380T}).

Our candidate J1639 is also coincident with the \textit{Gaia} DR3 variable source 5931215012555755008, which---like J0821 (Section~\ref{subsec:J0821disc}) and J1517 (Section~\ref{subsec:J1517disc})---is classified as an eclipsing binary in the variability catalog of \citet{2022gdr3.reptE..10R}. Using its parallax measurement of $0.55\pm0.08 \ \mathrm{mas}$ and adopting the distance prior from \citet{2021AJ....161..147B}, we estimate a distance of $D=1.9_{-0.2}^{+0.6} \ \mathrm{kpc}$.

No X-ray counterpart to J1639 was found in the latest catalogs from \textit{Chandra}, \textit{Swift}, \textit{XMM-Newton}, or \textit{eROSITA}. The most stringent upper limit on its X-ray luminosity comes from archival \textit{Swift}/XRT observations. These were carried out between 2011 May 19 and 2014 July 26 with a total exposure of $7.2 \ \mathrm{ks}$, which places a $3\sigma$ upper limit on the count rate of $1.5\times10^{-3} \ \mathrm{ct} \ \mathrm{s}^{-1}$ at J1639 location. Using a photon index of $\Gamma=1.2$ and a hydrogen column density of $N_{\mathrm{H}}=4.1\times10^{21} \ \mathrm{cm}^{-2}$, we derived a $3\sigma$ upper limit on the unabsorbed X-ray flux of $1.2\times10^{-13}\ \mathrm{erg} \ \mathrm{cm}^{-2} \ \mathrm{s}^{-1}$ (0.3--10 keV). From the parallax distance, we obtain an upper limit on the X-ray luminosity of $5.0\times10^{31} \ (D/1.9 \ \mathrm{kpc})^2 \ \mathrm{erg} \ \mathrm{s}^{-1}$, consistent with typical RB values. Therefore, we cannot rule out the presence of X-ray emission in J1639, at the level of the less luminous known RBs.

As shown in Figure~\ref{fig:J1639_FoV&lc}(a), J1639 lies inside the 95\% confidence ellipse of the $\gamma$-ray pulsar-like source 4FGL~J1639.3$-$5146, strengthening its spider association. The 0.1--100 GeV energy flux of the 4FGL object is $(2.4\pm0.2)\times10^{-11}\ \mathrm{erg} \ \mathrm{cm}^{-2} \ \mathrm{s}^{-1}$, from which we derive a $\gamma$-ray luminosity of $L_{\gamma}=9.9\times10^{33} \ (D/1.9 \ \mathrm{kpc})^2 \ \mathrm{erg} \ \mathrm{s}^{-1}$. This estimate is compatible with the typical MSPs luminosities 
measured in $\gamma$-rays.

The source 3FGL~J1639.4$-$5146, associated with our 4FGL target, was previously classified as a pulsar candidate by \citet{2016MNRAS.461.1062F}. In their study, they identified a radio counterpart, TGSSADR~J163923.8$-$514634, which lies within the 95\% confidence ellipses of both 3FGL and 4FGL but is located at a different position (R.A. = 16:39:23.83, decl. = $-$51:46:34.1) from our optical candidate. This radio source also coincides with an optical object cataloged in the USNO-B1.0 survey \citep{2003AJ....125..984M}. However, no additional information is provided by \citet{2016MNRAS.461.1062F} and we do not detect any variability from the USNO source in our data, with rms amplitudes of $0.01$--$0.02 \ \mathrm{mag}$ and magnitude $r'\simeq18$. Therefore, this earlier identification does not affect the validity of J1639 as a spider candidate.

\subsection{4FGL~J1748.8--3915} \label{subsec:J1748disc}
The optical variability and photometric temperature $T_{\mathrm{eff}}=5100\pm600 \ \mathrm{K}$ of J1748, along with the identification of its likely X-ray counterpart, make this system a compelling RB candidate (see Figure~\ref{fig:J1748_FoV&lc}).

We associate J1748 with the \textit{Gaia} DR3 source 5958230425510812928. Its parallax measurement of $0.36 \pm 0.20 \ \mathrm{mas}$ yields a distance estimate of $D=3.9^{+2.2}_{-1.1} \ \mathrm{kpc}$, based on the prior from \citet{2021AJ....161..147B}.

The field was observed by \textit{Swift}/XRT over several epochs between 2013 July 24 and 2018 June 2, for a total of $5.6 \ \mathrm{ks}$ of exposure. An X-ray point source was detected at R.A. = 17:48:54.1, decl. = -39:17:40, with a positional uncertainty of $6.1''$. This is just $6.2''$ away from the optical position of J1748, and therefore compatible with being the X-ray counterpart to our candidate. The \textit{Swift} 2SXPS catalog \citep{2020ApJS..247...54E} reports an unabsorbed 0.3--10 keV flux of $(8 \pm 6) \times 10^{-14} \ \mathrm{erg} \ \mathrm{cm}^{-2} \ \mathrm{s}^{-1}$ for this source. Adopting the parallax distance previously estimated, we infer an X-ray luminosity of $L_{\mathrm{X}} = 1.5 \times 10^{32} \ (D / 3.9 \ \mathrm{kpc})^2 \ \mathrm{erg} \ \mathrm{s}^{-1}$, fully consistent with typical values observed in Galactic RBs.

The match between our optical variable, the X-ray source and the pulsar-like \textit{Fermi} source 4FGL~J1748.8$-$3915 supports the identification of this system as a RB candidate. The 0.1--100 GeV energy flux cataloged from 4FGL is $(6.4\pm0.7)\times10^{-12}\ \mathrm{erg} \ \mathrm{cm}^{-2} \ \mathrm{s}^{-1}$, which translates into a $\gamma$-ray luminosity of  $L_{\gamma}=1.2\times10^{34} \ (D/3.9 \ \mathrm{kpc})^2 \ \mathrm{erg} \ \mathrm{s}^{-1}$. This luminosity falls well within the characteristic range observed in known MSPs \citep[$\sim10^{32}$–$10^{34} \ \mathrm{erg} \ \mathrm{s}^{-1}$;][]{2023ApJ...958..191S}.

\subsection{4FGL~J2056.4+3142} \label{subsec:J2056disc}
\begin{figure}[ht!]
\centering
\includegraphics[width=\columnwidth]{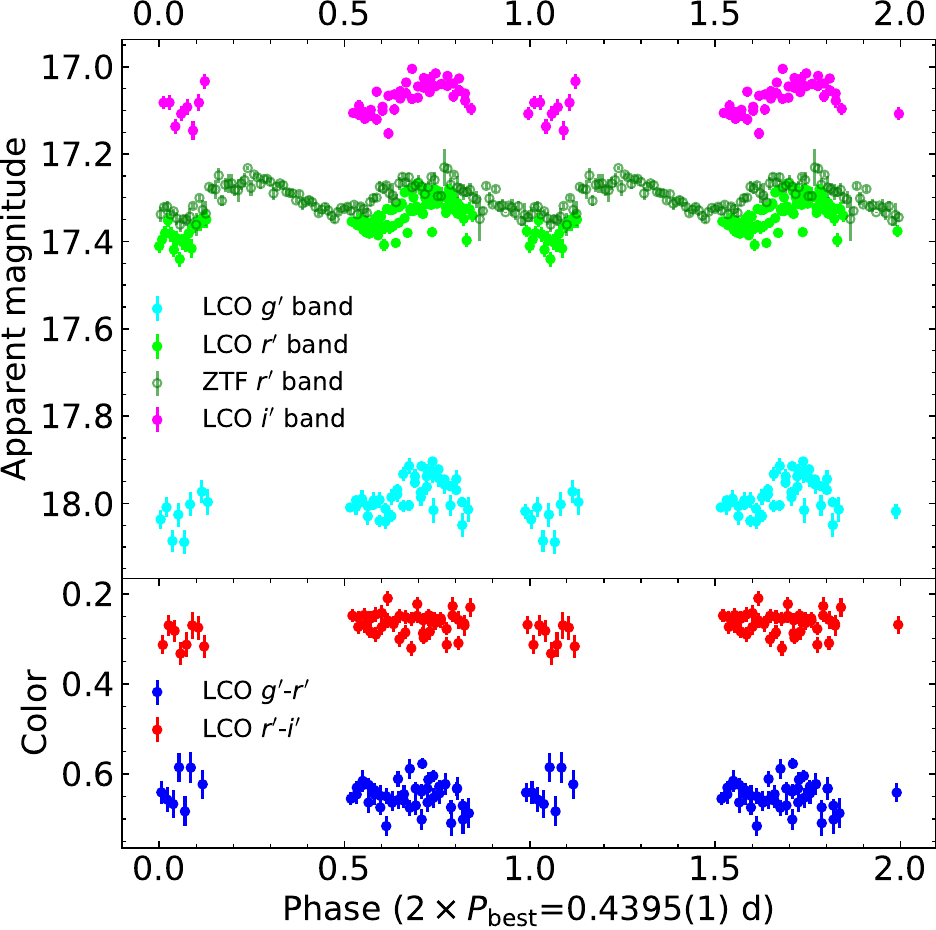}
\caption{\textit{Top panel}: LCO/Sinistro and P48/ZTF light curves of J2056 folded using the assumed orbital period $P_{\mathrm{orb}}=0.4395 \ \mathrm{d}$ and reference epoch $T_{0}=60606.1330 \ \mathrm{MJD}$, with two cycles shown for clarity. \textit{Bottom panel}: Folded color curves of J2056 folded on the same period.}
\label{fig:J2056_lcdoubleP}
\end{figure}
The light curve shape, stable colors and mean temperature $T_{\mathrm{eff}} = 5400 \pm 200 \ \mathrm{K}$ of J2056 (see Section~\ref{subsec:J2056res}) indicate a likely RB companion experiencing weak to no irradiation.
Folding the ZTF and LCO light curves over the presumed orbital period $P_{\mathrm{orb}}=2\times P_{\mathrm{best}}=0.4395\pm0.0001 \ \mathrm{d}$, we find two minima at superior and inferior conjunctions and two maxima near $\phi=0.25$ and $\phi=0.75$ (Figure~\ref{fig:J2056_lcdoubleP}), consistent with ellipsoidal modulation.

J2056 coincides with the \textit{Gaia} DR3 variable source 1864896760105313536. From its parallax measurement of $0.25\pm0.07 \ \mathrm{mas}$ and the distance prior of \cite{2021AJ....161..147B}, we infer a distance of $D=3.5_{-0.6}^{+0.9} \ \mathrm{kpc}$.

No X-ray counterpart for J2056 was found in the latest \textit{Chandra}, \textit{Swift}, \textit{XMM-Newton}, or \textit{eROSITA} catalogs. The tightest upper limit on its X-ray flux is provided by a short \textit{XMM-Newton} slew-mode exposure lasting $17 \ \mathrm{s}$. Assuming a photon index $\Gamma = 1.2$ and a hydrogen column density $N_{\mathrm{H}} = 1.5 \times 10^{21} \ \mathrm{cm}^{-2}$, we estimate an upper limit on the unabsorbed 0.2--12 keV flux of $5.0 \times 10^{-12} \ \mathrm{erg} \ \mathrm{cm}^{-2} \ \mathrm{s}^{-1}$. This translates to an upper limit on the X-ray luminosity of $7.7 \times 10^{32} \ (D / 3.5 \ \mathrm{kpc})^2 \ \mathrm{erg} \ \mathrm{s}^{-1}$---not stringent enough to exclude X-ray emission from J2056 at typical RB luminosities.

As shown in Figure~\ref{fig:J2056_FoV&lc}(a), J2056 is located inside the 95\% confidence region of the pulsar candidate 4FGL~J2056.4$+$3142, further supporting its identification as a RB system. The 0.1–100 GeV energy flux reported by the 4FGL catalog is $(1.1 \pm 0.1) \times 10^{-11} \ \mathrm{erg} \ \mathrm{cm}^{-2} \ \mathrm{s}^{-1}$, corresponding to a $\gamma$-ray luminosity of $L_{\gamma} = 1.6 \times 10^{33} \ (D / 3.5 \ \mathrm{kpc})^2 \ \mathrm{erg} \ \mathrm{s}^{-1}$---well within the typical range observed from MSPs.

\subsection{PSR~J1824--0621, PSR~J2116+3701 and the RB candidate 4FGL~J1702.7--5655} \label{subsec:J1824J2116J1702disc}
\begin{figure}[ht!]
\centering
\includegraphics[width=\columnwidth]{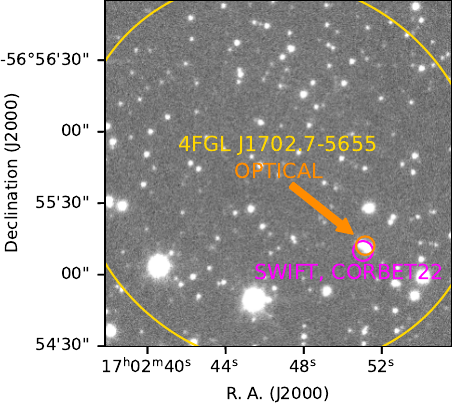}
\caption{Best-seeing \textit{r'}-band image of 4FGL~J1702.7$-$5655 observed with LCO/Sinistro. The 95\% \textit{Fermi}-LAT error ellipse is shown in yellow. The \textit{Swift} X-ray counterpart reported by \citet{2022ApJ...935....2C} is marked in magenta, and the blended optical source coincident with the X-ray position is highlighted in orange.}
\label{fig:J1702_FoV}
\end{figure}

While this work was ongoing, a MSP-white dwarf binary and an isolated pulsar were discovered in two of the \textit{Fermi} sources included in our survey: 4FGL~J1824.2$-$0621 \citep{2023MNRAS.518.1672M} and 4FGL~J2116.2$+$3701 \citep{2023MNRAS.524.5132D}, respectively. Additionally, one of our fields, 4FGL~J1702.7$-$5655, was proposed as a RB candidate by \citet{2022ApJ...935....2C}. Below, we briefly comment on the non-detection of their variable optical counterparts in our observations.

PSR~J1824$-$0621 was initially detected in the Commensal Radio Astronomy FAST Survey (CRAFTS) \citep{2018IMMag..19..112L}, with a position matching that of 4FGL~J1824.2$-$0621. \citet{2023MNRAS.518.1672M} identified it as a MSP–He white dwarf binary with an orbital period of $100.9 \ \mathrm{d}$. As expected for a wide-orbit system with a period outside our search range, we find no optical variable at its position, nor do we detect an optical counterpart in our data. In the ZTF survey, the source lies near the sensitivity limit and is detected only in two \textit{r'}-band images, with a mean magnitude of $r' = 20.7 \pm 0.3$.

PSR~J2116$+$3701 was discovered by \citet{2023MNRAS.524.5132D} using the \textit{Canadian Hydrogen Mapping Experiment} (CHIME) radio telescope and identified as a young isolated pulsar with a spin period of $0.14 \ \mathrm{s}$. Its location is inside the 95\% confidence error ellipse of the \textit{Fermi} source 4FGL~J2116.2$+$3701. As expected for an isolated pulsar, we do not detect any optical variable at this source location. No optical counterpart is found in our data or in ZTF, where the faintest sources in this field have magnitudes of $g' = 21.1$ and $r' = 20.9$.

4FGL~J1702.7$-$5655 was identified as a RB candidate system by \citet{2022ApJ...935....2C}, who found significant $\gamma$-ray modulation at a period of $\sim5.85 \ \mathrm{hr}$. The modulation includes narrow eclipses interpreted as the companion star eclipsing the $\gamma$-ray emission from the MSP, which has been also observed in some spider systems, particularly RBs \citep{2023NatAs...7..451C}. Furthermore, \citet{2022ApJ...935....2C} detected a possible X-ray counterpart to this \textit{Fermi} source in \textit{Swift}/XRT observations, located within the 95\% error ellipse at R.A. = 17:02:51.01, decl. = $-$56:55:09.1, with an uncertainty of $4.2''$. As shown in the \textit{r'}-band best-seeing image of our data for this field (Figure \ref{fig:J1702_FoV}), we find two blended optical sources matching the X-ray position, which cannot be reliably resolved using standard photometric extraction techniques. For the unresolved combined object, we estimate average magnitudes of $g'=15.6 \pm 0.3$, $r'=15.4 \pm 0.3$, and $i'=15.2 \pm 0.3$. Folding its optical light curve on the $\gamma$-ray orbital period of $0.2438033(11) \ \mathrm{d}$ estimated by \citet{2022ApJ...935....2C}, we find no orbital modulation in any band. This suggests that the true optical counterpart of this RB candidate is fainter than $r'\simeq15.5$ and is hidden within the unresolved blend, making it undetectable in our data.

\subsection{Gaia color--magnitude selection for spiders} \label{subsec:GAIAcolmag}
\begin{figure*}[ht!]
\centering
\includegraphics[width=1.65\columnwidth]{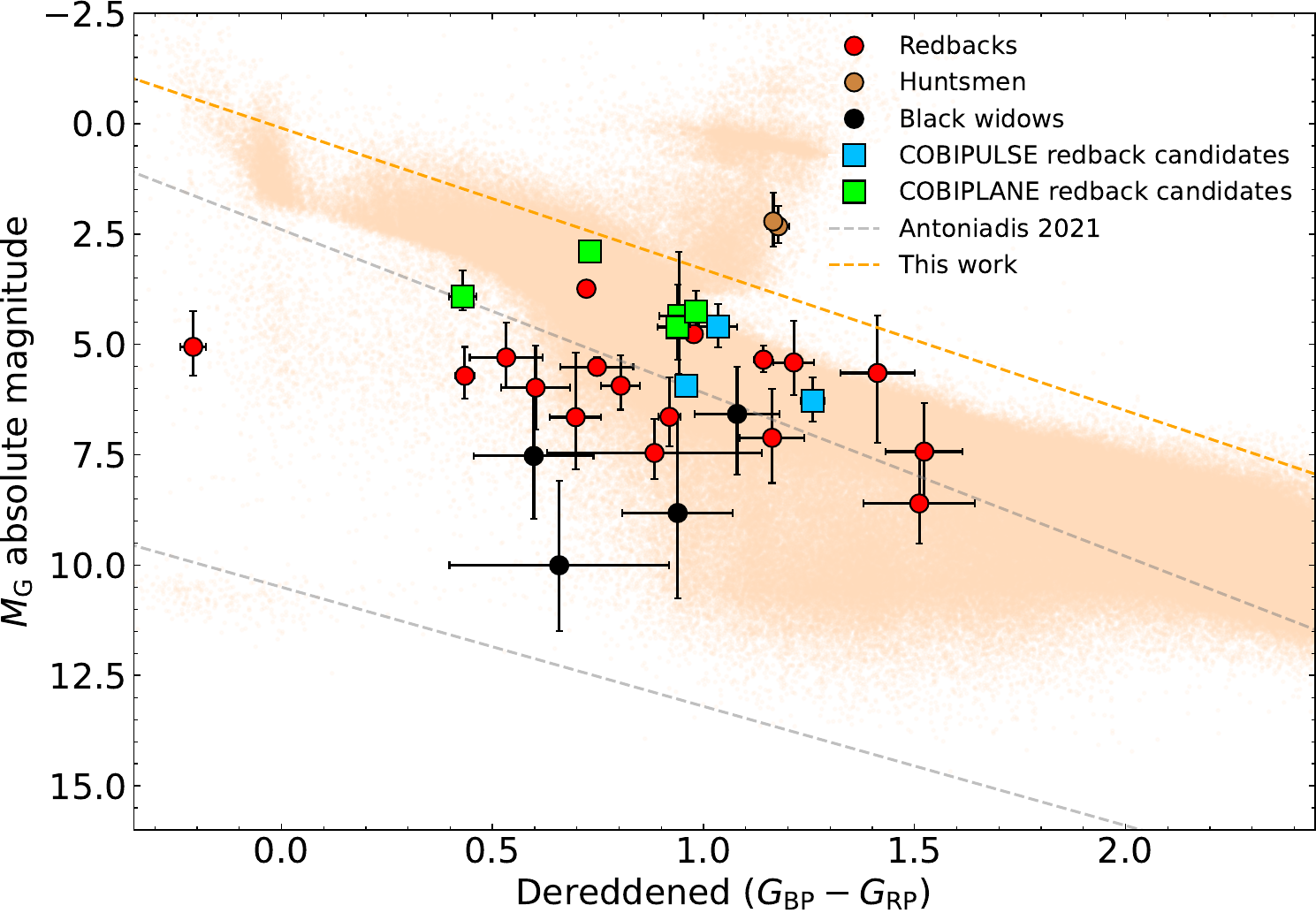}
\caption{Optical counterparts of confirmed spiders, with the error-to-parallax ratio between 0 and 1, overplotted on three million \textit{Gaia} stars within $500 \ \mathrm{pc}$, shown as shaded orange points, in the \textit{Gaia} $(G_{\mathrm{BP}}-G_{\mathrm{RP}})$ color--$M_{\mathrm{G}}$ magnitude diagram. RBs, huntsmen, and BWs are shown as red, brown, and black circles, respectively. RB candidates identified from COBIPULSE and COBIPLANE surveys are plotted as sky-blue and green squares. Grey dashed lines indicate the region boundaries identified by \citet{2021MNRAS.501.1116A}. All spiders---except the rare huntsmen---and our candidates fall within a broader area delimited by an orange dashed contour.}
\label{fig:Gaiacmd}
\end{figure*}

We now compare the optical colors and absolute magnitudes of the spider candidates discovered from COBIPULSE \citep{2024ApJ...977...65T} and COBIPLANE surveys with those of confirmed Galactic field spiders in the \textit{Gaia} Hertzsprung--Russell diagram.

The latest release of SpiderCat\footnote{\url{https://astro.phys.ntnu.no/SpiderCAT}} reports \textit{Gaia} optical counterparts to 35 spider systems out of the 84 in our Galaxy
\citep{2025ApJ...994....8K}. We prune this sample to 23 systems with reliable parallax measurements, retaining only those with error-to-parallax ratios between 0 and 1. The same filtering is applied to the four COBIPULSE and five COBIPLANE spider candidates, resulting in the exclusion of only 3FGL~J2117.6$+$3725-B \citep{2024ApJ...977...65T}. We obtain the intrinsic \textit{Gaia} magnitudes by correcting the observed $G$, $G_{\mathrm{BP}}$ and $G_{\mathrm{RP}}$ magnitudes with the respective extinction coefficients derived from the $V$-band values \citep{2005AJ....130..659A,2021MNRAS.508.1788A} through the transformation relations of \citet{2021A&A...649A...3R}. Then, the dereddened $G$ magnitudes have been converted to absolute $M_{\mathrm{G}}$ magnitudes using the parallax distances, estimated following the same method described in Section~\ref{sec:discussion}.

Figure~\ref{fig:Gaiacmd} shows the $(G_{\mathrm{BP}}-G_{\mathrm{RP}})$ color--$M_{\mathrm{G}}$ absolute magnitude diagram for our selected sample of spiders. Red, brown, and black circles mark the positions of RBs, huntsmen, and BWs, respectively, overplotted on a background of three million \textit{Gaia} stars located within $500 \ \mathrm{pc}$ and with error-to-parallax ratios below 1, shown as shaded orange points. According to \citet{2021MNRAS.501.1116A}, eclipsing MSPs generally lie between the main sequence and the white dwarf cooling track, with upper and lower boundaries of $M_{\mathrm{G}}=3.7(G_{\mathrm{BP}}-G_{\mathrm{RP}})+2.4$ and $M_{\mathrm{G}}=2.7(G_{\mathrm{BP}}-G_{\mathrm{RP}})+10.5$ (the region between dashed grey lines in Figure~\ref{fig:Gaiacmd}). However, the two huntsman systems \citep{2016ApJ...820....6C,2025ApJ...980..124S}, along with the optical counterparts to five confirmed RBs (PSRs~J0212$+$5321, \citealt{2023ApJ...952..150P}; J1306$-$4035, \citealt{2018MNRAS.473..116K}; J1628$-$3205, \citealt{2012arXiv1205.3089R}; J1723$-$2837, \citealt{2004MNRAS.355..147F}; and J1803$-$6707, \citealt{2023MNRAS.519.5590C}) are all located above this
region.

To accommodate these systems, we extend the color--magnitude region associated with spiders to:
\begin{equation}
M_{\mathrm{G}}>3.2(G_{\mathrm{BP}}-G_{\mathrm{RP}})+0.1
\label{eq:spidersregion}
\end{equation}
This broader area, outlined by a dashed orange line in Figure~\ref{fig:Gaiacmd}, includes all confirmed spiders in our cleaned sample, with the exception of the rare huntsmen. The COBIPULSE and COBIPLANE candidates, plotted as sky-blue and green squares, fall in the upper part of this region, consistent with their classification as RB candidates. 
As expected from earlier results with COBIPULSE, this $1$-m class optical survey is best suited to discover the brightest RBs, while being less effective at detecting the intrinsically fainter BWs. In general, Equation~\ref{eq:spidersregion} provides a more conservative criterion for identifying potential spider candidates in the \textit{Gaia} color–-magnitude diagram.
\section{Summary and Conclusions} \label{sec:conclusions}
The COBIPLANE survey was designed to find new spider MSPs by targeting their variable optical counterparts, including Galactic latitudes as low as $\pm$3$^\circ$. To this end, we conducted multi-band photometric monitoring of 30 promising \textit{Fermi}-4FGL pulsar candidates, selected for their characteristic $\gamma$-ray properties. 
This systematic search led to the discovery of five spider candidates associated with 4FGL~J0821.5$-$1436 (J0821), 4FGL~J1517.9$-$5233 (J1517), 4FGL~J1639.3$-$5146 (J1639), 4FGL~J1748.8$-$3915 (J1748), and 4FGL~J2056.4$+$3142 (J2056). Each of these sources exhibits sub-day optical flux modulations with peak-to-peak amplitudes $\gtrsim 0.2 \ \mathrm{mag}$, and companion star temperatures in the $4000$--$6000 \ \mathrm{K}$ range, consistent with RB MSP systems.

\begin{itemize}

\item J0821 shows low-amplitude ($\sim0.2~\mathrm{mag}$), double-peaked light curves and stable colors, indicative of ellipsoidal modulation from a non-irradiated companion at $6300 \pm 300 \ \mathrm{K}$. We determine an orbital period of $0.41576 \pm 0.00006 \ \mathrm{d}$. Archival \textit{XMM-Newton} data yield an upper limit on its X-ray luminosity of $4.2 \times 10^{33} \ (D/3.1 \ \mathrm{kpc})^2 \ \mathrm{erg} \ \mathrm{s}^{-1}$ (0.2--12 keV), consistent with a faint or undetected counterpart.

\item J1517 is another likely non-irradiated RB, with sinusoidal optical modulations of $0.3$--$0.4 \ \mathrm{mag}$ and nearly constant colors, consistent with a companion temperature of $5100 \pm 500 \ \mathrm{K}$. Its orbital period is $0.305 \pm 0.002 \ \mathrm{d}$, and archival \textit{Swift}/XRT data yield an X-ray luminosity upper limit of $2.7 \times 10^{32} \ (D/3.9 \ \mathrm{kpc})^2 \ \mathrm{erg} \ \mathrm{s}^{-1}$, which does not exclude intrabinary shock emission.

\item J1639 is characterized by asymmetric light curves with color maxima aligned with orbital phase $\phi = 0.5$, pointing to mild irradiation of the companion ($6700 \pm 300 \ \mathrm{K}$). Its orbital period is $0.204 \pm 0.007 \ \mathrm{d}$, and the observed modulation amplitude of $0.5$--$0.6 \ \mathrm{mag}$ is consistent with a low orbital inclination. \textit{Swift}/XRT observations constrain its X-ray luminosity to $<5.0 \times 10^{31} \ (D/1.9 \ \mathrm{kpc})^2 \ \mathrm{erg} \ \mathrm{s}^{-1}$.

\item J1748 displays variability with amplitudes of $0.3$--$0.4 \ \mathrm{mag}$ and a companion temperature of $5100 \pm 600 \ \mathrm{K}$. Its light curves are affected by blending with nearby brighter stars. Nevertheless, an associated X-ray source detected by \textit{Swift}/XRT, with a luminosity of $1.5 \times 10^{32} \ (D/3.9 \ \mathrm{kpc})^2 \ \mathrm{erg} \ \mathrm{s}^{-1}$ (0.3--10 keV), supports its classification as a RB MSP.

\item J2056 shows low-amplitude ($\sim0.2 \ \mathrm{mag}$) light curves with flat color profiles, consistent with a non-irradiated companion at $5400 \pm 200 \ \mathrm{K}$. The derived orbital period is $0.4395 \pm 0.0001 \ \mathrm{d}$. \textit{XMM-Newton} data constrain its X-ray luminosity to $<7.7 \times 10^{32} \ (D/3.5 \ \mathrm{kpc})^2 \ \mathrm{erg} \ \mathrm{s}^{-1}$, consistent with other RB systems.

\end{itemize}

We also compared the dereddened \textit{Gaia} colors and absolute magnitudes of the COBIPULSE \citep{2024ApJ...977...65T} and COBIPLANE RB candidates with those of confirmed Galactic spiders. All candidates fall within a region of the color–magnitude diagram that is broader than the one proposed by \citet{2021MNRAS.501.1116A} and also includes the optically brightest known RBs. This supports the classification of our systems as spider candidates. The boundary defined in Equation~\ref{eq:spidersregion} includes all spiders in the filtered sample (except rare huntsmen) and highlights that our survey is most sensitive to RBs with magnitudes of $r'\lesssim21$.

We provide sub-arcsecond precise sky positions for the five new RB candidates, enabling multi-wavelength counterpart searches and targeted radio and $\gamma$-ray observations to detect pulsations. 
If all are confirmed as MSPs, COBIPLANE will have provided crucial support to expanding the known spider population. The likely orbital periods narrow down the search parameter space, facilitating the use of acceleration techniques. Follow-up X-ray observations are needed to strengthen the spider associations of these candidates. Phase-resolved optical spectroscopy will allow the estimation of system parameters through combined modeling of light curves and radial velocities.

\section*{acknowledgments}
This project has received funding from the European Research Council (ERC) under the European Union’s Horizon 2020 research and innovation programme (grant agreement No. 101002352, PI: M. Linares). We thank M. Kennedy for discussions on periodicity search methods and for suggesting the use of the ZTF public data release to extend our dataset. We also thank M. Satybaldiev for his assistance with the method for estimating periods uncertainties. We acknowledge the \textsc{astrosource} software developers, in particular M. Fitzgerald for a discussion on ensemble photometry. P.A.M.P. acknowledges support from grant RYC2021-031173-I funded by MCIN/AEI/ 10.13039/501100011033 and by the 'European Union NextGenerationEU/PRTR’. This article makes use of observations made in the Observatorios de Canarias del IAC with the STELLA telescope operated on the island of Tenerife by the Leibniz Institute for Astrophysics Potsdam (AIP) in the Observatorio del Teide. This work is also based on observations from the Las Cumbres Observatory global telescope network, which were performed using the Sinistro camera at the 1-meter LCOGT network of autonomous telescopes. This research has made use of the NASA/IPAC Infrared Science Archive, which is funded by the National Aeronautics and Space Administration and operated by the California Institute of Technology. This work has also made use of data from the Zwicky Transient Facility (ZTF). ZTF is supported by the National Science Foundation under Grants No. AST-1440341 and AST-2034437 and a collaboration including current partners Caltech, IPAC, the Weizmann Institute for Science, the Oskar Klein Center at Stockholm University, the University of Maryland, Deutsches Elektronen-Synchrotron and Humboldt University, the TANGO Consortium of Taiwan, the University of Wisconsin at Milwaukee, Trinity College Dublin, Lawrence Livermore National Laboratories, IN2P3, University of Warwick, Ruhr University Bochum, Northwestern University and former partners the University of Washington, Los Alamos National Laboratories, and Lawrence Berkeley National Laboratories. Operations are conducted by COO, IPAC, and UW. ZTF is supported by the National Science Foundation under Grants No. AST-1440341 and AST-2034437 and a collaboration including current partners Caltech, IPAC, the Weizmann Institute for Science, the Oskar Klein Center at Stockholm University, the University of Maryland, Deutsches Elektronen-Synchrotron and Humboldt University, the TANGO Consortium of Taiwan, the University of Wisconsin at Milwaukee, Trinity College Dublin, Lawrence Livermore National Laboratories, IN2P3, University of Warwick, Ruhr University Bochum, Northwestern University and former partners the University of Washington, Los Alamos National Laboratories, and Lawrence Berkeley National Laboratories. Operations are conducted by COO, IPAC, and UW. This publication makes use of data products from the Two Micron All Sky Survey, which is a joint project of the University of Massachusetts and the Infrared Processing and Analysis Center/California Institute of Technology, funded by the National Aeronautics and Space Administration and the National Science Foundation. This publication makes use of data products from the Wide-field Infrared Survey Explorer, which is a joint project of the University of California, Los Angeles, and the Jet Propulsion Laboratory/California Institute of Technology, funded by the National Aeronautics and Space Administration. This research has made use of data products from the Pan-STARRS1 Surveys (PS1), which have been made possible through contributions by the Institute for Astronomy at the University of Hawaii, the Pan-STARRS Project Office, the Max-Planck Society and its participating institutes. This work has made use of data products from the European Space Agency (ESA) mission Gaia (\url{https://www.cosmos.esa.int/gaia}), processed by the Gaia Data Processing and Analysis Consortium (DPAC, \url{https://www.cosmos.esa.int/web/gaia/dpac/consortium}). Funding for the DPAC has been provided by national institutions, in particular the institutions participating in the Gaia Multilateral Agreement. NNG05GF22G issued through the Science Mission Directorate Near-Earth Objects Observations Program. This research or product makes use of public auxiliary data provided by ESA/Gaia/DPAC/CU5 and prepared by Carine Babusiaux. This work also used data products from the ATLAS project. The ATLAS project is primarily funded by NASA under grant number NN12AR55G, and it is operated by the Institute for Astronomy at the University of Hawaii. We also used data and/or software provided by the High Energy Astrophysics Science Archive Research Center, which is a service of the Astrophysics Science Division at NASA and Goddard Space Flight Center. This work made use of data supplied by the UK Swift Science Data Centre at the University of Leicester. This research used also data from the 4XMM-DR13s catalog, created from observations obtained with XMM-Newton, an ESA science mission with instruments and contributions directly funded by ESA Member States and NASA.

\section*{data availability}
The raw STELLA and LCO images with bias and flats frames used for data reduction can be obtained by contacting M. Turchetta. ZTF data are public and can be obtained through the IRSA light curve service \url{https://irsa.ipac.caltech.edu/docs/program_interface/ztf_lightcurve_api.html}.

%
\vspace{10mm}
\facilities{STELLA:1.2m/WiFSIP, LCO:1m/Sinistro, IRSA, PO:1.2m, Fermi, CXO, eROSITA, Swift (XRT), XMM.}

\software{SEP \citep{2016zndo....159035B}, 
          Source Extractor \citep{1996A&AS..117..393B}, astrosource \citep{2021JOSS....6.2641F}, VizieR \citep{2000A&AS..143...23O}, HEASoft/HI4PI \citep{2016A&A...594A.116H}, astropy \citep{2013A&A...558A..33A,2018AJ....156..123A}, astroquery \citep{2019AJ....157...98G}, SIMBAD \citep{2000A&AS..143....9W}.}
          
\bibliography{sample631}{}
\bibliographystyle{aasjournal}



\appendix
\restartappendixnumbering
\section{COBIPLANE fields of view} \label{sec:appA}
\begin{figure*}[ht!]
\gridline{\leftfig{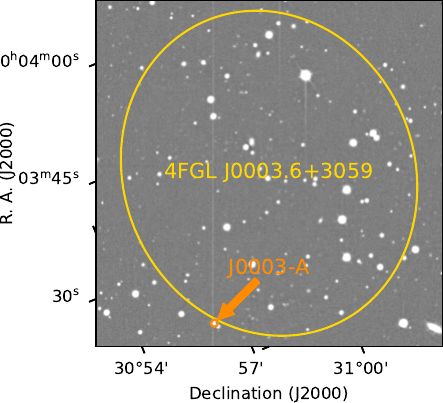}{0.43\textwidth}{}
          \leftfig{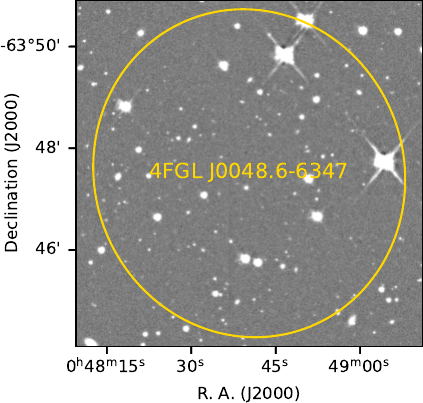}{0.41\textwidth}{}
          }
\gridline{\leftfig{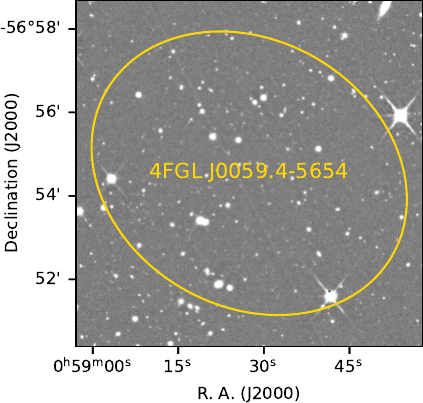}{0.43\textwidth}{}
          \leftfig{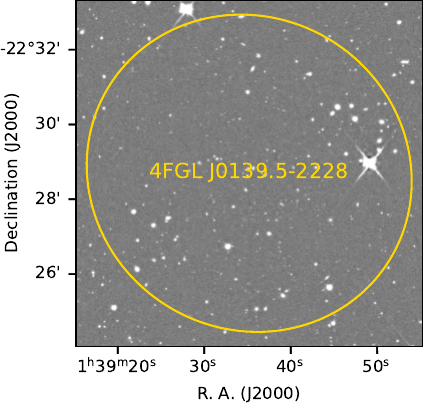}{0.43\textwidth}{}
          }
\gridline{\leftfig{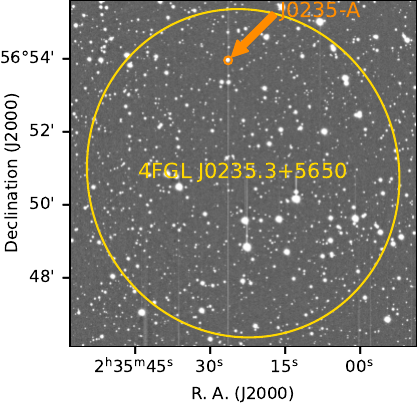}{0.43\textwidth}{}
          \leftfig{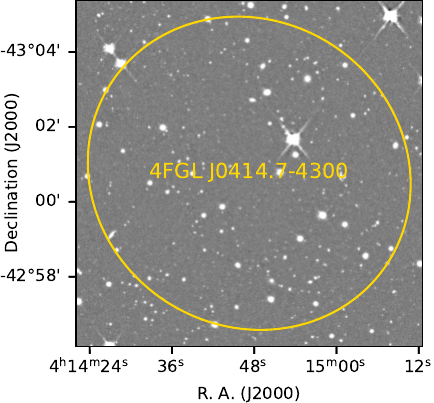}{0.445\textwidth}{}
          }
\caption{Combined \textit{r'}-band images of the \textit{Fermi}-4FGL fields observed in this survey.}\label{fig:FoVs}
\end{figure*}
\begin{figure*}[ht!]
\gridline{\leftfig{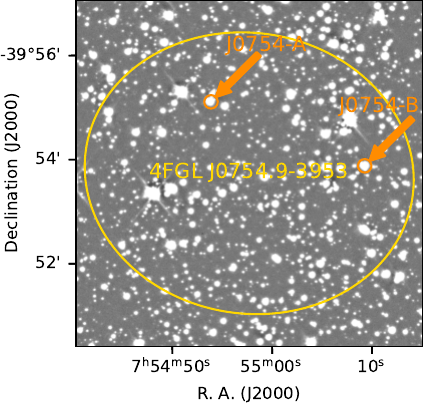}{0.43\textwidth}{}
          \leftfig{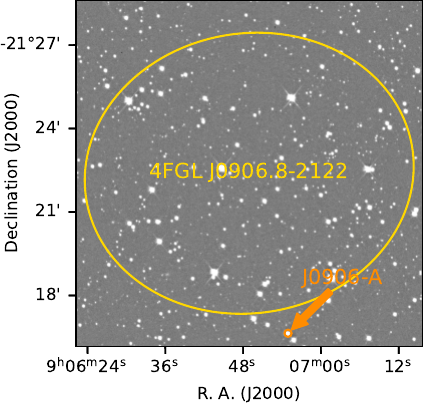}{0.43\textwidth}{}
          }
\gridline{\leftfig{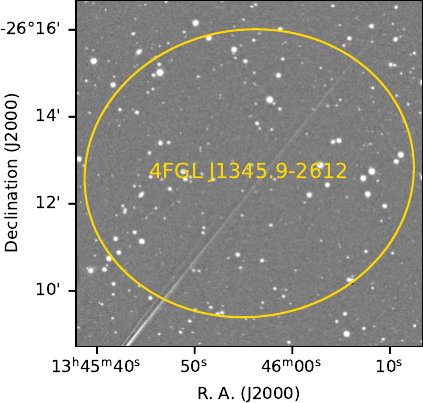}{0.43\textwidth}{}
          \leftfig{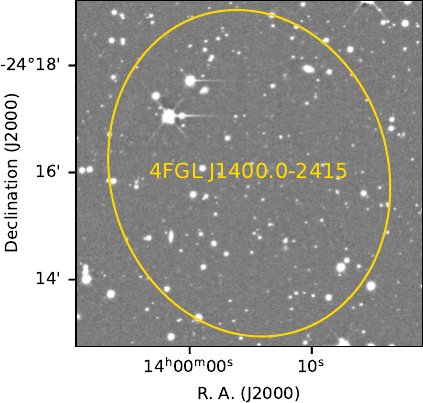}{0.43\textwidth}{}
          }
\gridline{\leftfig{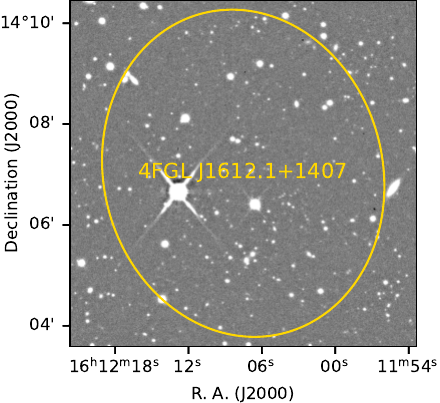}{0.43\textwidth}{}
          \leftfig{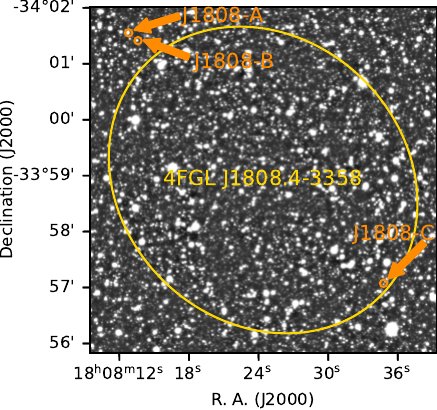}{0.43\textwidth}{}
          }
\caption{Continued.}
\label{fig:FoVs_cont1}
\end{figure*}
\begin{figure*}[ht!]
\gridline{\leftfig{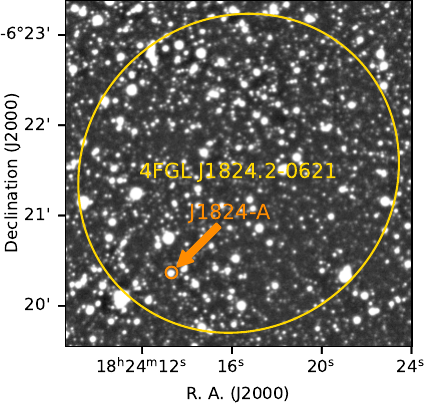}{0.42\textwidth}{}
          \leftfig{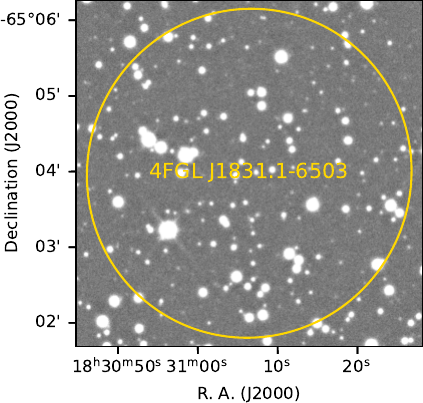}{0.42\textwidth}{}
          }
\gridline{\leftfig{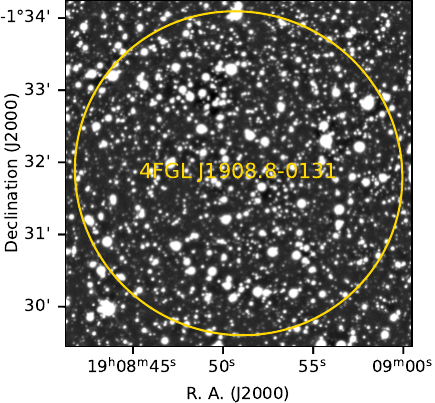}{0.43\textwidth}{}
          \leftfig{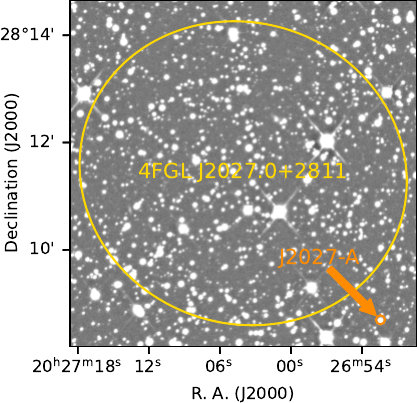}{0.412\textwidth}{}
          }
\gridline{\leftfig{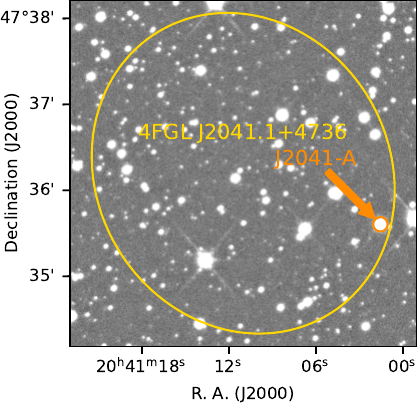}{0.41\textwidth}{}
          \leftfig{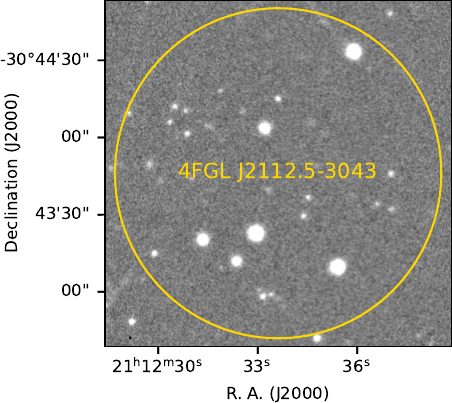}{0.44\textwidth}{}
          }
\caption{Continued.}
\label{fig:FoVs_cont2}
\end{figure*}
\begin{figure*}[ht!]
\gridline{\leftfig{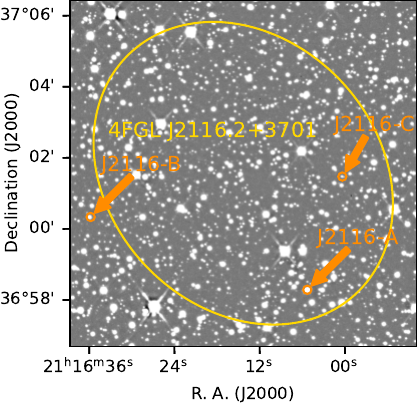}{0.42\textwidth}{}
          \leftfig{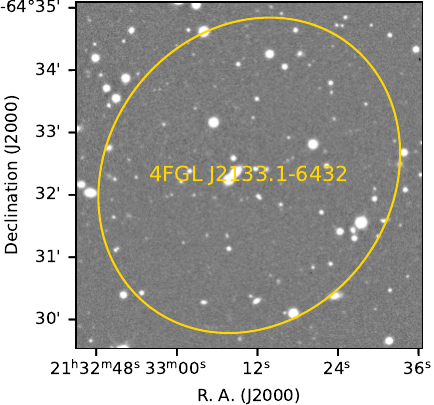}{0.44\textwidth}{}
          }
\gridline{\leftfig{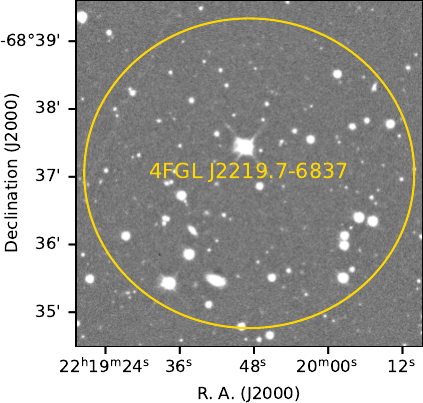}{0.42\textwidth}{}
          \leftfig{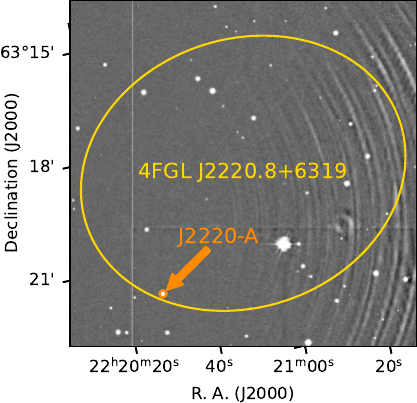}{0.415\textwidth}{}
          }
\gridline{\leftfig{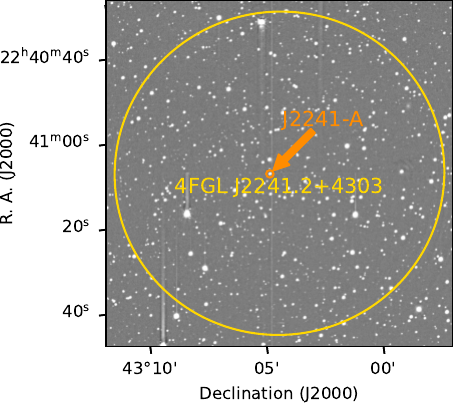}{0.43\textwidth}{}
         \leftfig{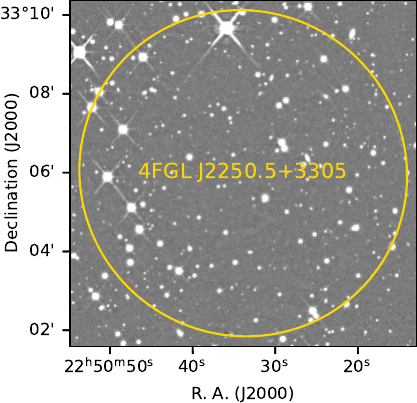}{0.4\textwidth}{}
          }
\caption{Continued.}
\label{fig:FoVs_cont3}
\end{figure*}
In Figures \ref{fig:FoVs}--\ref{fig:FoVs_cont3} we report the combined \textit{r'}-band images for each field observed with the STELLA and LCO optical telescopes. We exclude the 4FGL~J0821.5$-$1436, 4FGL~J1517.9$-$5233, 4FGL~J1639.3$-$5146, 4FGL~J1702.7$-$5655, 4FGL~J1748.8$-$3915, and 4FGL~J2056.4$+$3142 fields, as they are already shown in Figures~\ref{fig:J0821_FoV&lc}(a), \ref{fig:J1517_FoV&lc}(a), \ref{fig:J1639_FoV&lc}(a), \ref{fig:J1702_FoV}, \ref{fig:J1748_FoV&lc}(a), and \ref{fig:J2056_FoV&lc}(a), respectively. We plot the 4FGL $95\%$ error ellipses in yellow and the 21 periodic variables
that we did not attribute to spiders in orange (see Appendix \ref{sec:appC} for details).
\section{Selection of optical variables} \label{sec:appB}
\restartappendixnumbering
\begin{figure*}[ht!]
\gridline{\leftfig{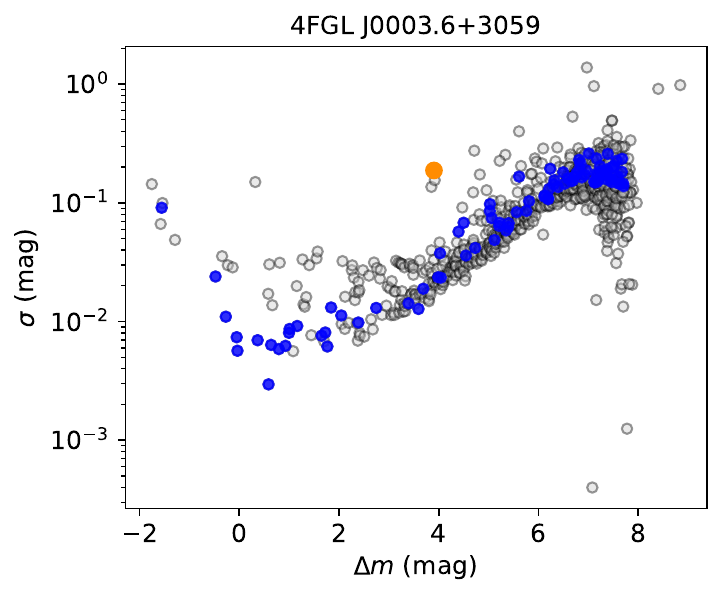}{0.48\textwidth}{}
          \leftfig{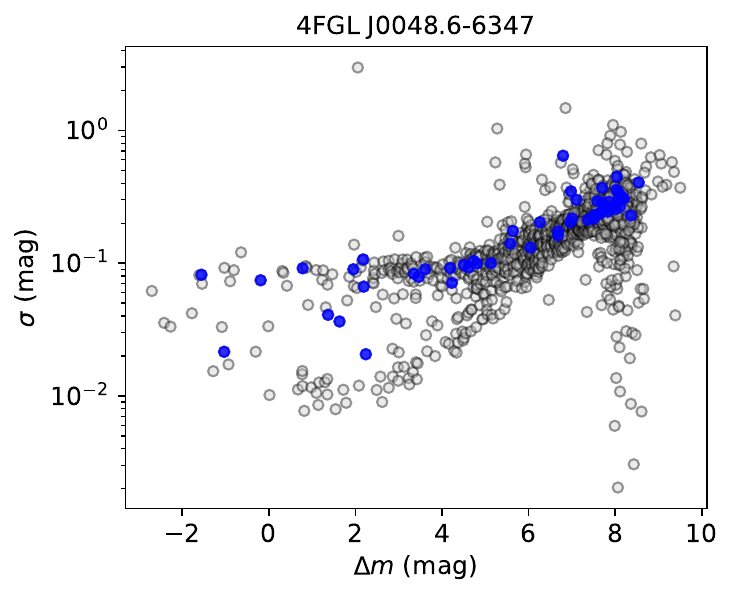}{0.48\textwidth}{}
          }
\gridline{\leftfig{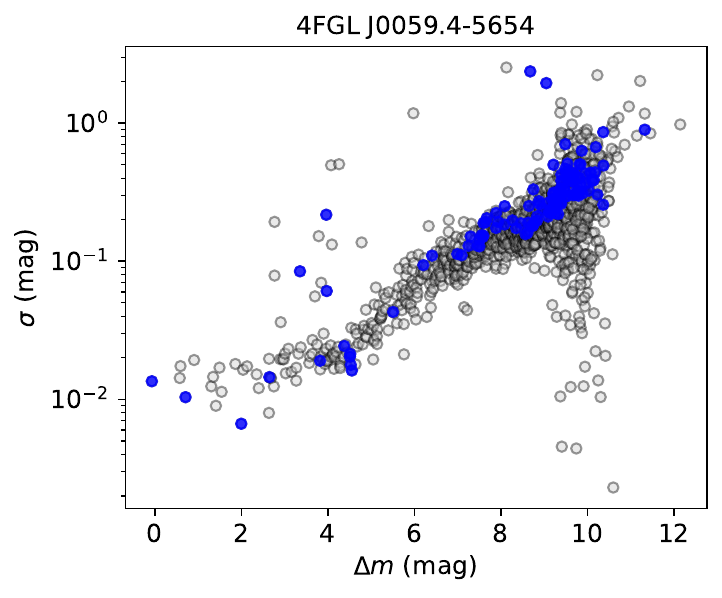}{0.48\textwidth}{}
          \leftfig{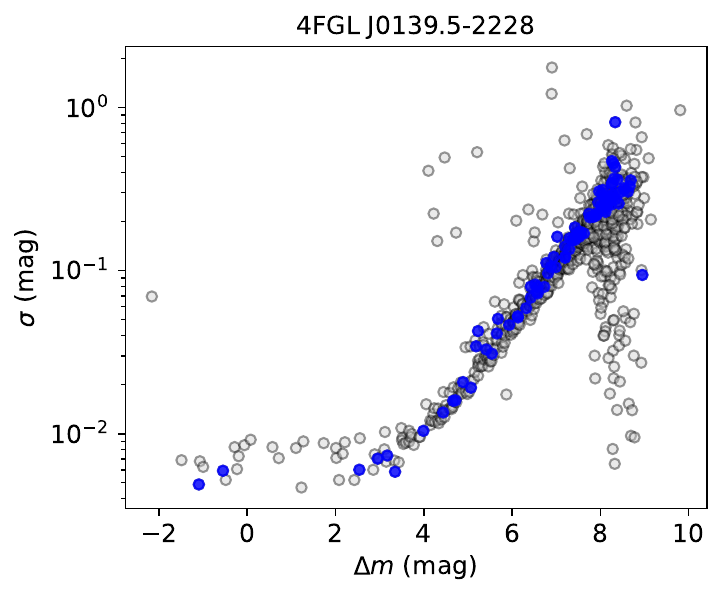}{0.48\textwidth}{}
          }
\gridline{\leftfig{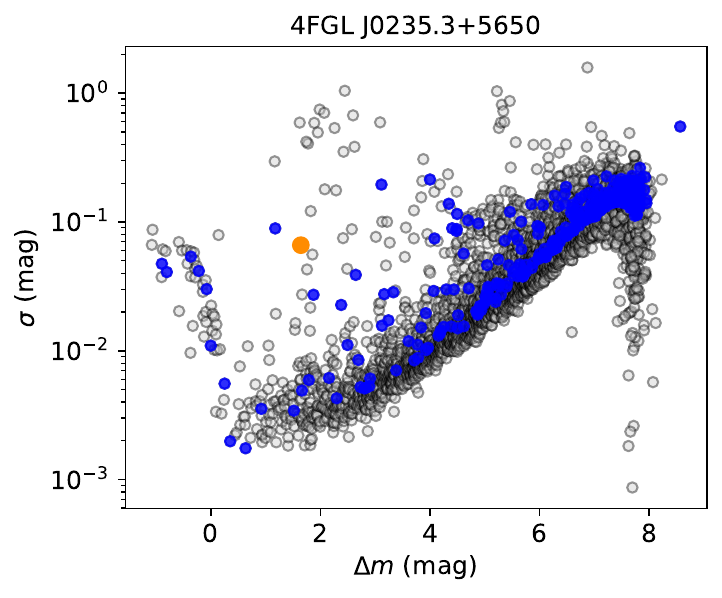}{0.48\textwidth}{}
          \leftfig{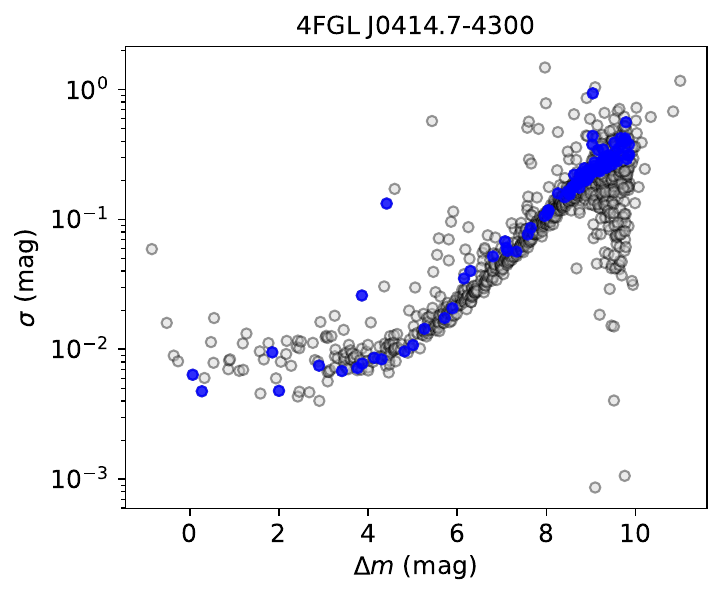}{0.48\textwidth}{}
          }
\caption{$\sigma$ vs $\Delta m$ plots for each of the COBIPLANE fields of view.}
\label{fig:sigmavsdm}
\end{figure*}
\begin{figure*}[ht!]
\gridline{\leftfig{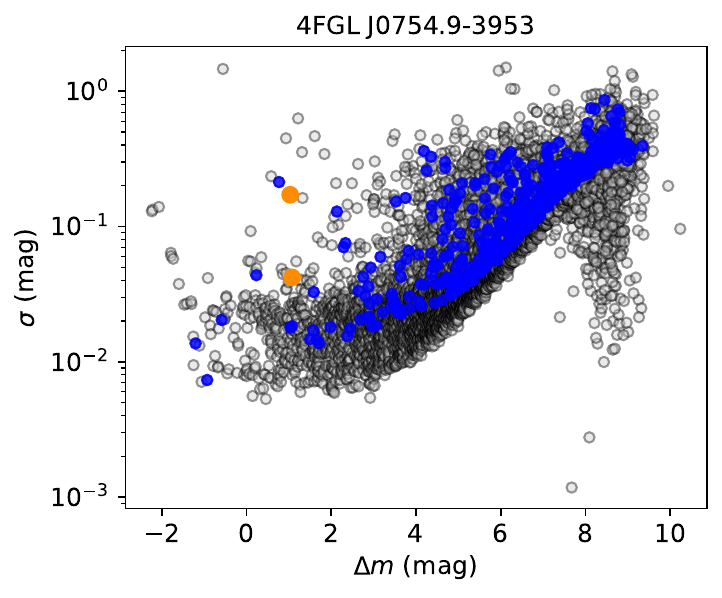}{0.48\textwidth}{}
          \leftfig{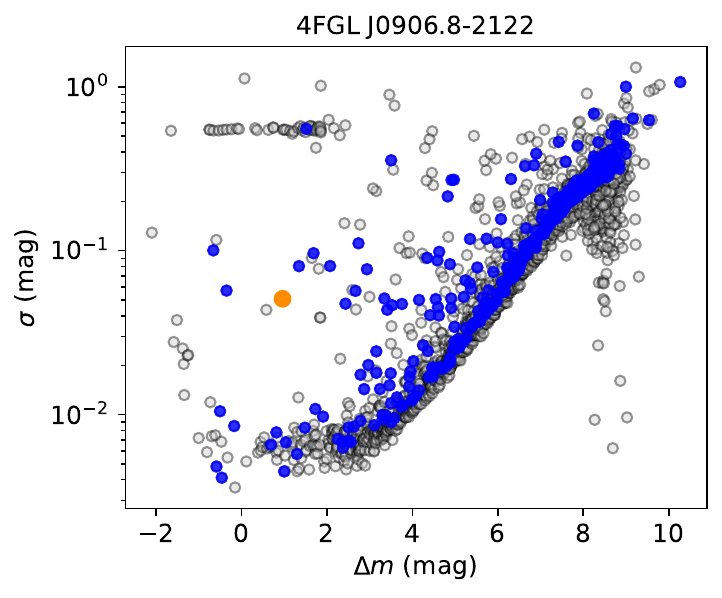}{0.48\textwidth}{}
          }
\gridline{\leftfig{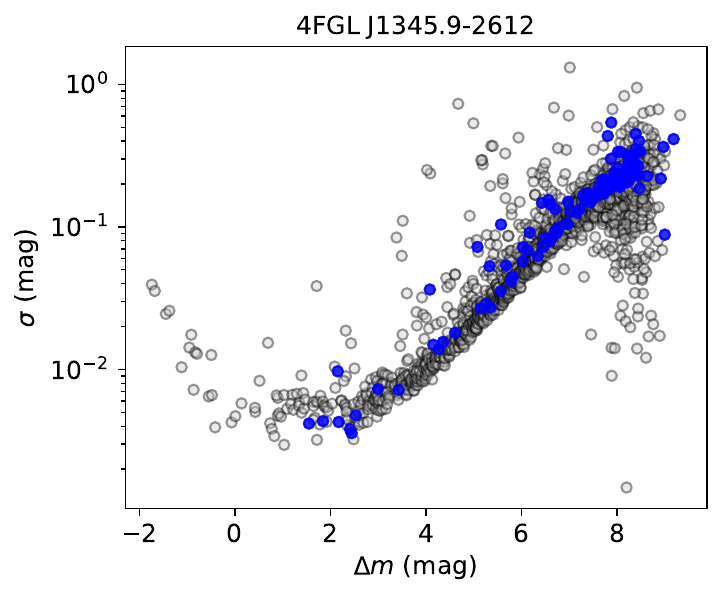}{0.48\textwidth}{}
          \leftfig{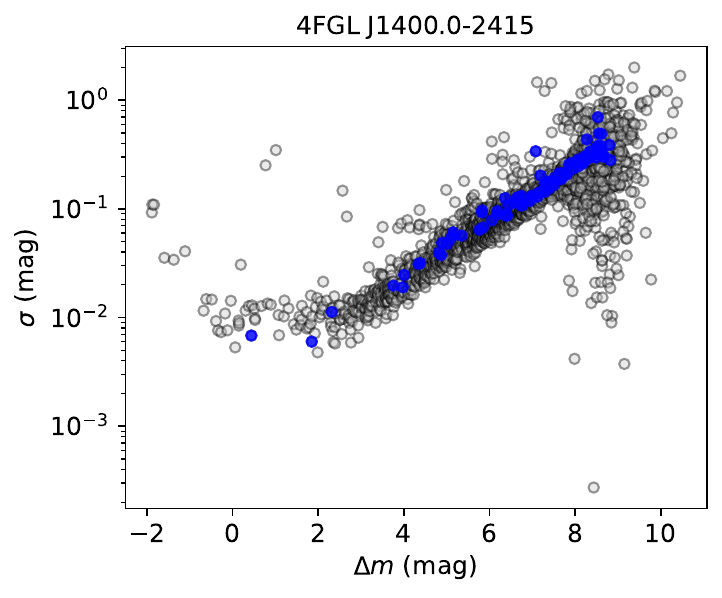}{0.48\textwidth}{}
          }
\gridline{\leftfig{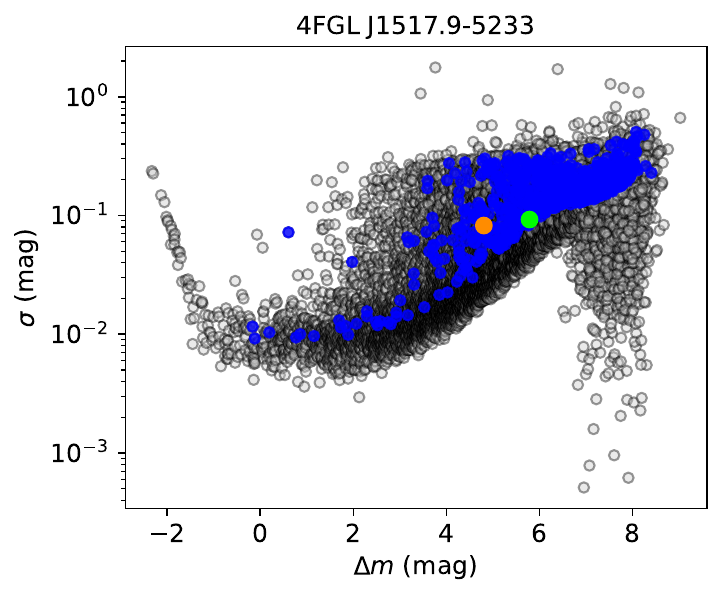}{0.48\textwidth}{}
          \leftfig{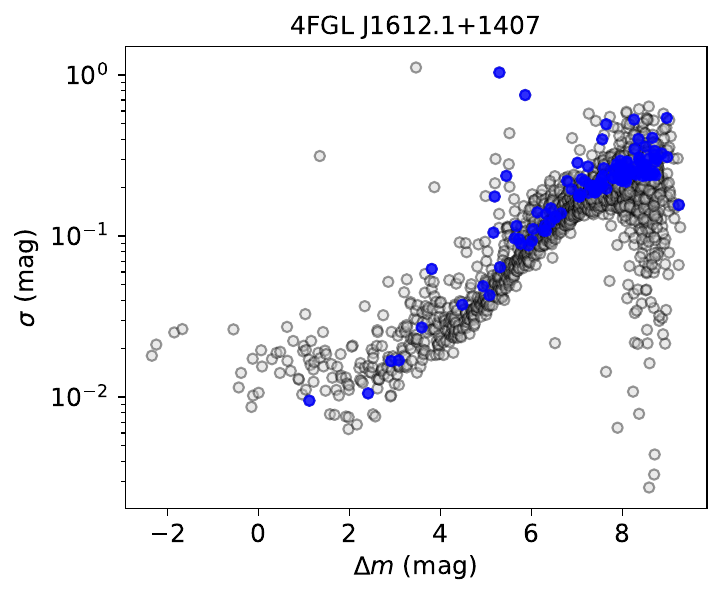}{0.48\textwidth}{}
          }
\caption{Continued.}
\label{fig:sigmavsdm_cont1}
\end{figure*}
\begin{figure*}[ht!]
\gridline{\leftfig{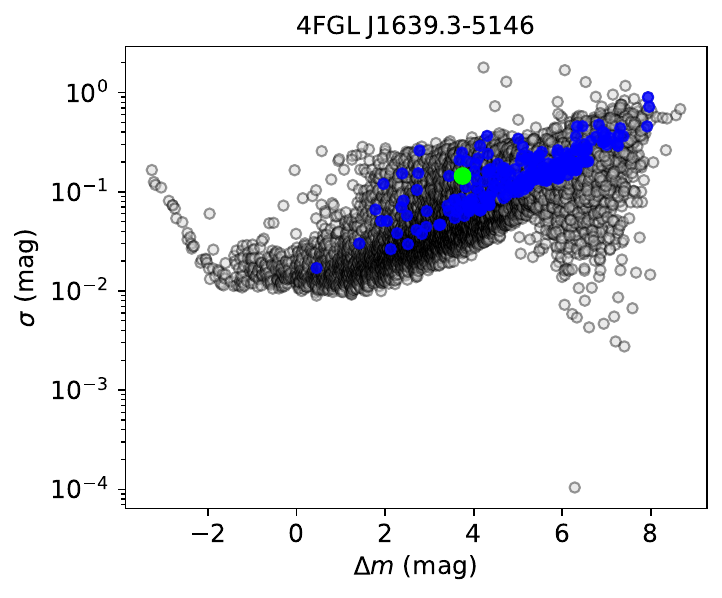}{0.48\textwidth}{}
          \leftfig{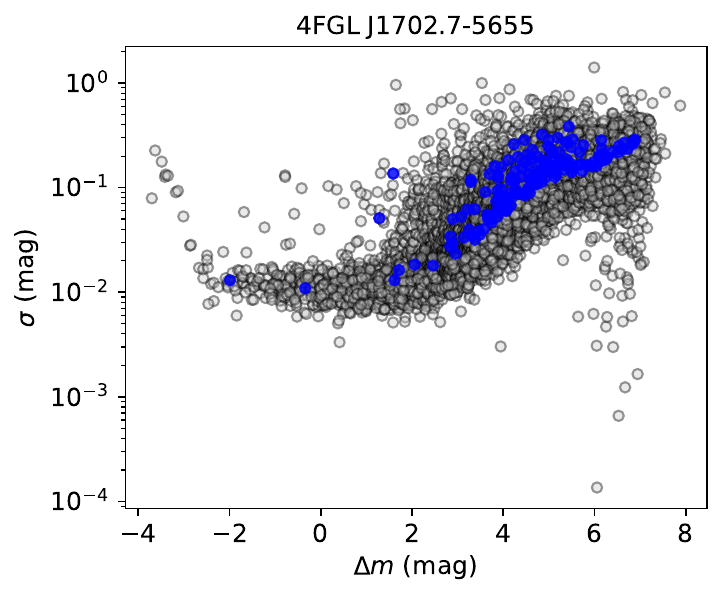}{0.48\textwidth}{}
          }
\gridline{\leftfig{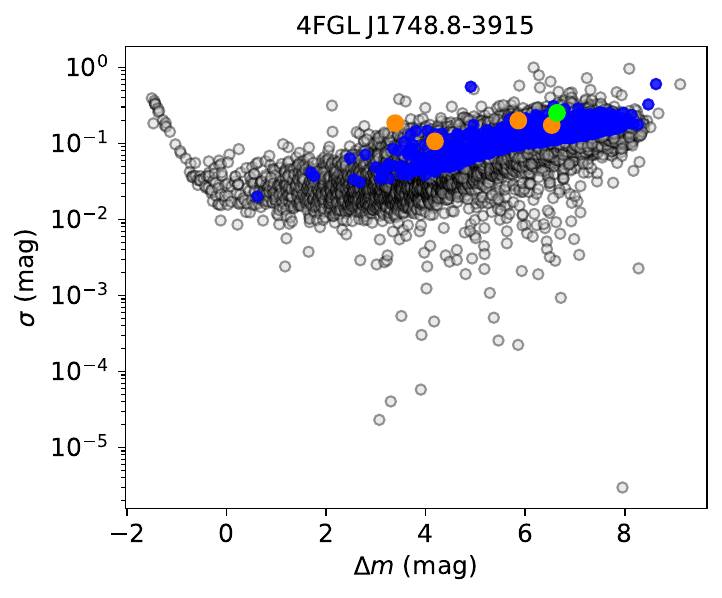}{0.48\textwidth}{}
          \leftfig{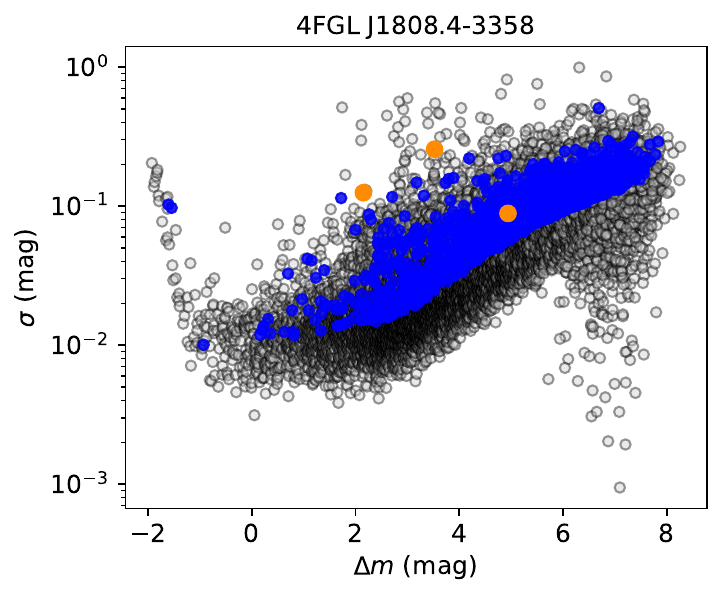}{0.48\textwidth}{}
          }
\gridline{\leftfig{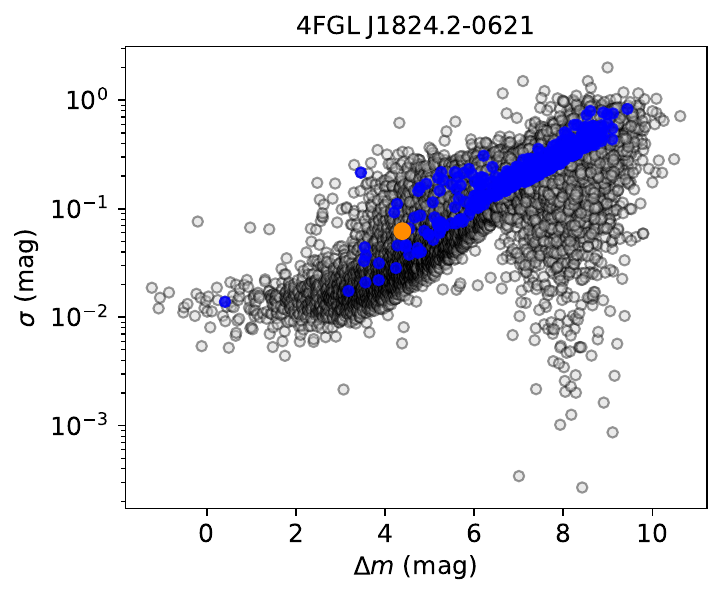}{0.48\textwidth}{}
          \leftfig{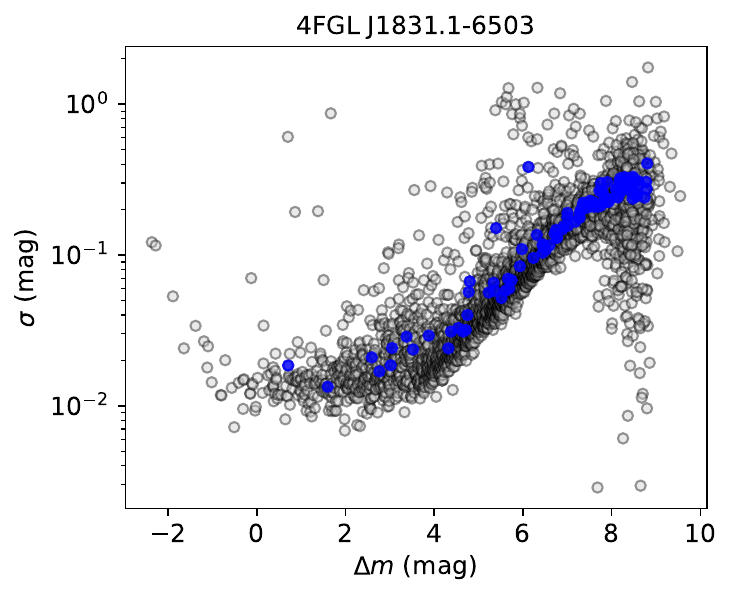}{0.48\textwidth}{}
          }
\caption{Continued.}
\label{fig:sigmavsdm_cont2}
\end{figure*}
\begin{figure*}[ht!]
\gridline{\leftfig{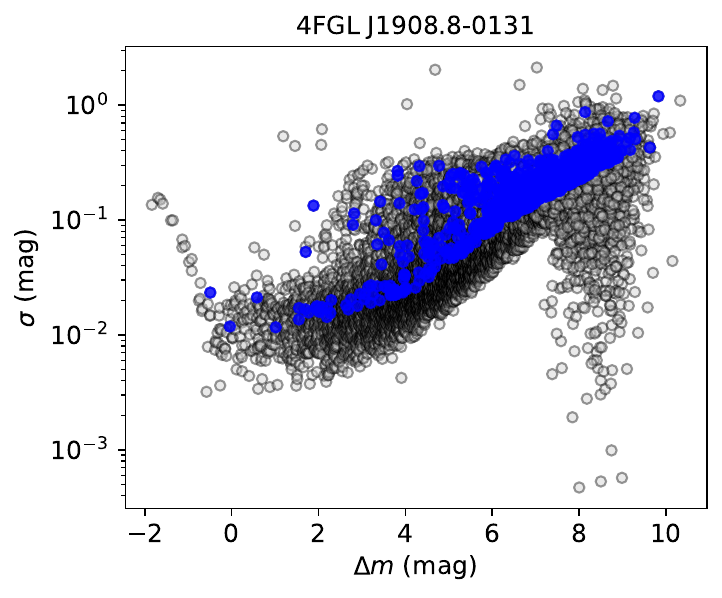}{0.48\textwidth}{}
          \leftfig{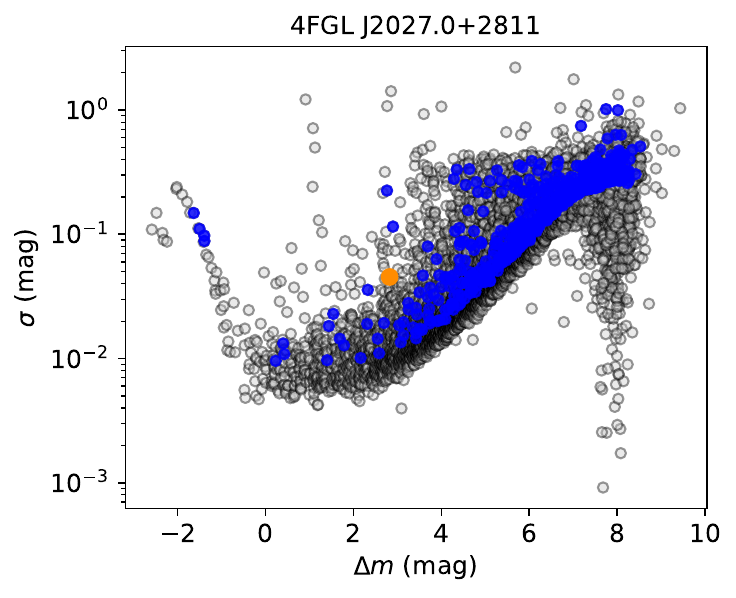}{0.48\textwidth}{}
          }
\gridline{\leftfig{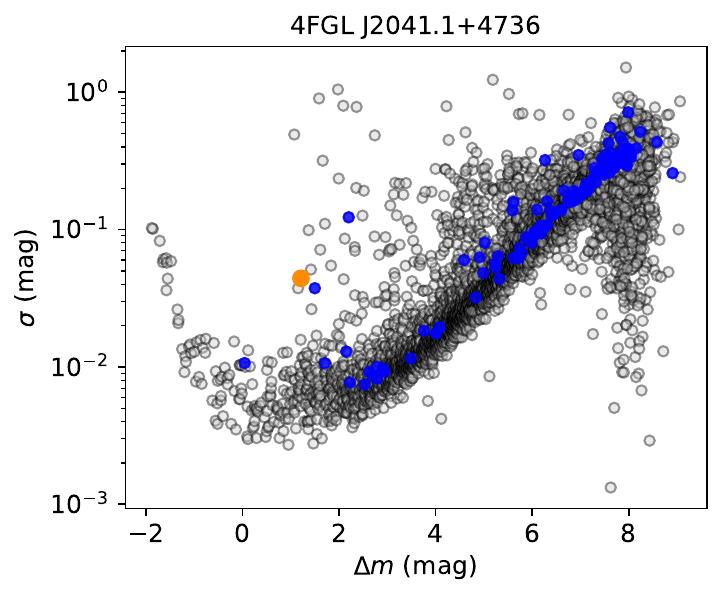}{0.48\textwidth}{}
          \leftfig{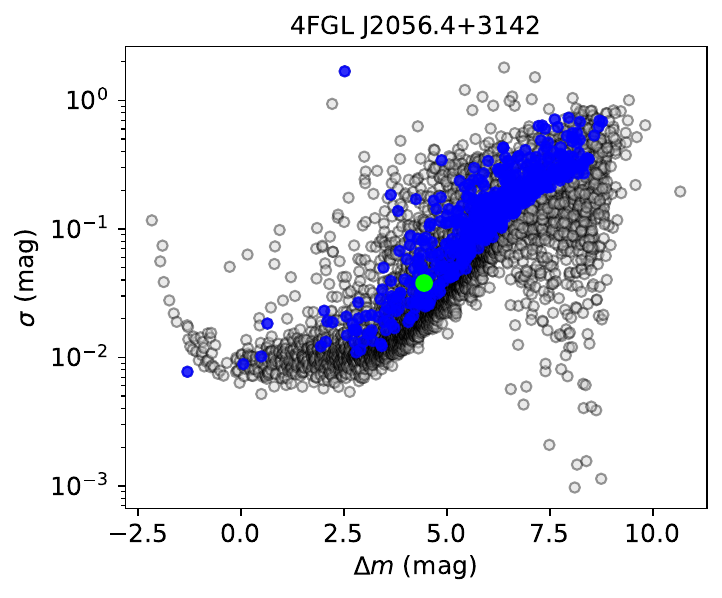}{0.48\textwidth}{}
          }
\gridline{\leftfig{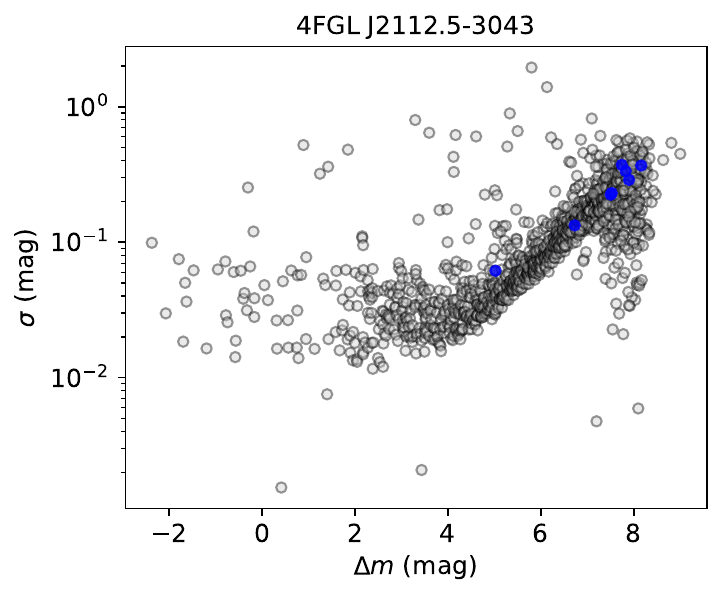}{0.48\textwidth}{}
          \leftfig{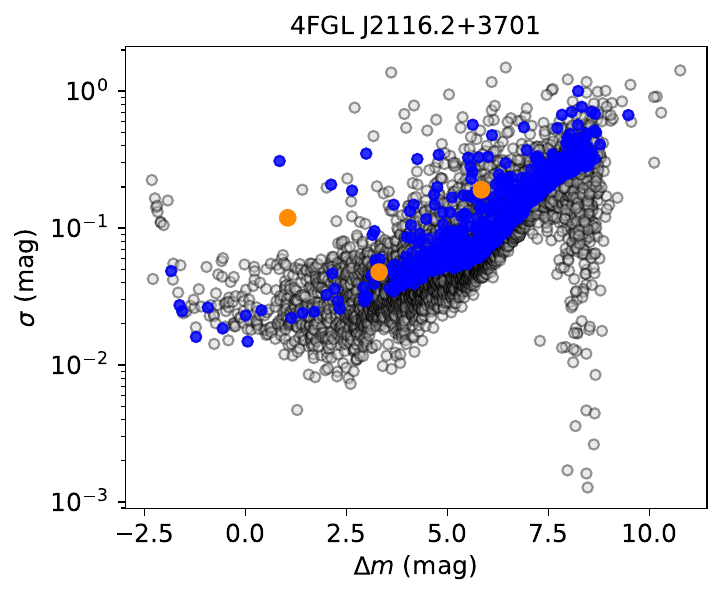}{0.48\textwidth}{}
          }
\caption{Continued.}
\label{fig:sigmavsdm_cont3}
\end{figure*}
\begin{figure*}[ht!]
\gridline{\leftfig{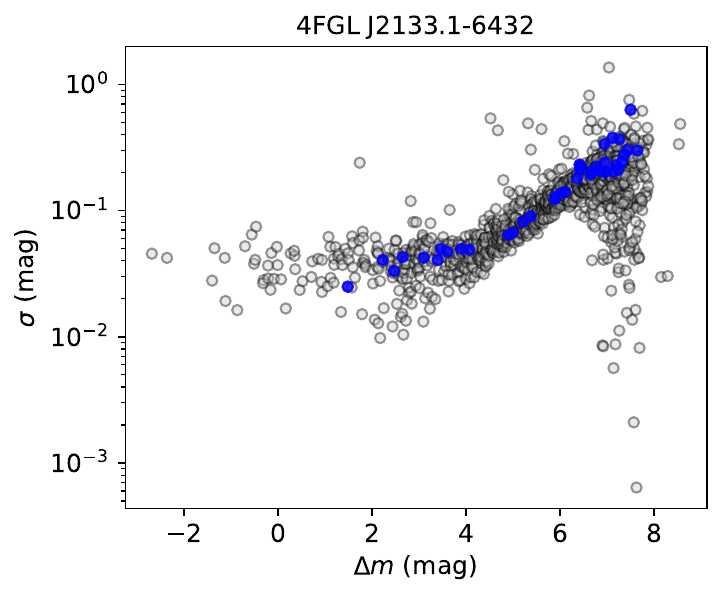}{0.48\textwidth}{}
          \leftfig{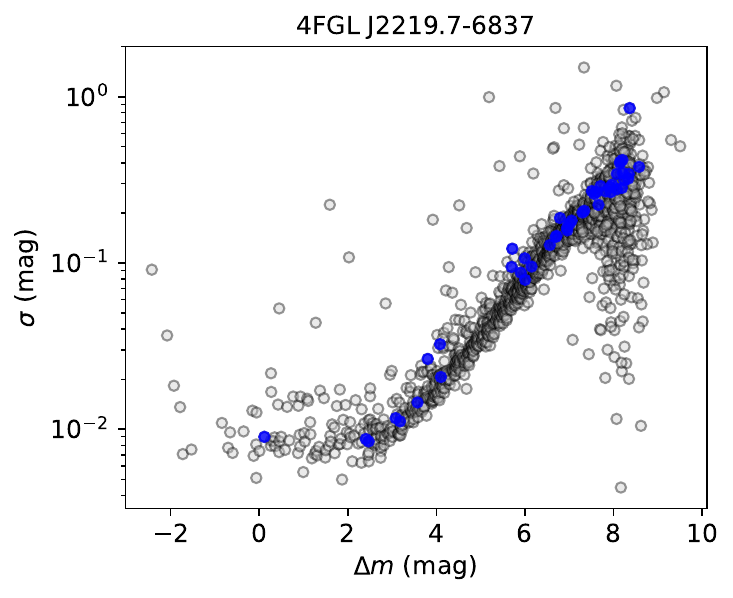}{0.48\textwidth}{}
          }
\gridline{\leftfig{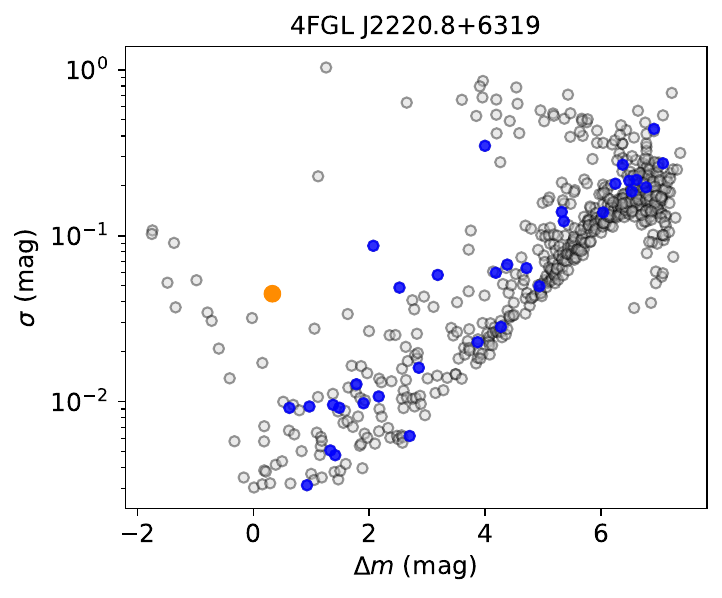}{0.48\textwidth}{}
          \leftfig{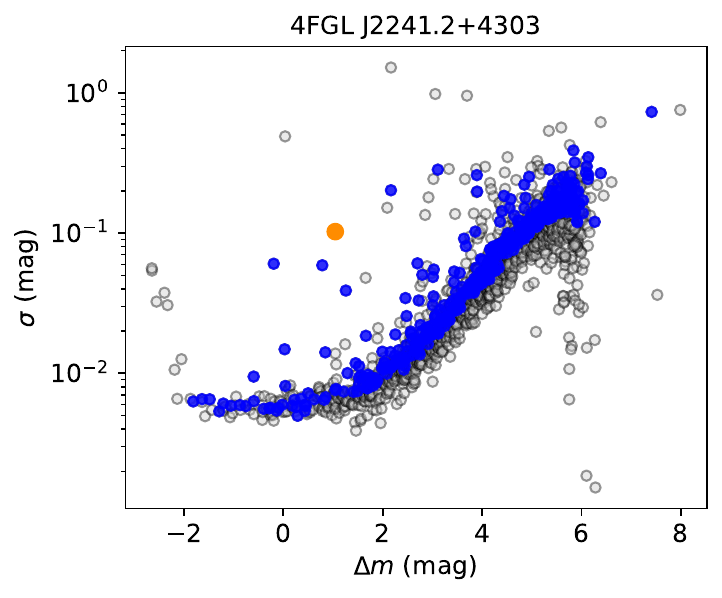}{0.48\textwidth}{}
          }
\gridline{\leftfig{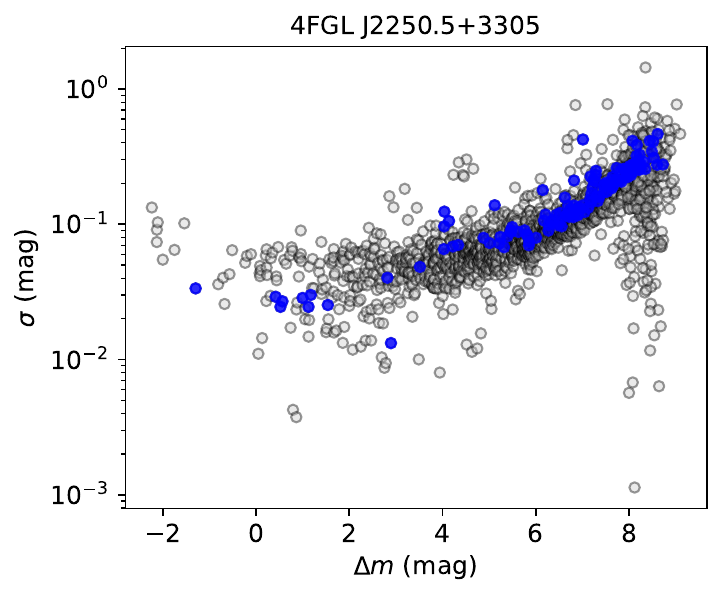}{0.48\textwidth}{}
          }
\caption{Continued.}
\label{fig:sigmavsdm_cont4}
\end{figure*}

Figures~\ref{fig:sigmavsdm}--\ref{fig:sigmavsdm_cont4} show the plots of light curve standard deviation ($\sigma$) versus differential magnitude ($\Delta m$) for all optical sources detected in each of the 29 COBIPLANE fields of view observed with STELLA/WiFSIP and LCO/Sinistro. We indicate in blue all the ``photometric variables" (see Section \ref{subsec:variablesandperiods}),
in orange the optical periodic variables not associated to spiders, and in green the variables classified as spider candidates.

\section{Periodic variables} \label{sec:appC}
\restartappendixnumbering 
\begin{table*}
\raggedright
    \caption{Optical location, best estimate of the photometric period from our analysis (with uncertainty in brackets), and classification with corresponding period measurement from external catalogs for each periodic variable.}
    \setlength{\tabcolsep}{3.0pt}
    \begin{tabular}{lcccccc}
    \hline\hline
        Name & R.A. (J2000) & Decl. (J2000) & Error radius & Photometric period & Catalog ID\tablenotemark{a} & Catalog period\\
       4FGL & (h:m:s) & ($^{\circ}$: $'$: $''$) & ($''$) & (d) & (ID, CLASS) & (d)\\ 
        \hline
    \hline
        J0003.6$+$3059-A & 00:03:20.8 & $+$30:56:14 & 1.3 & 0.2727(5) & \textit{ATO} J000.8365$+$30.9372, CBH & 0.272664\\
        J0235.3$+$5650-A & 02:35:26.56 & $+$56:53:57.9 & 0.7 & 0.1758(5) & \textit{ATO} J038.8607$+$56.8994, PULSE & 0.173232\\
        J0754.9$-$3953-A & 07:54:54.04 & $-$39:55:05.1 & 0.8 & 0.4292(5) & \textit{Gaia} 5537441892385539840, ECL & 0.597480\\
        J0754.9$-$3953-B & 07:55:09.40 & $-$39:53:49.8 & 0.8 & 0.1849(5) & \textit{Gaia} 5537430248742509696, DGS & 0.185593\\
        J0906.8$-$2122-A & 09:06:54.92 & $-$21:16:35.0 & 0.9 & 0.8314(5) & \textit{ATO} J136.7288$-$21.2765, IRR & 0.828152\\
        J1517.9$-$5233-A & 15:18:01.75 & $-$52:32:20.6 & 0.6 & 0.4069(5) & \textit{Gaia} 5888147073336947072, ECL & 0.407438\\
        J1748.8$-$3915-A & 17:48:39.9 & $-$39:13:25 & 1.0 & 0.2965(5) & \textit{Gaia} 5958233998924110080, ECL & 0.592856\\
        J1748.8$-$3915-B & 17:48:33.1 & $-$39:14:51 & 1.0 & 0.3686(5) & \textit{Gaia} 5958232482763347712, RR & 0.370129\\
        J1748.8$-$3915-C & 17:48:50.9 & $-$39:15:44 & 1.0 & 0.2856(5) & \textit{Gaia} 5958230700388895488, ECL & 0.286250\\
        J1748.8$-$3915-D & 17:48:38.2 & $-$39:17:22 & 1.0 & 0.2979(5) & \textit{Gaia} 5958229149892560256, ECL & 0.411052\\
        J1808.4$-$3358-A & 18:08:12.89 & $-$34:01:32.5 & 0.7 & 0.3660(5) & \textit{Gaia} 4039558372734596864, RR & 0.577096\\
        J1808.4$-$3358-B & 18:08:13.70 & $-$34:01:24.2 & 0.7 & 0.2965(5) & \textit{Gaia} 4039558372734609152, RR & 0.477768\\
        J1808.4$-$3358-C & 18:08:34.77 & $-$33:57:02.6 & 0.7 & 0.2127(5) & \textit{Gaia} 4039560404327618560, ECL & 0.291034\\
        J1824.2$-$0621-A & 18:24:13.29 & $-$06:20:21.6 & 0.8 & 0.3566(5) & \textit{ATO} J276.0553$-$06.3393, dubious & 0.356590\\
        J2027.0$+$2811-A & 20:26:52.44 & $+$28:08:38.5 & 0.9 & 0.8326(5) & \textit{ATO} J306.7184$+$28.1440, DBF & 0.832998\\
        J2041.1$+$4736-A & 20:41:01.51 & $+$47:35:33.9 & 0.7 & 0.9442(5) & \textit{ATO} J310.2562$+$47.5926, CBF & 0.944338\\
        J2116.2$+$3701-A & 21:16:05.3 & $+$36:58:15 & 1.1 & 1.2482(5) & \textit{ATO} J319.0219$+$36.9709, dubious & 2.494694\\
        J2116.2$+$3701-B & 21:16:35.8 & $+$37:00:20 & 1.1 & 0.3526(5) & \textit{ATO} J319.1490$+$37.0054, NSINE & 0.352593\\
        J2116.2$+$3701-C & 21:16:00.3 & $+$37:01:26 & 1.1 & 0.2821(5) & \textit{Gaia} 1868610081445474048, ECL & 0.283122\\
        J2220.8$+$6319-A & 22:20:29.04 & $+$63:21:51.7 & 0.8 & 0.7275(5) & \textit{Gaia} 2205445880926964224, ECL & 0.727484\\
        J2241.2$+$4303-A & 22:41:12.0 & $+$43:03:55 & 1.2 & 0.3762(5) & \textit{ATO} J340.2998$+$43.0652, CBF & 0.376172\\
        \hline
    \end{tabular}
    \tablenotetext{a}{ID and classification of the periodic variables as found in catalogs (refer to the main text for details).
    }
    \tablecomments{Classes of optical variables attributed by the \textit{ATLAS} catalog:
     \\
     \textbf{CBF}=close binary, full period identified, contact or near-contact eclipsing binary star;
     \\
     \textbf{CBH}=close binary, half period identified, contact or near-contact eclipsing binary star;
     \\
     \textbf{DBF}=distant binary, full period identified, detached eclipsing binary star;
     \\
     \textbf{PULSE}=pulsating star showing the classic sawtooth light curve, identified as RR Lyrae, $\delta$ Scuti stars, or Cepheids;
     \\
     \textbf{NSINE}=sinusoidal variables with much residual noise or with evidence of additional variability not captured in the fit. Many spotted rotators with evolving spots likely fall into this class.
     \\
     \textbf{IRR}=``irregular" variable, this class serves as "catch-all" bins for objects that do not seem to fit into any specific category;
     \\
     \textbf{dubious}=star might not be a real variable.
     \\
     Classes of optical variables attributed by the \textit{Gaia} DR3 catalog:
     \\
     \textbf{ECL}=eclipsing binary star of type beta Persei (Algol).
     \\
     \textbf{DGS}=Set of variable types: delta Scuti, gamma Doradus, and SX Phoenicis.
     \\
     \textbf{RR}=RR Lyrae stars of the following types: fundamental-mode, first-overtone, double mode and anomalous double mode.
     }
    \label{tab:periodicresults}
\end{table*}

\begin{figure*}[ht!]
\gridline{\leftfig{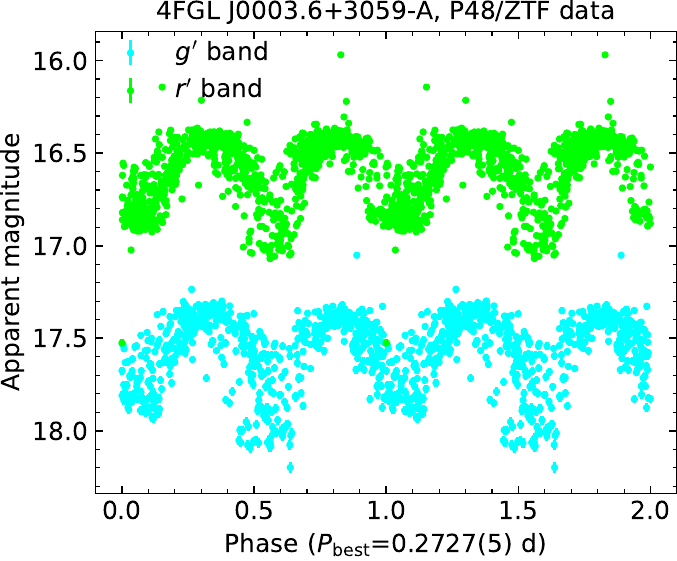}{0.46\textwidth}{}
          \leftfig{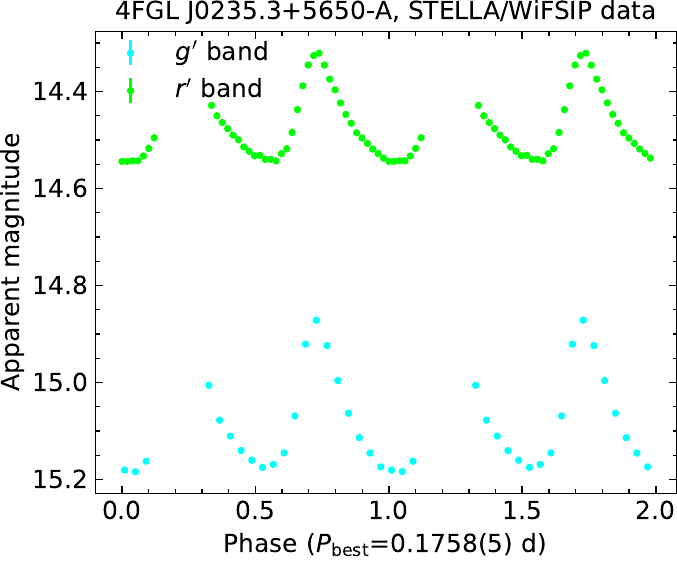}{0.46\textwidth}{}
          }
\gridline{\leftfig{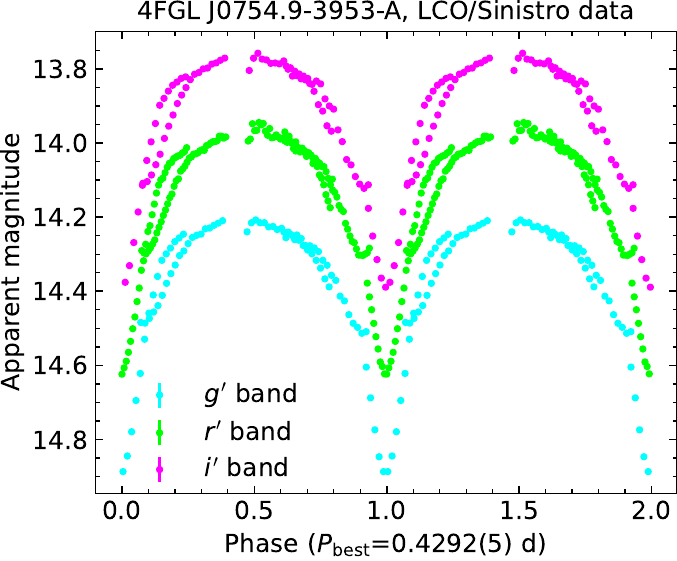}{0.46\textwidth}{}
          \leftfig{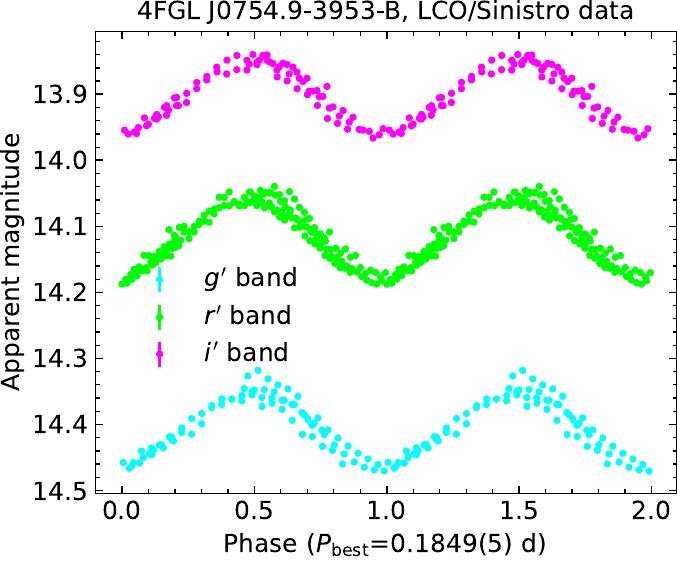}{0.46\textwidth}{}
          }
\gridline{\leftfig{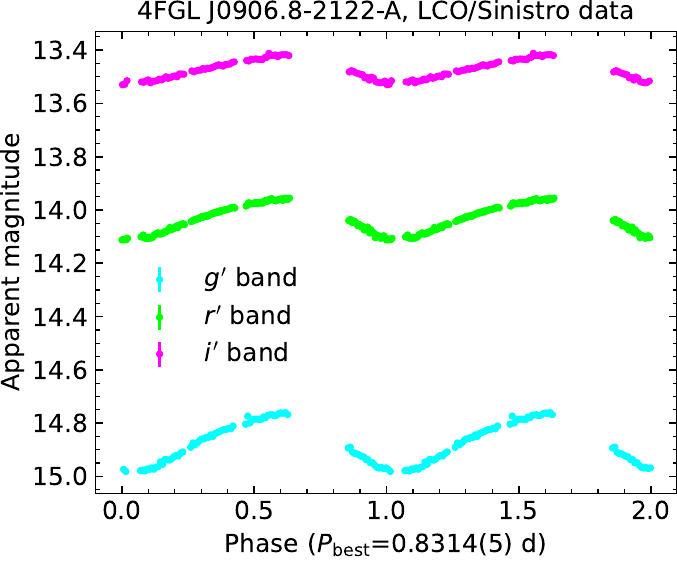}{0.46\textwidth}{}
          \leftfig{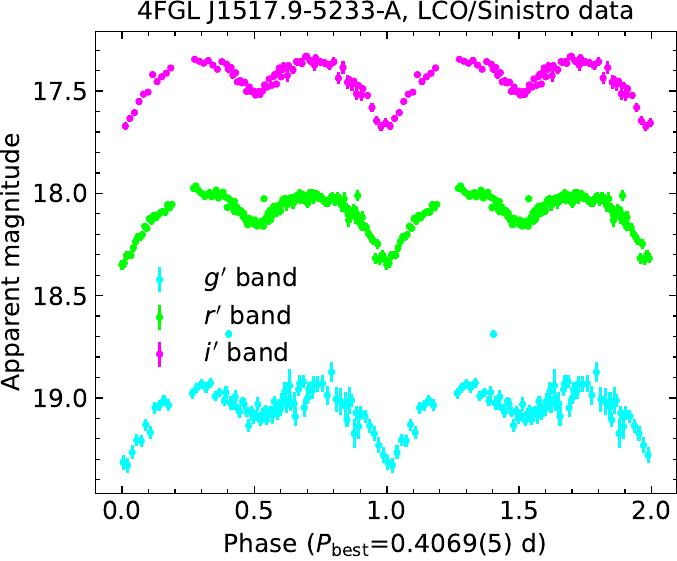}{0.46\textwidth}{}
          }
\caption{Phase-folded optical light curves for each of the 21 COBIPLANE periodic variables.}
\label{fig:pervarlightcurves}
\end{figure*}
\begin{figure*}[ht!]
\gridline{\leftfig{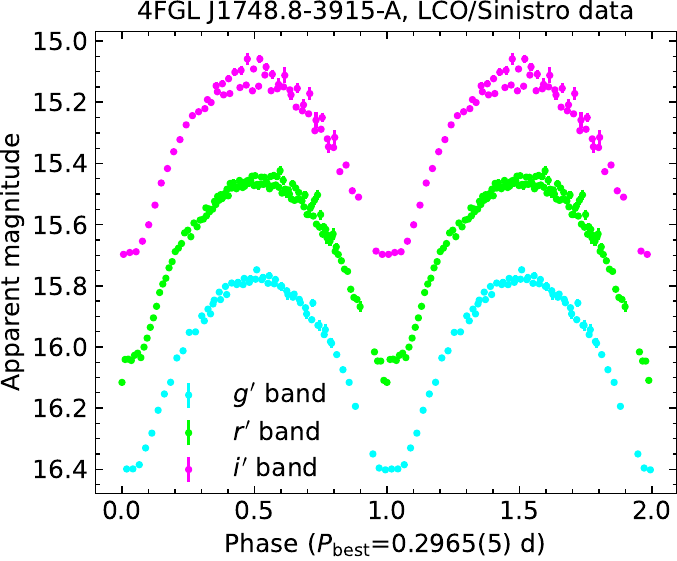}{0.46\textwidth}{}
          \leftfig{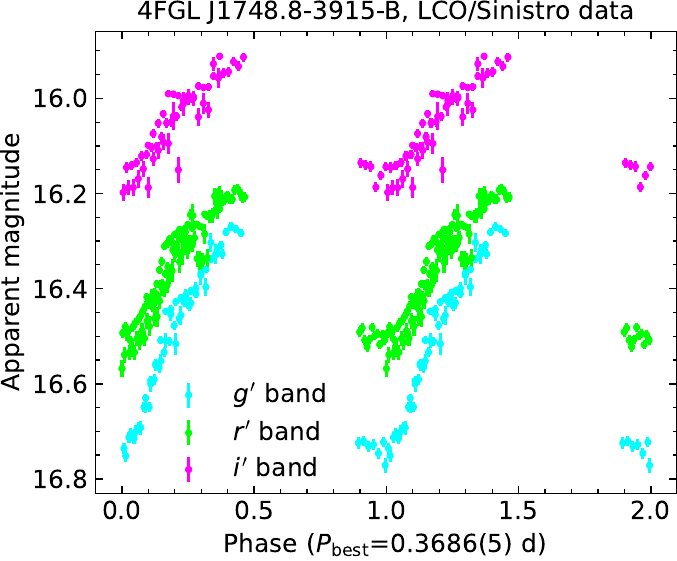}{0.46\textwidth}{}
          }
\gridline{\leftfig{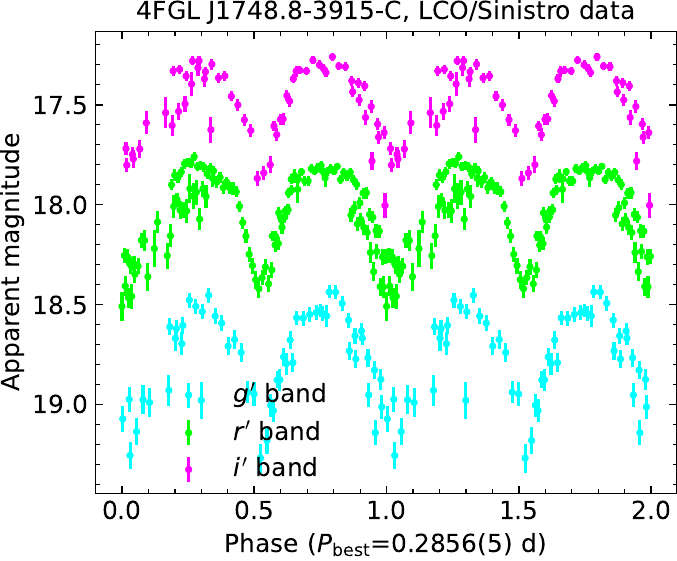}{0.46\textwidth}{}
          \leftfig{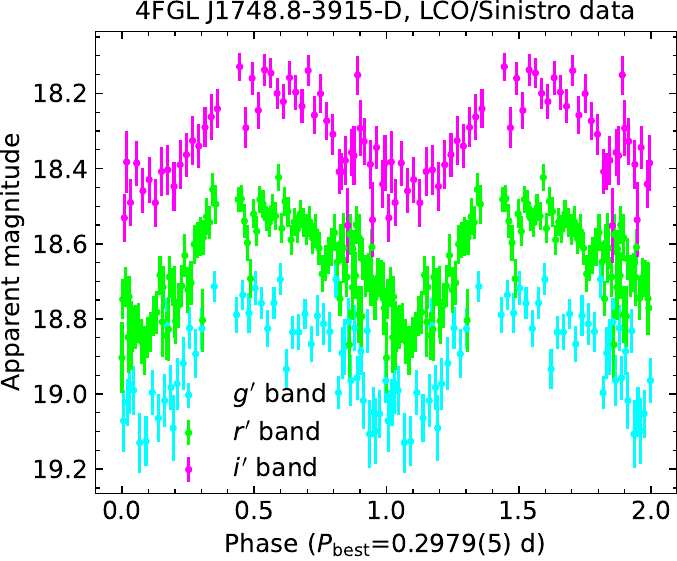}{0.46\textwidth}{}
          }
\gridline{\leftfig{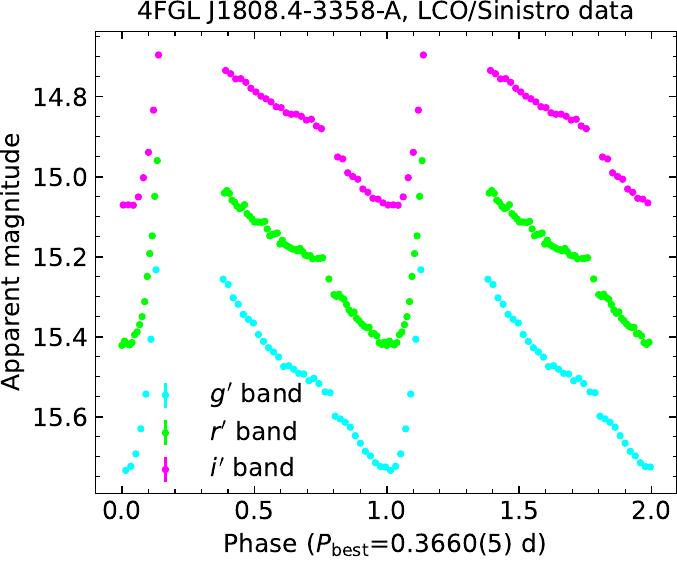}{0.46\textwidth}{}
          \leftfig{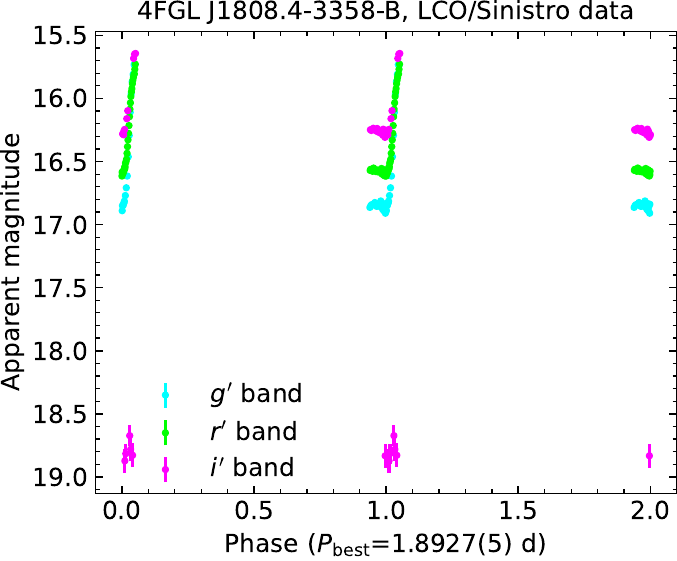}{0.46\textwidth}{}
          }
\caption{Continued.}
\label{fig:pervarlightcurves_cont1}
\end{figure*}
\begin{figure*}[ht!]
\gridline{\leftfig{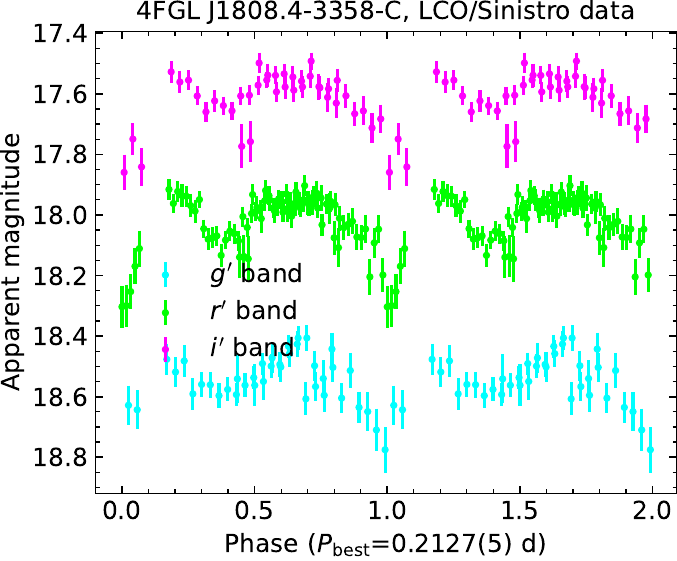}{0.46\textwidth}{}
          \leftfig{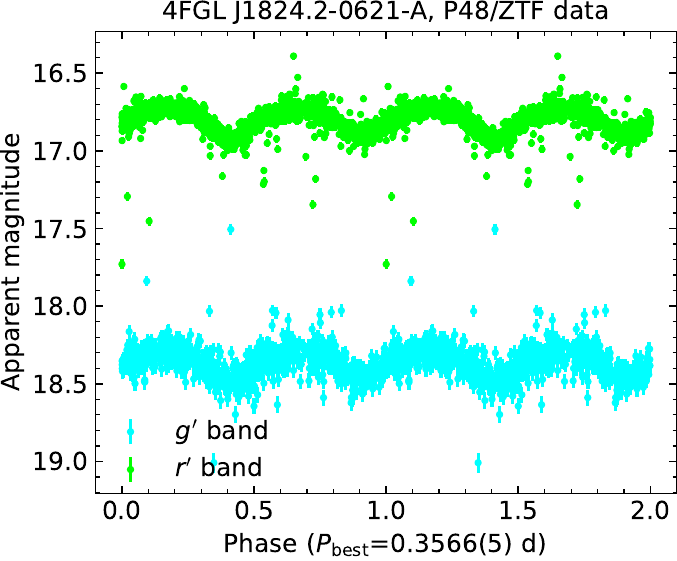}{0.46\textwidth}{}
          }
\gridline{\leftfig{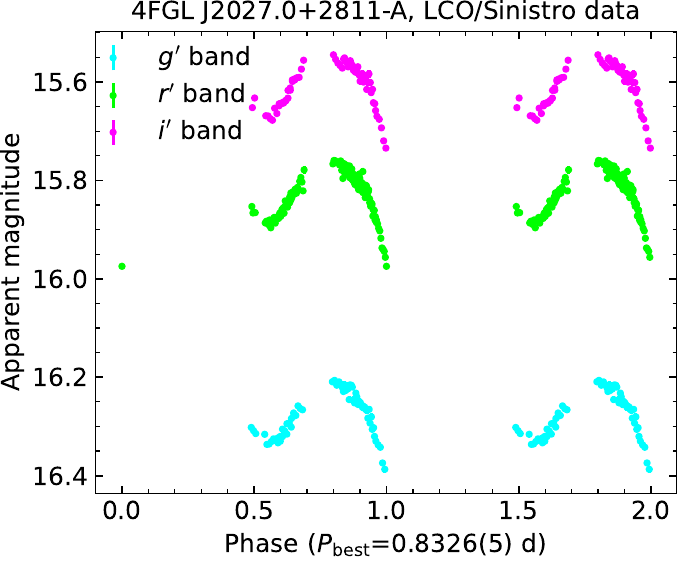}{0.46\textwidth}{}
          \leftfig{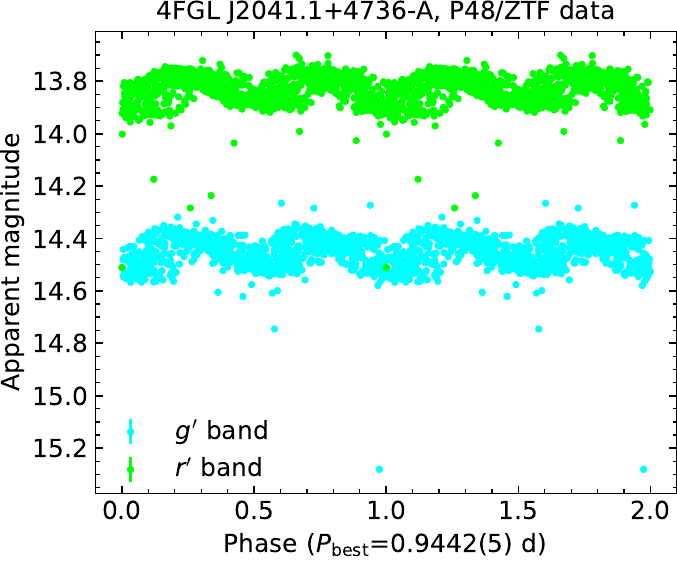}{0.46\textwidth}{}
          }
\gridline{\leftfig{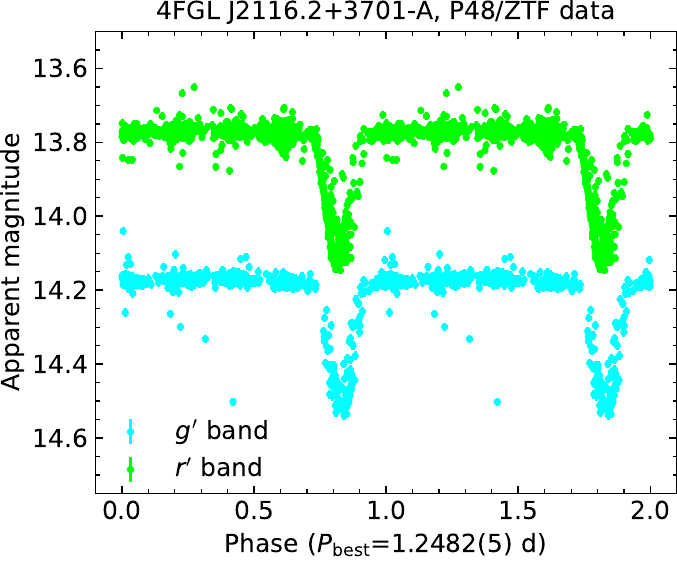}{0.46\textwidth}{}
          \leftfig{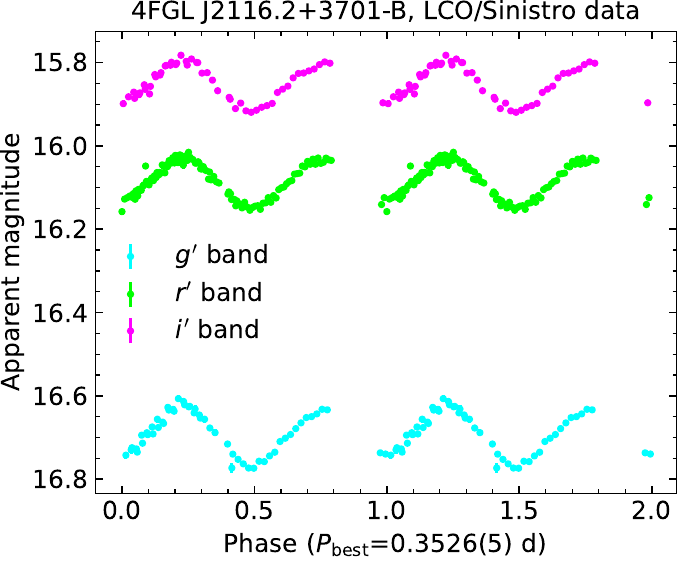}{0.46\textwidth}{}
          }
\caption{Continued.}
\label{fig:pervarlightcurves_cont2}
\end{figure*}
\begin{figure*}[ht!]
\gridline{\leftfig{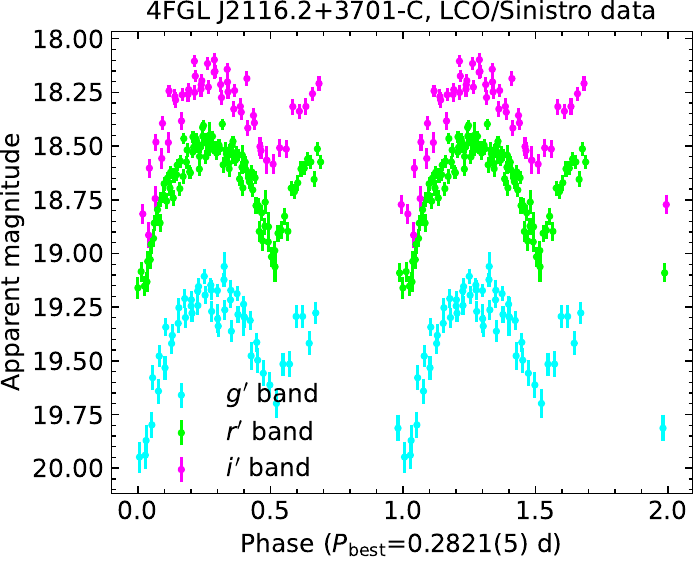}{0.46\textwidth}{}
          \leftfig{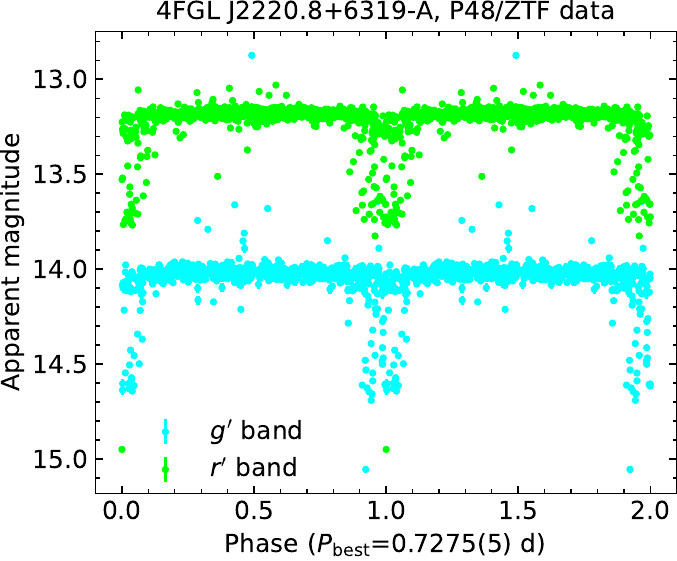}{0.46\textwidth}{}
          }
\gridline{\leftfig{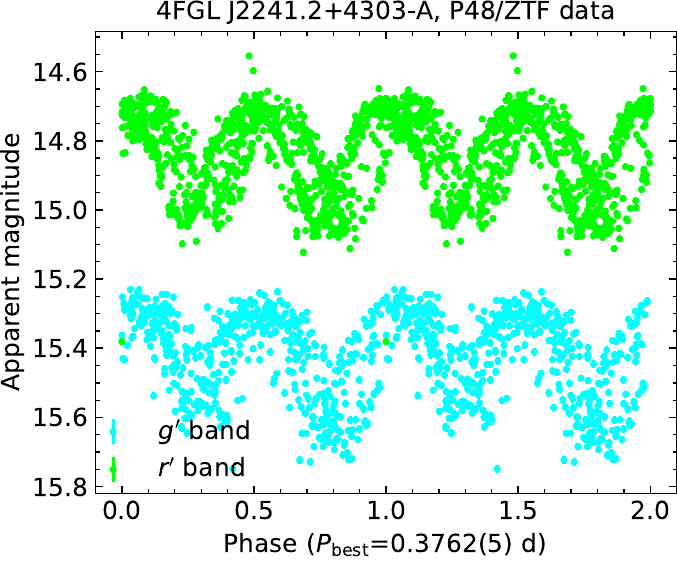}{0.46\textwidth}{}
         }
\caption{Continued.}
\label{fig:pervarlightcurves_cont3}
\end{figure*}

Table \ref{tab:periodicresults} lists the optical locations and photometric periods estimated for the 21 periodic variables identified in COBIPLANE. For comparison, we also include the periods measured from other variable catalogs, if any previous identification was present from \textit{ATLAS} \citep{2018AJ....156..241H} or \textit{Gaia} DR3 \citep{2022gdr3.reptE..10R}. We also show in Figure \ref{fig:pervarlightcurves}--\ref{fig:pervarlightcurves_cont3} the optical light curves for each periodic variable, phase-folded with the photometric period found either from our data (STELLA or LCO) or from ZTF, in case this was needed to cover a full orbital cycle of the system.

\end{document}